\pdfoutput=1

\documentclass[12pt,a4paper]{report}

\usepackage{cite}
\usepackage{fancyvrb} 
\usepackage{graphicx} 
\usepackage{makeidx}
\usepackage{url} 
\usepackage{listings}
\usepackage{amsmath}
\usepackage{subfiles}
\usepackage{paralist}
\usepackage{paralist}
\usepackage{listings}
\usepackage{setspace}
\usepackage{pdfpages}
\usepackage[T1]{fontenc}
\usepackage[utf8x]{inputenc}
\usepackage[linesnumbered,ruled,vlined,boxed]{algorithm2e}
\usepackage[noend]{algpseudocode}
\usepackage{graphicx}
\usepackage{caption}
\usepackage[noend]{algpseudocode}
\usepackage{bm}

\renewcommand{\and}{\hspace{.4cm}}

\fvset{baselinestretch=1}

\lstdefinestyle{mystyle}{
    basicstyle=\footnotesize\singlespacing,
    breakatwhitespace=false,
    breaklines=true,
    captionpos=b,
    keepspaces=true,
    numbers=left,
    numbersep=5pt,
showspaces=false,
    showstringspaces=false,
    showtabs=false,
    tabsize=1
}

\begin{document}


\begin{titlepage}
\begin{centering}
\begin{LARGE}
\textbf{Augmented Understanding and Automated Adaptation of Curation Rules}\par
\end{LARGE}

\vspace{10mm}
\begin{large}
\textbf{Alireza Tabebordbar}\\
\vfill
\end{large}
\vspace{15mm}
A thesis in fulfilment of the requirements for the degree of\\
\vspace{2mm}
\textbf{\large Doctor of Philosophy}
\par
\vspace{30mm}
\centering
\includegraphics[scale=0.7]{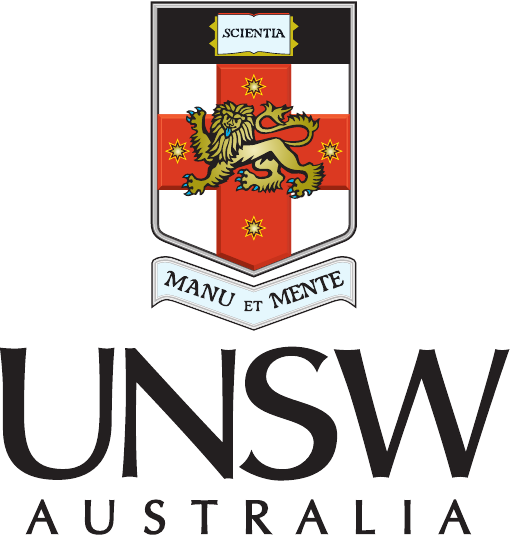}
\vspace{5mm}
\par
School of Computer Science and Engineering\\
Faculty of Engineering
\par
\par
\vspace{10mm}
March 2020
\par
\end{centering}

\end{titlepage}


\chapter*{Acknowledgements}
\addcontentsline{toc}{chapter}{Acknowledgements}
Firstly, I would like to express my special thanks to my Ph.D. supervisor Dr.~Amin~Beheshti. Amin was not only a knowledgeable and an expert scientist in the field of data science and Artificial Intelligence, but also supportive, loyal, honest, trustworthy, and a true friend. Amin is a credible and effortless research academic, who supported me throughout my study and help my growth as a Ph.D. research student. Thank you for all your supports and comments, and I really enjoyed working with you during these years.
\newline

I would like to express my appreciation to my supervisor, Prof. Boualem Benatallah, who is a passionate scientist, and an excellent forward thinker. I gained valuable insight from his comments during the last three years.\newline

I gratefully thank my co-supervisor, Dr. Hamid Reza Motahari-Nezhad, for his insightful comments on my study. Hamid is an excellent and inspiring scientist and I really appreciated the opportunity to have your suggestions during my study.\newline

I like to also express my sincere appreciation to UNSW workers, especially ICT for providing equipment to facilitate my research.\newline

I would like to thank my sponsor, Data to Decisions Cooperative Research Centre (D2D CRC), for funding my study during the last three and half years.\newline

I would like to thank Reza Nouri for his technical support and the configurations he has made for running my codes.\newline

I would like to appreciate the UNSW learning centre for providing advanced academic writing courses and helping me to improve my writing skills.\newline

\chapter*{Abstract}
\addcontentsline{toc}{chapter}{Abstract}

 Over the past years, there has been many efforts to curate and increase the added value of the raw data. Data curation has been defined as activities and processes an analyst undertakes to transform the raw data into contextualized data and knowledge. Data curation enables decision-makers and data analyst to extract value and derive insight from the raw data. However, to curate the raw data, an analyst needs to carry out various curation tasks including, extraction linking, classification, and indexing, which are error-prone, tedious and challenging. Besides, deriving insight require analysts to spend a long period of time to scan and analyze the curation environments. This problem is exacerbated when the curation environment is large, and the analyst needs to curate a varied and comprehensive list of data. To address these challenges, in this dissertation, we present techniques, algorithms and systems for augmenting analysts in curation tasks. We propose: ~(1) a feature-based and automated technique for curating the raw data. ~(2) We propose an autonomic approach for adapting data curation rules. ~(3) We provide a solution to augment users in formulating their preferences while curating data in large scale information spaces. ~(4) We implement a set of APIs for automating the basic curation tasks, including Named Entity extraction, POS tags, classification, and etc.

In this dissertation, we automate many of tedious and time-consuming curation tasks and creates a Knowledge Lake (i.e., contextualized data lake) to augment analysts in deriving insight and extracting value. We assist analysts to adapt data curation rules in dynamic curation environments. Our solution, autonomic-ally learns the optimal modification for rules using an online learning algorithm. We present a novel approach for augmenting user comprehension of curation environments. We explain techniques for formulating user preferences in large and varied environments. We discuss how summarization techniques help users to understand curation environments without scanning and synthesizing a large amount of data. We present a system, which allows users to retrieve their information using a set of high-level concepts such as persons, locations, and topics.

We conduct different experiments to highlight the applicability of our solutions: (1)~We discuss how our proposed feature-based approach significantly enhances users in curating data and extraction of knowledge. We study both scalability and precision of our approach in curating social data. (2)~We show how our solution can learn to curate data without needing analysts. We present the performance of our adaptation technique in adapting curation rules. We compare our results with systems relying on analysts and compare the precision and recall of our solution with analysts. 
(3)~We introduced our system, namely ConceptMap, which aids users to comprehend the information space without constantly scanning or querying the information space. Our results show ConceptMap can significantly lower the user's workload in understanding a curation environment and extracting value. Our results prove that ConceptMap can significantly lower the user's workload and time in understanding the data.
 
\chapter*{Publications}
\addcontentsline{toc}{chapter}{Publications}

\begin{itemize}
    \item A Tabebordbar, A Beheshti, B Benatallah, and M C Barukh, \textbf{Adaptive rule adaptation in unstructured and dynamic environments}, International Conference on Web Information Systems Engineering, Springer, 2019, pp. 326–340.
    
    \item A Tabebordbar, A Beheshti, and B Benatallah, \textbf{Conceptmap: A conceptual approach for formulating user preferences in large information spaces}, International Conference on Web Information Systems Engineering, Springer, 2019, pp. 779–794. (Selected as the top five paper among 250 submissions)
    
    \item A Tabebordbar and A Beheshti, \textbf{Adaptive rule monitoring system}, 2018 IEEE/ACM 1st International Workshop on Software Engineering for Cognitive Services (SE4COG), IEEE, 2018, pp. 45–51 (Best paper award).
    
     \item A Tabebordbar, A Beheshti, B Benatallah, and M C Barukh, \textbf{Feature-based Rule Adaptation in Unstructured and Dynamic Environments}, Data Science and Engineering (DSE) Journal (2020).
     
    \item  A Tabebordbar, A Beheshti, B Benatallah, \textbf{Augmenting user's comprehension of curation environments using social exploratory search}.  \emph{World Wide Web Journal, 2020}, Accepted (minor revision).

    \item A Beheshti, A Tabebordbar, B Benatallah, and Reza Nouri, \textbf{On automating basic data curation tasks}, In companion proceedings of the 26th International Conference on World Wide Web (WWW), International World Wide Web Conferences Steering Committee, 2017, pp. 165–169.
    
    \item A Beheshti, B Benatallah, A Tabebordbar, H R Motahari-Nezhad, M C Barukh, and R Nouri, \textbf{Datasynapse: A social data curation foundry}, Distributed and Parallel Databases Journal (2018), 1–34.
    
    \item A Beheshti, A Tabebordbar, B Benatallah, \textbf{iStory: Intelligent Storytelling with Social Data}, In companion proceedings of the  International Conference on World Wide Web (Web) Conference, Taipei, 2020.
    
    \item A Beheshti, A Tabebordbar, B Benatallah, \textbf{Data curation APIs}, Tech. Report UNSWCSE-TR-201617, The University of New South Wales, Sydney, Australia, 2016.
    
    \item A Beheshti, K Vaghani, B Benatallah, and A Tabebordbar, \textbf{Crowdcorrect: a curation pipeline for social data cleansing and curation}, International Conference on Advanced Information Systems Engineering, Springer, 2018, pp. 24–38.
    
    \item A Beheshti, B Benatallah, R Nouri, and A Tabebordbar, \textbf{Corekg: a knowledge lake service}, Proceedings of the VLDB Endowment 11 (2018), no. 12, 1942–1945.

\end{itemize}

\tableofcontents

\chapter{Introduction}\label{Chapter1}

\section{Introduction, Background and Aims}
The expansion of Web, social media and sensors' data have made a deluge in the generation of the raw data. This data can be generated across various platforms and is available in different forms, from structured to unstructured, e.g., atomic data has not been processed for use. This availability of the raw data coupled with the continued improvement in capabilities of big data processing systems introduced a new era for deriving insight from the raw data. Data curation is a quintessential part of every big data processing system, which aims at transforming the raw data into contextualized data knowledge. 

Data curation may include processes and activities for principled and controlled data creation, maintenance, and management~\cite{cavanillas2016new}. Typically, a curation task consists of a set of mathematical, statistical, and computational models to help data curators in extracting actionable insight from the raw data~\cite{miller2014big}. This paradigm, often utilizes various big data processing sub-tasks, including machine learning algorithms (e.g., Bayesian and regression), enrichment (e.g., knowledge base and knowledge graph), annotation, summarization, and visualization. For example, consider a social media platform, e.g., Twitter~\cite{kwak2010twitter}, that enables users in expressing their opinions and receive feedback. A data curation system may analyze users' Tweets~\footnote{https://twitter.com/} to investigate their opinions about their community. The curation system may extract various information, e.g., keywords, part of speech, named entities, synonyms, and stems, from users' Tweets and link the extracted data to external knowledge bases to derive a deeper understanding of users' opinions regarding their communities~\cite{beheshti2018datasynapse}.   

Over the past years, different curation systems have been proposed to help organizations and data curators in transforming their raw data into knowledge. 
Trending applications include: improving government services~\cite{chen2012business,criado2013government}, predict intelligence activities~\cite{fader2014open,van2014datafication}, unravel human trafficking activities~\cite{burke2013introduction,allahbakhsh2012reputation,allahbakhsh2012analytic}, understand impact of news on stock markets~\cite{boudoukh2013news}, analysis of financial risks~\cite{ cont2011statistical,allahbakhsh2012detecting}, accelerate scientific discovery~\cite{towns2014xsede}, as well as to improve national security and public health~\cite{kamel2016instagram,heer2015predictive}. However, often to curate data, analysts~\footnote{In this dissertation, we use the term data curators and analysts interchangeably.} need to handle a large number of painstakingly difficult, error-prone, and time-consuming tasks. These challenges exacerbated in dynamic curation environments as curation algorithms typically fail to curate data and analysts need to continuously update their comprehension of curation environments to capture the salient aspect of data. Thus, in this dissertation, we focus on approaches for augmenting analysts in curating the data and augmenting their understanding of curation environments. 
 Overall, we can summarise our contributions as below:

\begin{enumerate}
    \item We propose an automated and feature-based framework for \emph{Extracting knowledge} from the raw data and developing insight. 
    \item We propose a learning algorithm for \emph{Adapting Data Curation Rules} in dynamic and constantly changing curation environments.
    \item We propose a system for augmenting user's \emph{Comprehension of Curation Environments} and lowering user's cognitive load in formulating her preferences.
\end{enumerate}

The rest of this chapter is organized as follows. We first introduce the central concepts discussed in this dissertation in Section~\ref{intro_premilinary}. Then, in Section~\ref{intro_issue}, we describe the key research issues tackled in this dissertation. Finally, we summarize our contributions in Section~\ref{intro_contribution}, and describe the organization of the dissertation in Section~\ref{intro-organize}.

\section{Preliminaries}
\label{intro_premilinary}

\subsection{Knowledge Extraction}

Data curation promotes contextualization of the raw data into knowledge by unravelling the hidden patterns and associations~\cite{miller2014big}. Data curation acts as a glue between the raw data and analysis and greatly assists analysts in interpreting the data and extracting value~\cite{beheshti2018datasynapse}.  

Curation of data starts with identifying open, social, and private data islands, and the processing elements that need to be used in the curation task. It divides each curation task into smaller sub-tasks and provides an end-to-end velocity by eliminating errors and diminishing bottlenecks and latency. A robust pipeline of curation tasks removes many barriers involved in curating data and provides a smooth, automated flow of data from one source to another. A data curation pipeline consists of various curation elements, including ingesting, cleansing, integration, transforming, and adding-value. In the followings, we briefly discuss different curation tasks that may involve in transforming the raw data into knowledge.
\begin{itemize}
    \item \textbf{Ingestion} is the process of obtaining data from different sources for immediate use and storage~\cite{dataingestion}. Data can be ingested as stream or batch. Stream processing systems capture data in real-time emitted from a source. While in batch processing systems data is imported in big chunks at a periodic interval. Examples of data ingestion systems are Apache Kafka~\footnote{https://kafka.apache.org/}, AirFlow~\footnote{https://airflow.apache.org/}, Amazon Kinesis~\footnote{https://aws.amazon.com/kinesis/}.
    
    \item \textbf{Cleansing} is the process of repairing or removing unwanted data from a dataset~\cite{datacleaning}. In many cases, data is incomplete, poorly formatted, or contain duplicated values. Data cleansing allows preparing data for processing by removing outliers.

    \item \textbf{Integration} aims at combining data from multiple sources into a central repository~\cite{mirza2016data}. The successful integration of data needs to address several challenges, including schema integration, detecting and resolving inconsistencies, removing duplicates and redundant values.
    
    
    \item \textbf{Transforming} aims at smoothing, summarising, generalizing, or normalizing the data. Transformation can remove noise from data and normalizes the data within a specified range, e.g., –1.0 to 1.0 or 0.0 to 1.0. 
    
    \item \textbf{Adding Value} focuses on deriving insight from data and consists of several activities, including:
    \begin{itemize}
        \item \textbf{Extraction} focuses on extracting actionable insight, e.g., named entities, part of speech, keywords, and synonym, from the raw data. Examples of extraction tools are, Stanford Core NLP~\cite{manning-EtAl:2014:P14-5} and NLTK~\cite{bird2009natural}.

        \item \textbf{Similarity} approximates the similar features or aspects between two data items using similarity metrics, such as edit distance~\cite{danielsson1980euclidean}, jaccard~\cite{niwattanakul2013using}, and TF-IDF~\cite{aizawa2003information}. 
        
        \item \textbf{Linking} links data items, e.g., named entities, part of speech tags, and keywords, to external knowledge sources for further enrichment and analysis. Example of existing knowledge bases Wikidata~\footnote{https://www.wikidata.org}, Google Knowledge Graph~\footnote{https://developers.google.com/knowledge-graph}, Geonames~\footnote{http://geonames.org/}.

        \item \textbf{ Summarising} focuses on identifying and grouping similar items within the data. Examples of summarization techniques include clustering, sampling, compression, and histograms.

    \end{itemize}

\end{itemize}

 Over the past years, several solutions~\cite{beheshti2017systematic,beheshti2018datasynapse,pu2015topic,sellam2015semi,cavanillas2016new,cheng2015flock} have been proposed to assist analysts in curating data through adopting different learning algorithms for deriving insight and extracting knowledge. Usually, relying on these solutions, an analyst investigates the curation environment and performs a feature extraction task to identify the content bearing features that best describe the data.  Example of such a curation system is Snorkel~\cite{ratner2017snorkel}, which relies on a set of user-defined learning functions to train a generative model and curate the data.

 
\subsection{Adapting Data Curation Rules}
Today, a large number of curation tasks are happening in dynamic and constantly changing environments. Example of such an environment is social media, e.g., Twitter and Facebook~\footnote{https://www.facebook.com}, where data generates as a never-ending and ever-changing stream~\cite{gc2015big}. 
In a dynamic curation environment, the curation system needs to be updated iteratively to remain applicable and precise. Let us go back to our example regarding capturing citizens opinions in their communities, which was introduced in the previous sections. A citizen may face a new problem in her community, e.g., broken light, traffic, and light rail delay, and create a new hashtag on social media to express her topic of interest. Consequently, the curation system needs to be updated to capture such changes to be applicable.



In the past years, several solutions~\cite{milo2018interactive,liu2010refining,volkovs2014continuous,he2016interactive,gc2015big,xie2017automatic} have been proposed to curate data in dynamic environments. Normally, these approaches rely on learning algorithms~\cite{liu2010refining,volkovs2014continuous,ratner2017snorkel,ratner2017snorkel2,beheshti2017coredb}, e.g., regression, naive Bayes, and SVM; to adapt a curation system with recent changes. For example, one may train an initial model to label the data relevant to her topic of interest. Then, over time the system will be updated with new data to capture changes in the curation environment. However, relying on pure algorithmic approaches for curating data suffer from several problems~\cite{gc2015big}:~(1) Algorithms are complex and difficult to interpret and require an expert for tuning and training,~(2) Algorithms are designed for a specific context and cannot be easily adapted to work in another context, and~(3) in many cases, algorithms require a large amount of training data, which may not be available or difficult to obtain.

In recent years, several solutions augmented algorithms with curation rules to curate data in dynamic and changing environments. 
These systems~\cite{bak2014rule,liu2010refining,milo2016rudolf,chiticariu2013rule,cheng2015flock,brooks2015featureinsight} relies on a set of hand-crafted rules and analysts for adapting rules (removes the imprecise rules or adds new ones) and maintain the curation system applicable overtime. The advantage of augmenting algorithms with rules are manifold:~(1) Writing rules are more straightforward than designing algorithms. A rule can be added to a curation system much faster than an algorithm~\cite{gc2015big},~(2) Correcting mistake for rules is faster than learning algorithms for analysts~\cite{gc2015big}, and~(3) Rules can consider cases that learning algorithms cannot yet cover. In cases that a curation system needs to curate data for a new topic, e.g., transportation and bus schedule, an analyst can easily add new rules to the system. However, algorithms need to be trained with new training data, which may not be available or difficult to obtain~\cite{domingos2012few}. 

Although coupling rules with learning algorithms enhance the performance of curation systems in curating data, still an analyst needs to continuously monitor rules' performance to identify and adapt the imprecise ones. Over the past years, several approaches~\cite{bak2014rule,milo2016rudolf,milo2018interactive,gc2015big,sun2014chimera} relied on interactive techniques for adapting rules. These systems adapt a rule by identifying the potential modifications by interacting with the analyst. In the next sections, we discuss how an analyst can be aided to comprehend the curation environment without iteratively scanning and querying the data.

\subsection{Data Comprehension}

Understanding of data involves processes and activities a user undertakes to explore the curation environment to describe and determine the quality of data. Typically, to understand the data a user requires to understand the data and re-represents the data in a format that allows planning, evaluation and reasoning~\cite{pirolli2005sensemaking}. Text-based queries are one of the main techniques that have been used to scan the curation environment, deriving insight, and extracting value~\cite{tabebordbar2019conceptmap}.

 From early days of computers, text-based queries have been used to explore and scan curation environments~\cite{hearst2009search}. Today, text-based queries and search button have become a universal user interface component across the operating systems and Web applications. Usually, when a user has a limited information need, text queries in-conjunction with search engines, e.g., Google or Bing, can adequately accommodate user's searches~\cite{hearst2009search}. The user expresses her information and provides a set of keywords or phrases, and a search engine returns results based on their relevancy to the user queries in a ranked list of items. However, when the aim of information seeking task is not to \emph{look up} a few or an individual document, the user needs to go beyond the current text-based queries to conduct her searches~\cite{marchionini1988finding}. Exploratory search refers to search activities that require \emph{learning and investigation}~\cite{marchionini2006exploratory}. In this context, the data curation process can help users to scan and comprehend the curation environment to retrieve items relevant to her information needs. 

Overall, users' behaviour in seeking their information needs can be divided into three steps~\cite{marchionini2006exploratory}: \emph{lookup, learn, and investigate}. 
 Followings, we discuss each of these steps in details:
\begin{itemize}
    \item \textbf{Lookup} is the essential character of a search task and has been widely supported by search engines and database management systems. Lookup tasks retrieve both discrete and structured objects such as names, statements, files, numbers or media. An example of a lookup search is retrieving fast and accurate records of data using a database management system. Mostly lookup searches are considered as "fact retrieval" or "question answering" search task~\cite{marchionini2006exploratory}. Lookup searches are also suitable for analytical search approaches that begin with a set of precisely designed queries and retrieve accurate results without the need for further comparison and examination~\cite{marchionini2006exploratory}. 
    \item \textbf{Learning Search} tasks involve multiple iterations and return sets of results that require additional processing and interpretation~\cite{hearst2009search}. These results can be generated in various formats, e.g., graphs, texts, videos, and maps, and often require user's judgement and comparison. Learning search tasks allows users to make sense of data and develop new knowledge. Bloom's taxonomy~\cite{forehand2010bloom} defines the aim of learning search tasks to achieve: knowledge acquisition, comprehension of concepts or skills, interpretation of ideas, and comparisons or aggregations of data and concepts. Social search is another type of learning search, where a user aimed at finding communities of interest in social media, e.g., Twitter, Facebook, and Instagram~\cite{brusilovsky2018social}. Overall, learning search aims at locating, analyzing and assessing similar results and much of users' time is devoted to examining and reformulating their queries. Learning search tasks can be embedded with lookup searches to guide the user to better locate the information and capture the salient aspect of data.
    \item \textbf{Investigative search} considers a much broader search space and requires multiple iterations that take place over very long periods of time~\cite{marchionini2006exploratory}. Investigative search results may critically be assessed before being integrated into personal and professional knowledge bases. These searches often include explicit annotation of the search results and may be done to support planning and forecasting or to transform the existing data into new data or knowledge. Another usage of investigative search is to identify gaps in information and to avoid "dead-end alley"~\cite{garfieldnegative} in research. Investigative searches also can be used for alerting service profiles that need to be executed systematically and automatically. Serendipitous browsing~\cite{massis2011serendipitous} is another example of an investigative search. Investigative searching is more concerned with recall and aims at retrieving the maximum number of relevant results rather than minimizing the irrelevant results. These searches are not a good fit for today's Web search engines that are highly tuned to retrieve the most relevant results first. 

\end{itemize}

Over the past years different techniques~\cite{di2016rank,di2018study,marchionini2006exploratory,marchionini1988finding,patterson2001predicting,ruotsalo2018interactive} have been proposed to support user's comprehension of the curation environment. A large number of these approaches focused on lowering user's cognitive load through relying on visualization, e.g., bar charts, table, and stack bar. Visual encoding of a curation environment maps the data into a visual structure to enhance the user understanding of data ~\cite{di2016rank,gratzl2013lineup}. Visual encoding boosts the user's memory in absorbing information to better extract and locate their information needs. 

\section{Key Research Issues}
\label{intro_issue}
This section outlines the key research issues tackled in this dissertation. We intend to facilitate the curation of data in dynamic and constantly changing environments. We describe techniques to support analysts for transforming the raw data and deriving insight. Finally, we accentuate approaches for augmenting user's comprehension of the curation environment.

 \subsection{Transforming the Raw Data and Extracting Knowledge}
 
 One of the challenges exists in data curation systems is to effectively transform a large amount of structured/unstructured data ingested from different sources into contextualized data and knowledge. Usually, an analyst needs to examine the curation environment and write code for performing her curation tasks, which is painstakingly time-consuming and error-prone. Over the past years, several curation systems in both academia~\cite{krishnan2016towards,troncy2016linking} and industry~\cite{chiticariu2010systemt,sun2014chimera,bak2014rule} have been proposed to assist analysts in curating the data. These systems offer a set of tools or algorithms for helping analysts in curation tasks. Examples of such systems in the industry, include \emph{Talend}~\footnote{https://www.talend.com}, which offers services for integration, cleansing and masking a large amount of data. \emph{Informica}~\footnote{https://www.informatica.com}, is an Extraction-Transform-Load (ETL) tool and comes with a variety of components, including data quality, data replica, data management, and data virtualization. \emph{Alteryx}~\footnote{https://www.alteryx.com}, which comes with several elements for discovering, preparing and analyzing the raw data. \emph{Alation}~\footnote{https://www.alation.com}, is an interactive data curation tool for data annotation, and data governance, which contribute user knowledge in curation data. Although, data curation tools lower analysts burden in curating the raw data, using current solutions, analysts require extensive knowledge of the curation environment to extract and identify features that adequately describes their curation needs. Feature extraction has been proven to be painstakingly time-consuming and error-prone, as analysts need to spend an extensive period of time to scan and analyze the data within the curation environment.
\subsection{Rule Adaptation in Dynamic Curation Environments}

 Rule-based systems have been used increasingly to augment machine learning-based algorithms for annotating data in unstructured and continuously changing environments. Rules can alleviate many of the shortcomings inherent in pure algorithmic approaches. However, to couple rules with a learning algorithm:~(1) There is a need for an analyst to craft and adapt rules. Adapting rules is challenging and error-prone as the analyst needs to spend an extended period to identify the potential modifications that make a rule applicable and precise. This problem exacerbated in dynamic environments as rule adaptation is not a one-shot rule modification task, and the analyst needs to adapt the rule over time, and~(2) Typically, an analyst adapts a rule at the syntactic level, e.g., keywords and regular expression. Adapting a rule using syntactic level features limits the ability of the rule in annotating items when a curation system needs to curate a varied and comprehensive list of data. 
 \subsection{Comprehension of Curation Environments}
 
 In a large curation environment, often an analyst needs to iteratively investigate the data to retrieve items relevant to her topic of interest. Investigating the curation environment is both time-consuming and challenging as the user needs to issue different queries to retrieve items relevant to her information needs. In recent years, several visualization techniques~\cite{wall2018podium,hearst1995tilebars,di2016rank,di2018study,gratzl2013lineup} have been proposed to enhance user's understanding of data in large curation environment. These techniques augment user comprehension of the curation environment with various visualization elements such as line charts~\cite{gratzl2013lineup}, tilebars~\cite{hearst1995tilebars}, or tables~\cite{wall2018podium}. Although, relying on visual elements lowers user's cognitive load in absorbing information, using current techniques a user needs to explicitly specify her preferences for curation systems in forms of keywords or phrases. Text-based queries need to iteratively scan the curation environment and fails to retrieve user's information needs when the user is seeking for a varied and comprehensive list of items. 

\section{Contributions Overview}
\label{intro_contribution}
In the previous sections, we discuss different challenges for curating data. In this section, we explain our solutions to address those challenges, in particular~(1) We propose an automated and feature-based data curation foundry for transforming the raw data and deriving insight,~(2) We propose an adaptive approach for adapting data curation rules in dynamic and changing environments, and~(3) we propose a conceptual system for augmenting a user's comprehension of curation environments.

\subsection{Automated and Feature-Based Data Curation}

To enhance analysts in curating data and reducing the time of curation tasks, we introduced \emph{Knowledge Lake}~\cite{beheshti2018corekg} and \emph{automated data curation}~\cite{beheshti2017automating} services. The proposed solution offloads analysts from many of time-consuming and error-prone curation tasks and allows analysts to transform the raw social media data (e.g., a Tweet in Twitter) into contextualized knowledge without spending a large amount of time. The Knowledge Lake offers a customizable feature extraction service to harness desired features from diverse data sources by leveraging a cross-document co-reference resolution technique. The curation services provide a microservice-based architecture ~\footnote{publicly available on GitHub supporting networks such as Twitter, Facebook, and LinkedIn} that offloads analysts from many of time-consuming curation tasks. Additionally, we introduce a simple rule language to facilitate the interaction of analysts with the Knowledge Lake in querying the data and performing the analytical tasks. 
\subsection{Adaptive Rule Adaptation in Dynamic Curation Environments}
 In a dynamic curation environment, there is a need for an analyst to adapt curation rules to keep them applicable and precise. Rule adaptation is both time-consuming and error-prone. Thus, we propose an autonomic approach for adapting curation rules. We utilize a \emph{Bayesian multi-armed-bandit} algorithm~\cite{russo2017tutorial}, an online learning algorithm, which determines the adequate forms of a curation rule by gathering feedback from the curation environment over time. To frame the problem as a Bayesian multi-armed bandit algorithm, we propose a reward and demote schema. The schema rewards a rule if it identifies the rule correctly tagged~\footnote{A tag is a label, e.g., "Mental Health", a rule assigns to a curated item, e.g., "Tweet", to describe the item.} an item, and at the same time demotes a rule if it identifies an item incorrectly tagged by the rule. Over time, the algorithm by observing the accumulated rules reward and demote learns a better adaptation for rules~\cite{tabebordbar2019adaptive}. 
 
 Besides, we propose a technique to adapt rules at the conceptual level, e.g., topic, rather than syntactic level. Conceptual level adaptation boosts rules to annotate a larger number of items. 
 
\subsection{Augmenting User's Comprehension of Curation Environments}
 Understanding of data allows users to formulate their information needs better when seeking for information in large curation environments~\cite{wall2018podium,peltonen2017topic}. Thus, to enhance users comprehension of data, we propose a method that provides a conceptual summary of curation environments and allows users to specify their preferences implicitly as a set of concepts. Our approach lowers users' cognitive load in ranking and exploring data in a curation environment. Contrary to previous techniques that allow users to formulate their preferences explicitly, e.g., keywords and phrases. Our approach focuses on creating a conceptual summary of the curation environment to help users understand the data and relate it to their preferences. Hence, we focus on boosting users' cognitive skill in understanding the data and formulating that understanding to extract information relevant to their topic of interest. We do this by taking advantage of deep learning and a Knowledge Lake to provide a conceptual summary of the information space. Users can specify her preferences implicitly as a set of concepts without the need to iteratively investigate the information space. It provides a 2D Radial Map of concepts where users can rank items relevant to their preferences through dragging and dropping. Our experiment results show that our approach can help users to formulate their preferences better when they need to retrieve a varied and comprehensive list of information across a large curation environment~\cite{tabebordbar2019conceptmap}.
 
\section{Dissertation Structure}
\label{intro-organize}
The remainder of this dissertation organized as follows. We start with presenting the current state of the art on data curation in Chapter~\ref{Chapter2}. We explain in more depth how a curation system can aid analysts to transform the raw data and extract knowledge. We continue our discussion on curating data in dynamic and changing environments. We discuss different components of data curation rules and techniques for enriching and adapting rule. We wrap up the chapter with a discussion on the sensemaking of curation environments and how users can be aided to comprehend the data while formulating their preferences. 

In Chapter~\ref{chapter5}, we discuss our proposed solution for transforming the data and extracting knowledge. We discuss related works and our proposed solution to build a Knowledge Lake. We explain steps for constructing Knowledge Lake and how it enhances analysts in feature extraction. Next, we discuss curation services and how it aids analysts in curation tasks. Finally, we discuss the usage scenario and results to illustrate the usability of our approach.

In Chapter 4, we present our proposed solution for adapting data curation rules. We discuss related works and present a case study to demonstrate the usage of our approach. Then, we, explain how online learning can learn to adapt a curation rule without relying on analysts. Finally, we wrap up the chapter with results and conclusion.

In Chapter~\ref{Chapter4}, we present our proposed solution for augmenting user's comprehension of curation environments. We offer a data visualization system that utilizes deep-learning and a Knowledge Lake to provide a visual summary of curation environments. We discuss our proposed approach and how it generates different types of data summarise. Then, we discuss the components of our system and how it interacts with a user to formulate her preferences. We conclude the chapter with experiments and conclusion.

In Chapter~\ref{Chapter6}, we present a software prototype for automating data curation tasks. The proposed system facilitates the data curation process and enhances the productivity of researchers and developers in transforming their raw data into curated data. The curation APIs enable developers to easily:~(1) add features - such as extracting keyword, part of speech, and named entities (e.g., persons, locations, organisations, companies, products, diseases, and drugs),~(2) providing synonyms and stems for extracted information items leveraging lexical knowledge bases for the English language (e.g., WordNet~\cite{esuli2006sentiwordnet}),~(3)  linking extracted entities to external knowledge bases,~(4) discovering similarity among the information items,~(5) classifying, sorting and categorizing data into various types, forms or any other distinct class, and ~(6) indexing structured and unstructured data. 

Finally, in Chapter~\ref{conclusion_7}, we present concluding remarks of this dissertation and discuss possible directions for future work.


\chapter{Background and State of the Art}\label{Chapter2}

In this chapter, we discuss the state of the art in data curation models, and accentuate techniques for transforming the raw data, adapting data curation rules, and augmenting user comprehension of curation environments.

This chapter is organized as follows: In Section~\ref{intro_2}, we briefly introduce some of the challenges that exist in data curation systems. Then, we discuss data curation models and accentuate techniques for extracting value from the raw data (Section~\ref{transform_2}). In Section~\ref{dynamic_2}, we discuss solutions on adapting data curation rules in dynamic and constantly changing environments. Finally, in Section~\ref{conceptmap_2}, we discuss techniques for enhancing user comprehension and sensemaking of a curation environment, before concluding the chapter in Section~\ref{conclusion_2}.

\section{Introduction}
\label{intro_2}
Over the past years, there has been increasing recognition to curate and increase the added value of raw data. Data curation increase the visibility of data, and support enterprises to outperform their peers in output and productivity~\cite{akers2014building,cox2015research,buys2015data,locher2016starting,macmillan2014data,olendorf2012beyond}. Today, many companies and enterprises realized the importance of data curation for deriving insight and extracting value. However, the expansion of data generation platforms, e.g., sensors, social media, and Web, have made curating and analyzing data more challenging. Many enterprises and companies are struggling to implement practices and policies for curating and organizing their raw data. As an ongoing and emerging field, data curation lacks clear answers to several fundamental problems, including:~(1) Lack of a cohesive and robust framework to support analysts in transforming and increasing the value of data,~(2) Lack of supports for curating data in dynamic and changing environments, and ~(3) Lack of systems to support analysts in comprehending and analyzing curation environments. In this chapter, we aim at discussing the above problems after digging into data curation models and activities associated with it.

\section{Data Curation}
\label{data_curation_2}
Data curation defined as the activities a user undertakes to preserve the value of data~\cite{locher2016starting,mohr2015data}. Digital Curation Centre~\footnote{http://www.dcc.ac.uk/digital-curation/glossary} (DCC)~\cite{higgins2008dcc} is one of the communities that attempts to define data curation under a unite terminology. In general, data curation is defined as processes and activities related to the long-term management of data throughout its lifecycle to extract value and derive insight. In the following, we discuss two frameworks provided to establish a baseline for data curation activities and processes.

\subsection{Data Curation Frameworks} 
In this section, we briefly discuss two frameworks proposed for framing data curation:~(1) Digital Curation Center (DCC) model~\cite{higgins2008dcc}, and~(2) Open Archival Information System (OAIS) model~\cite{asopen}. The former provides a holistic view of activities and actions associated with each stage of curation tasks. The latter aims at providing a conceptual framework for curating and preserving the data.

\noindent\textbf{1. Digital Curation Center (DCC) Model}

\begin{figure}[t]
\hspace*{1.2cm}
\includegraphics[width=5in, height= 4in]{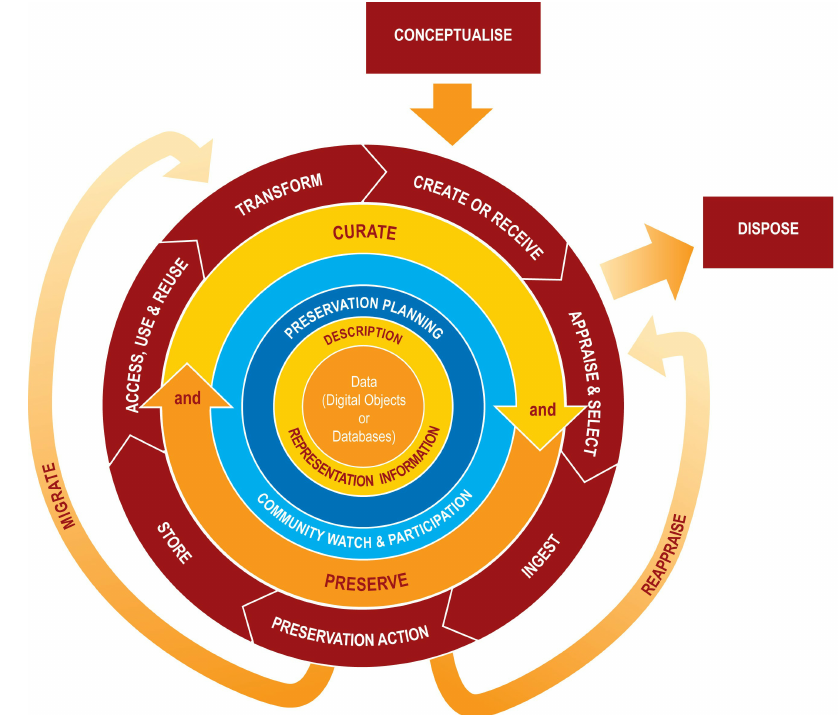}
\caption{Overview of Digital Curation Center (DCC) model (Source:~\cite{higgins2008dcc})}
\label{fig:current_approach_dcc}
\end{figure}

 \noindent DCC is a curation model to identify and assess the risk associated with managing data. It depicts the relationships between different stages of a curation task and provides a set of recommendations for transforming and preserving the value of data. 
 The model categorizes the curation tasks into three different actions:~(1) Lifecycle actions and~(2) Sequential actions, and~(3) Occasional actions. The lifecycle actions encompass activities for describing and representation of information. Sequential actions cover activities for ingestion, transformation, and conceptualization of information. Finally, occasional actions consider activities for disposing and migrating information. Figure ~\ref{fig:current_approach_dcc} shows different stages of data within the DCC curation model.

 \noindent\textbf{2. Open Archival Information System Model}
 
\noindent Open Archival Information System (OAIS) is a curation model that acts as a starting point' for building a sustainable pipeline for curating data and extraction of value~\cite{asopen}. The OAIS model defines terminologies to enhance the common understanding of curation tasks between data curators, and standards for better preservation, development, and assessment of data. 
 
 OAIS model is made up of two components: A functional model and an information model. The functional model defines activities for ingesting and preserving the data, which can be fulfilled either by humans or by machines such as computer systems. The information model defines different activities for dissemination and understanding of data, and specifies how the different types of information can relate to each other and how they are structured. 
 
In the next sections, we focus on techniques proposed for transforming and representation of data (refer to the DCC curation model). In particular, we discuss how an analyst can be aided to transform the raw data and extract knowledge. We, then discuss techniques for adapting data curation rules in dynamic and constantly changing environments, and how to enhance user's comprehension of curation environments.

\section{Transforming the Raw Data and Extracting Knowledge}
\label{transform_2}
As data grows and diversifies, many organizations realized that traditional methods of managing information are becoming difficult and outdated. Thus, there is a need for solutions to effectively leverage the implication of the new data generation platforms for organizations and enterprises to transform their raw data and extract knowledge. Over the past years, different technologies have been proposed to manage the data grow and augment analysts in deriving insight and making the decision. \emph{Data Warehouse}~\cite{kimball2000data} and \emph{Data Lake}~\cite{pasupuleti2015data} are the most common and widely used technologies for managing and transforming the data. 

\begin{enumerate}
    \item \textbf{Data Warehouse: } A data warehouse is a database optimized to analyze relational data produced by transactional systems. The data in a data warehouse is structured, with a predefined schema to enable users performing fast and effective information retrieval. Typically, data stored in a data warehouse is cleaned, enriched, and transformed, so it can act as the `single source of truth'~\cite{chute2010enterprise} that users can trust.
    \item \textbf{Data Lake: } A data lake is a centralized repository that allows storing structured and unstructured data~\cite{beheshti2017coredb}. Data lake stores data as-is, without having to schema the data, and can provide the ability to perform different types of analytics, including visualizations, big data processing, real-time analytics, and machine learning. 
\end{enumerate}

Following, we discuss each of these technologies and how they contribute to augment analysts in transforming the raw data and deriving insight.
\subsection{Data Warehouse}
Combining sparse data collected from different sources into a comprehensive and central repository provides several advantages for businesses and enterprises to derive insight~\cite{ProcessAnalytics,POLAP,Beheshti2016,casewall}. For example, in a sales system, a data warehouse might incorporate customer information from several sources, including a company's point-of-sale systems, mailing lists, and comment sections. Alternatively, it might include employees' data, including time cards, demographic data, and salary information~\cite{araque2006application}, allowing the company to analyze the customers and employees interactions.

Over the past years, a large number of works~\cite{croset2016flexible,10.1093/bioinformatics/btw579,mayfield2009cross,marquez2013coreference,yeastmind2012,wang2011fast,wick2009entity} leveraged data warehouse technology for their curation tasks. A large number of these works focused on integrating a curation result with data warehouses. For example, Croset et al.~\cite{croset2016flexible}, proposed a graph-based method to identify and remove the erroneous records of a curation process and their integration with a data warehouse. OntoBrowser ~\cite{10.1093/bioinformatics/btw579}, is a collaborative and continuous data warehousing system for mapping experts reported terms to ontologies. The system designed to facilitate continuous data integration and mapping tasks in an evolutionary ecosystem.
 YeastMine~\cite{yeastmind2012}, is a data warehouse system with a multifaceted search and retrieval interface. YeastMine allows data curators to search and retrieve a diverse set of genes using query customization. 
 
Another line of works mainly aims at leveraging the conceptual aspect of data warehouses for developing insight. These approaches rely on annotating and enriching curation results with information collected from data warehouses. For example, Sellam et al.~\cite{sellam2015semi}, introduced an automated data warehouse exploration system by detecting the fundamental aspect of data using approximation and greedy search. Karlgren et al.~\cite{karlgren2014semantic}, proposed an incremental system for learning the semantic component of data to distinguish the topical impact of different terms within a data warehouse. The system examines the local context of terms and their neighbourhood to identify the semantic quality. Beheshti et al.~\cite{beheshti2018processatlas}, introduced a framework for scalable graph-based OLAP analytics over process execution data. The system facilitates the analytics of OLAP systems through summarising the process graph and providing multi-views of data at different levels of granularity. Besides, many works (e.g.,~\cite{friedman1974projection,tatu2010automated,eades2017shape}) have relied on visualization to aid analysts in developing insight and detecting the best view of multidimensional datasets. These systems mainly focused on providing a 2-dimensional scatter-plot of data or analyzing and materializing every possible 2D view of the data.

\subsection{Data Lake} 
The data lake analogy aims at handling and storing multiple types of data without changing their formats. Data lakes provide a high degree of flexibility and scalability for companies and businesses that require to manage a large amount of data. According to Aberdeen et al.~\cite{ab2017angling}, the average company is seeing the volume of their data grow at a rate that exceeds 50\% per year. Additionally, these companies are managing an average of 33 different data sources in their analysis. Thus, the need for data lakes is inevitable to respond to the rapid growth of data volume and complexity. In the next section, we accentuate on opportunities data lakes bring to manage the complexity of data.

\noindent\textbf{Data Lake Opportunities:}
A data lake empowers companies to apply more advanced and sophisticated techniques for transforming the raw data, developing insight and supporting decision-makers. The data lake architecture boosts scalability in handling the growth of data, so companies can adapt their strategies with changes in the business environment~\cite{beheshti2017coredb}. Besides, data lakes provide the flexibility to support analysts on a variety of sophisticated analyses within an adequate timeframe.

Over the past years, a large number of works leveraged the data lake concept for transforming the raw data or extracting value (e.g., ~\cite{beheshti2017coredb,beheshti2018datasynapse,beheshti2017automating,sun2014chimera,personality2vec,BehavioralAnalytics,ProcessAtlas,iRecruit,iProcess,iCOP,iSheets,beheshti2016galaxy}). A large body of these works relies on data processing and analysis algorithms, including machine learning-based algorithms for information extraction~\cite{chiticariu2013rule}, item classification~\cite{jajuga2012classification}, record linkage~\cite{marquez2013coreference}, clustering~\cite{blei2003latent}, and sampling~\cite{MapReduce}. For example, CoreDB~\cite{beheshti2017coredb}, a data lake service, offers a single REST API to organize, index and query data and metadata. CoreDB manages multiple database technologies and offers a built-in design for security and tracing. AsterixDB~\cite{astrixdb}, is a BDMS (Big Data Management System) with a rich feature set and is well-suited for social data storage and analysis. AsterixDB provides facilities, including data modelling, query language, indexing, and transactions. Orchestrate~\cite{orchstrate}, provides a cloud-agnostic service to unify all queries needed for creating interactive applications, such as geospatial, time-series, graph, full-text search, and key-value queries. 

Another line of works, (e.g.,~\cite{alizadeh2019capturing,sarkar2011community,sarkar2009design,beheshti2018datasynapse}), has focused on coupling algorithmic approaches and data lake technologies for organizing the raw data and extracting insight. For example, to curate social media data (e.g., a text in Twitter), a machine learning algorithm can be used to cluster Tweets based on their topical similarity. Then, results can be displayed using different visualization elements~\cite{iStory}, e.g., bar charts and bubble graphs, to assist analysts in identifying the content bearing topics~\cite{tabebordbar2019conceptmap}. CiViC~\cite{alizadeh2019capturing}, also is a real-time data processing system, which clusters citizens' opinions through analyzing social media data, e.g., Twitter and Facebook, and news agencies' comments. The system relies on several machine learning algorithms and a data lake to store and analyze citizens' opinions regarding their communities. 

Extraction-Transform-Load (ETL) systems also have been used mainly for managing the user data and deriving insight. For example, Apache UIMA~\footnote{https://uima.apache.org/} is an ETL system. Which facilitates the analysis of unstructured data, and provides a common platform for analytics. PowerCenter~\footnote{https://www.informatica.com/au/products/data} is a unified enterprise ETL platform for accessing, discovering, and integrating data. SAP~\footnote{https://www.sap.com/australia/index.html} is a service-based ETL tool, that provides pervasive and extensible support for analyzing text, big data, social, and spatial data. IBM InfoSphere Information
 Server~\footnote{https://www.ibm.com/au-en/analytics/information-server} is a data integration platform that helps to understand, cleanse, transform and deliver data relevant to business initiatives.

\subsection{Knowledge Lake }
A Knowledge Lake~\cite{beheshti2018corekg,IntelligentKG,beheshti2018datasynapse} defined as a contextualized data lake. It is made up of a set of facts, information, and insights extracted from the raw data using data curation techniques~\cite{beheshti2018corekg} such as extraction, linking, summarization, annotation, enrichment, classification and more. In particular, a Knowledge Lake is a centralized repository containing inexhaustible amounts of data that is readily available to perform analytical activities. Knowledge Lake provides the foundation for deriving insight by automatically curating the raw data into a data lake. Beheshti et al.~\cite{beheshti2018corekg}, introduced an open-source data and Knowledge Lake service, which offloads analysts from many of curation tasks for deriving insight and extracting value.  In another work, Beheshti et al.~\cite{beheshti2018datasynapse}, proposed a generalized social data curation foundry for transforming the social data. The system relies on a Knowledge Lake to extract features and uncover hidden patterns of data. Tabebordbar et al.~\cite{tabebordbar2019conceptmap}, introduced a system which utilizes a Knowledge Lake for augmenting user' comprehension of curation environments and formulating their preferences. In Chapter~\ref{chapter5}, we explain how a Knowledge Lake can aid users to transform and extract knowledge from the raw data.

\subsection{Automated Data Curation}
\label{automated_data_curation}
Typically, for transforming the raw data into knowledge and deriving insight, an analyst may need to perform various curation tasks. These tasks not only are time-consuming and challenging, but the analyst also needs to have extensive knowledge of data curation and the curation environment for accomplishing the curation task. Automated data curation aims at offloading analysts from many tedious and challenging curation tasks~\cite{beheshti2017automating,WISE12,caise13,tagging,BigCDCR}, such as:
\begin{enumerate}
\item \textbf{Extraction:} extracting features such as keyword, part of speech, and named entities (Persons, Locations, Organizations, Companies, Products, and more) from unstructured texts~\cite{caise13}.
\item \textbf{Enrichment:} enrich the extracted features by providing synonyms and stems leveraging lexical knowledge bases for the English language, such as WordNet~\cite{esuli2006sentiwordnet}.
\item \textbf{Linking:} links the extracted enriched features to external knowledge bases (such as Google knowledge~\cite{singhal2012introducing} graph and Wikidata~\cite{vrandevcic2014wikidata}) as well as the contextualized data islands. 
\item \textbf{Annotation:} annotates features using different similarity metrics, classification, and clustering algorithms~\cite{ratner2017snorkel}.
\end{enumerate}
Several works have been proposed for automating the curation tasks. For example, Alex et al.~\cite{alex2008automating}, proposed a system for automating the curation of biomedical research papers. The system utilizes different natural language processing techniques, including: named-entity/relation extraction and term identification to form a pipeline of curation tasks and extracting documents. Kurator~\cite{dou2012kurator}, is a data curation system, which automates data curation pipelines by proposing several services for constructing a workflow of curation tasks. The system provides curation services for modelling, execution, provenance~\cite{Provenance,BPM11,caise13}, and management of data curation tasks. Song
et al.~\cite{song2014towards}, proposed a declarative and semi-automated approach for workflow design. The system implements a set of data curation actors, including name validators, summary validators, and annotation validators, to assist data curators in curation processes. 

In Chapter~\ref{chapter5}, we explain how automation reduces many of analysts tedious and time-consuming curation tasks.

\section{Data Curation Rules}
 \label{dynamic_2}
One of the key principles in a curation task is the need to maintain the quality of data. Gartner~\footnote{https://www.gartner.com/en} estimates that at least 25 \% of data in the top companies is flawed. Extracting quality data has a significant impact on business outputs, particularly when it comes to the decision-making processes within organizations ~\cite{curry2010role}. The increasing availability of open data on the Web, and generation of data across different platforms, produces an unprecedented volume of data, which increases the challenges for curating quality data~\cite{howe2008big,brodie2010power}. This problem exacerbated when the curation environment is dynamic or changing constantly, e.g., in Twitter and Facebook. In such environments, the curation system needs to be updated continuously to capture changes to remain applicable.

Over the past years, many solutions coupled humans with knowledge bases and learning algorithms for curating data in dynamic and changing environments. These algorithms focused on identifying and removing residue information through continuously updating the curation using analysts or crowds feedback~\cite{beheshti2018crowdcorrect,allahbakhsh2014representation,ratingMuh}. For example, Volks et al~\cite{volkovs2014continuous}, proposed a declarative data cleaning system coupled with a probabilistic classifier to assist analysts in repairing inaccurate records in a database. He et al~\cite{he2016interactive}, introduced an interactive data cleaning system that rectifies errors in a database using humans' feedback and a set of generated SQL update queries. The system uses SQL update queries for repairing database fields. DataSynapse~\cite{beheshti2018datasynapse}, is a feature-based data curation pipeline, which utilizes several knowledge bases and a co-reference resolution technique for creating a Knowledge Lake and annotating the data. 
Ratner et al.~\cite{ratner2017snorkel}, proposed a learning system for the rapid generation of training data. The system relies on weak supervision and a set of user-defined learning functions to train a generative model and label the data. De et al.~\cite{de2016deepdive}, proposed DeepDive, a method for knowledge base construction by extracting information from unstructured text and tables. DeepDive relies on statistical inference and machine learning for extraction, cleaning, and integration of data into a knowledge base.

Another line of works (e.g.,~\cite{gc2015big,sun2014chimera,milo2016rudolf,milo2018interactive,bak2014rule,chiticariu2013rule}) relied on curation rules for curating data in dynamic and changing environments. Curation rules can annotate data within a curation environment to enhance the interpretability of data for both humans and machines. In the next section, we accentuate approaches that leverage rules for curating data. First, we introduce rule languages proposed for curating data. Then, We discuss techniques for adapting curation rules, after describing the rule enrichment techniques.

\subsection{Curation Rule Languages}
Over the past years, different rule languages~\cite{w3c-dataextraction,valenzuela2016odin,chiticariu2010systemt,marietto2013artificial} have been proposed for curating the data, such as SystemT, JADE, Odine, AQL, DEL, AIML, etc. Rule languages mainly rely on pattern-matching to extract user information needs. These languages extract information using a set of regular expressions or lexical tokens and return results that satisfying the user-specified patterns.  In the following, we review some of these rule languages.

\begin{enumerate}
    \item \textbf{Data Extraction Language (DEL)~\cite{w3c-dataextraction}:} is an XML based rule language for describing the data conversion process. DEL specifies how to extract and locate pieces of data from an input document. It outputs the resulting documents in a well-formed XML document and locates data fragments using the pattern matching and regular expressions. 
    
    \item\textbf{Odin Runes~\cite{valenzuela2016odin}:} is a grammar-based rule language that implies cascades of finite-state automata over both surface text and syntactic dependency graphs. The rule language aims at augmenting analysts in crafting rules through coupling both lexical and syntactic automata.
    
    \item \textbf{AQL~\cite{chiticariu2010systemt}:} is a SQL like rule language to extract semi-structured, and structured information from text. AQL is the primary component in many information extraction systems, including the InfoSphere~\cite{biem2010ibm} and BigInsights~\cite{birjali2017analyzing}. The syntax of AQL is similar to that of Structured Query Language (SQL), which is case insensitive and removes the need for regular expressions to format the Information Extraction tasks. However, AQL does not support SQL features like recursive queries and sub-queries but relies on extract statements for retrieving the information.

    \item \textbf{SystemT~\cite{chiticariu2010systemt}:} is a declarative rule language, which extracts information from both unstructured and semi-structured data using a SQL like syntax. SystemT has been used in a wide array of enterprise applications and many information extraction systems. 
    The rule language is made up of three components:~(1)\emph{AQL}, a declarative rule language with a similar syntax to SQL,~(2) \emph{Optimizer}, which generates high-performance algebraic execution plans for AQL statements, and~(3)\emph{Executing engine}, which executes the algebraic plans and performs information extraction over input documents. 


\end{enumerate}

\subsection{Curation Rule Enrichment}
Over the past years several rule enrichment techniques~\cite{gc2015big,sun2014chimera,tabebordbar2019conceptmap,uzuner2010extracting,spasic2010medication,fatemi2009using,tabebordbar2018adaptive} have been proposed to enhance data curation systems. These techniques mostly rely on algorithms, such as similarity, extraction, classification, linking, summarization, etc. For example, for enriching a rule that curates Tweets relevant to `mental health', it is possible to extract information, e.g., keywords and named entities, from Tweets and link them to a knowledge base and generate a graph of related entities to reveal hidden information in the data~\cite{beheshti2018datasynapse,graphSurvey,DREAM,graphperformance,Extendingsparql}. Following, we discuss different techniques proposed for enriching curation rules. 
 
\begin{enumerate}
\item \textbf{Knowledge-Graph Based Enrichment:}

Incorporating the information extracted from Knowledge Graphs (KGs) to enrich rules. For example, consider rule $R_1$: 
\begin{center}
    $R_1 = $ \textbf{IF} Tweets contains (`health') \textbf{AND} Tweets contains (`service') \textbf{THEN} tag as “MENTAL HEALTH”
\end{center}
This rule tags a Tweet with `mental health', if the Tweet contains `Health' and `Service' keywords. However, there exists a large number of Tweets relevant to mental health, which rule $R_1$ skips as those Tweets may not contain both `Health' and `Service' keywords. To alleviate this problem, an analyst may utilize an ontology e.g., WordNet, to enrich rule $R_1$ with its synonyms. Thus, modifies the rule to:
\begin{center}
    $R_1 = $ \textbf{IF} Tweets contains (`health'|`wellbeing'|`wellness') \textbf{AND} Tweets contains (`service') \textbf{THEN} tag as “MENTAL HEALTH”
\end{center}
Rule $R_1$ tags a Tweet, if it contains any of `health', `wellbeing', and `wellness' keywords and the `service' keyword.

Rule enrichment have gained lots of attention in recent years~\cite{zamanirad2017programming,gc2015big,sun2014chimera,tabebordbar2019conceptmap}. There are a large number of lexicons or knowledge bases, exist to enrich a set of rule, such as: WordNet~\cite{esuli2006sentiwordnet}, ConceptNet~\cite{speer2017conceptnet}, Wikimedia~\cite{bergsma2007wikimedia}, Google Knowledge Graph~\cite{singhal2012introducing}, BabelNet~\cite{navigli2012babelnet}, Yago~\cite{rebele2016yago}, KnowItAll~\cite{etzioni2004web}, DbPedia~\cite{auer2007dbpedia} (see Table~\ref{tbl:knowledge-bases}). Typically, an analyst enriches a rule by extracting keywords, phrases, or entities that are relevant to her information needs. Enrichment augments a rule to curate a larger number of items. For example,  Kaufmann et al.~\cite{kaufmann2006querix}, assists analysts in enriching rules by utilizing several ontologies and Natural Language Processing (NLP) techniques. Lopez et al.~\cite{lopez2012poweraqua}, propose PowerAcqua, a Question Answering (QA) system, which combines several knowledge sources to enrich queries to retrieve the information stored in heterogeneous knowledge resources. Zamanirad~\cite{zamanirad2017programming}, proposed an approach for synthesizing natural language expression, to determine the proper API call. The technique understands the user intention and knowledge over an enriched knowledge graph of APIs.

\begin{table}
\caption{A sample list of knowledge bases}
\hspace*{-0.2cm}
\includegraphics[]{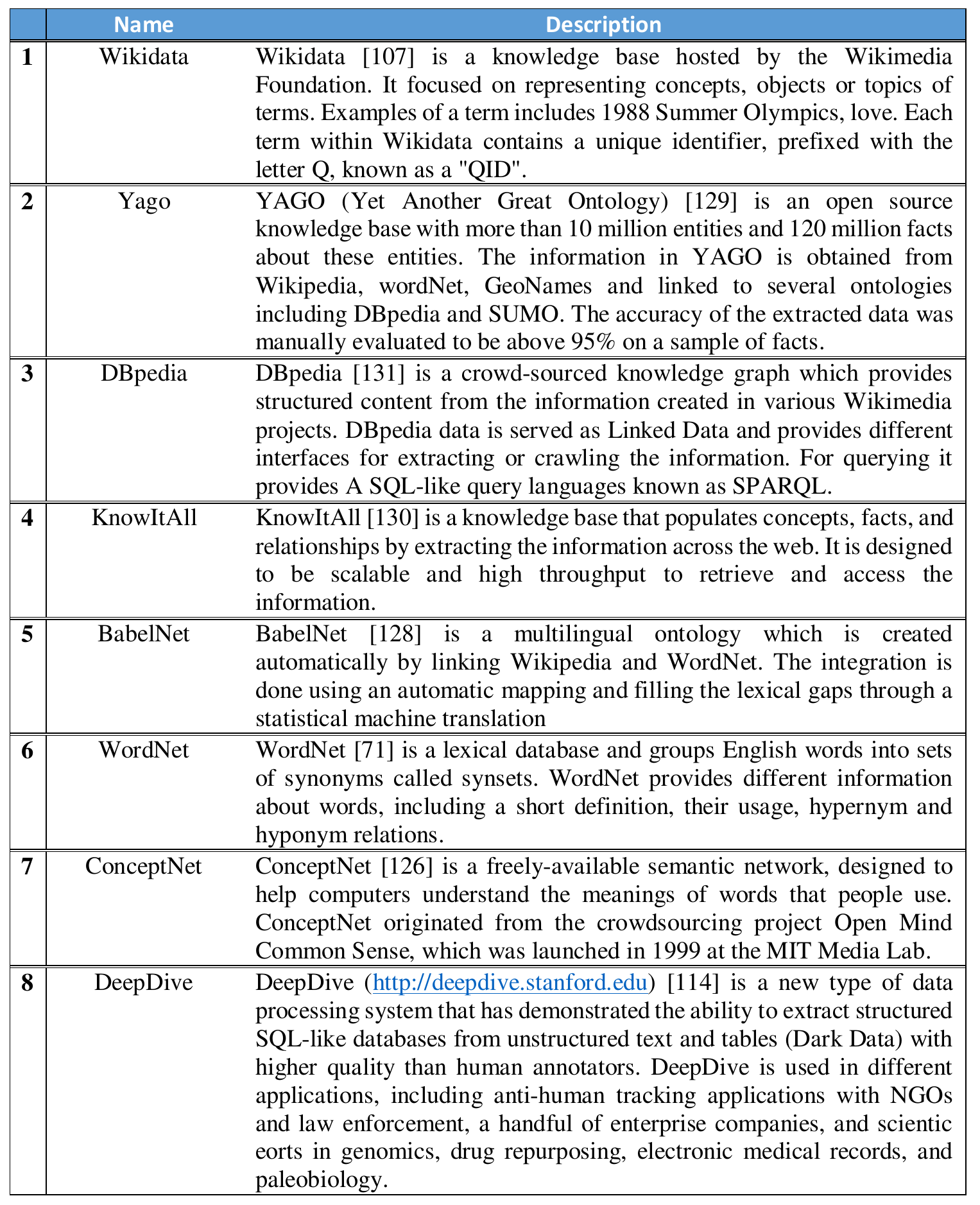}

\label{tbl:knowledge-bases}
\end{table}

\item \textbf{Similarity-Based Enrichment:}

Similarity-based enrichment is a syntactic level enrichment, which enriches items by quantifying the measure of similarity between the rule and data. There exist several similarity metrics for both string and numerical values. To measure the similarity between string values, metrics such as euclidean distance, Jaccard similarity, TF-IDF~\cite{salton1985advanced}, and cosine can be used to enrich a rule. To examine the similarity between numerical values, metrics like Hamming distance~\cite{norouzi2012hamming}, and Soundex~\cite{soundex2007}, can be utilized. One embodiment of similarity metrics in enriching rules is to classify data based on hashtags. As an example, consider a community advertising drastic weight-loss measures for youngsters. Suppose, initially the social media content circulated using the hashtag $\#thighgap$. Over time, a group of health advocates attempts to counteract these drastic and negative messages by writing rules that identify the posts containing $\#thighgap$, and posting materials that promote healthy weight choices. The supporters of drastic weight loss might be displeased and evolve their hashtag into misspelled versions, say hashtag $\#thyhgapp$. Using similarity metrics, the supporters of healthy weight-loss can enrich their rule to capture such changes.

\item \textbf{Pattern-Matching Based Enrichment:}

Pattern matching, or regular expressions, has been used for a long time to enrich rules (e.g., ~\cite{cayrol1982fuzzy,uzuner2010extracting,spasic2010medication,fatemi2009using}). As an illustrative embodiment, pattern matching based enrichment can be adopted to provide additional keywords or classifications by determining common keywords that co-occur with those specified in a rule. For example, Cayrol et al.~\cite{cayrol1982fuzzy}, proposed a fuzzy pattern matching for enhancing rules to extract information by considering the similarity between referents designated in the data and the pattern respectively. Irena Spasić et al.~\cite{spasic2010medication}, designed a system to enrich curation rules by exploiting the morphological and lexical aspect of data.  Ozlem Uzuner et al.~\cite{uzuner2010extracting}, proposed a hybrid system (rules, machine learning) for extracting medical text information. The system relies on pattern-matching based enrichment to extract phrases, and eliminate the irrelevant information, then uses the collected information to train learning algorithms and extracting the information. Fatemi et al.~\cite{fatemi2009using}, proposed an approach to enrich the representation of video content through a combination of semantic concepts and their co-occurrences. The approach leverages an existing partial set of semantic concepts for video archives and exploiting their relationships using association rules.

\end{enumerate}

\subsection{Rule Refinement:}
Rule refinement~\footnote{In this dissertation, we use the terms refinement and adaptation interchangeably.} is the process of modifying a rule to make the rule better suited to the curation environment~\cite{tabebordbar2019adaptive}. For example, consider an analyst and is interested in curating Tweets relevant to `mental health'. The analyst may examine the curation environment and, after a scanning a set of Tweets, crafts the rule $R_1$. This rule curates items that contain both `health' and `service' keywords.
\begin{center}
    $R_1 = $ \textbf{IF} Tweets contains (`health') \textbf{AND} Tweets contains (`service') \textbf{THEN} tag as “MENTAL HEALTH”
\end{center}
However, after curating a set of items, the analyst may identify the rule is imprecise and needs adaptation. Typically, to adapt a rule, an analyst examines different modifications to determine the optimal one. For example, after several changes, the analyst may adapt the rule $R_1$ to 
\begin{center}
    $R^{\prime}_1 = $ \textbf{IF} Tweets contains (`health') \textbf{AND} Tweets contains (`mental') \textbf{THEN} tag as “MENTAL HEALTH”
\end{center}
Rule adaptation is time-consuming and error-prone, which has been studied in several areas, including information retrieval~\cite{chiticariu2013rule,liu2010refining}, fraud detection~\cite{milo2016rudolf,milo2018interactive}, and database integration~\cite{volkovs2014continuous}. A large number of works for adapting rules relied on a ground truth of manually annotated items ~\cite{volkovs2014continuous,sun2014chimera,bak2014rule,liu2010refining,milo2016rudolf,milo2018interactive}. In these solutions, an analyst uses a ground truth to determine whether an 
adaptation could improve the rule precision or not. For example, Milo et al.~\cite{milo2018interactive}, used grand truth for assessing the performance of rules in a fraud detection system. Liu et al.~\cite{liu2010refining}, relied on a ground truth for assisting analysts in adapting rules and assessing the impact of her modifications. However, these solutions have focused on adapting rules that operate in a structured and more static environment, where the grand truth doesn't need to be updated frequently.

Although relying on ground truth can reduce the analyst burden in determining the performance of rules, in environments where the distribution of data is changing, e.g., social media, the analyst needs to iteratively adapt a rule to keep the rule applicable and precise~\cite{Trustworthiness,BigSocialData}. Thus, for adapting rules in dynamic and changing environments coupled crowd workers and analysts (e.g.,~\cite{gc2015big,bak2014rule,gc2015big}). For example, Sun et al.~\cite{sun2014chimera}, coupled analysts and crowd workers in adapting rules. The approach relies on workers to verify items curated with rules and the analyst for determining the optimal modifications for the rule. Bak et al.~\cite{bak2014rule}, proposed a voting technique for validating rules performance in information extraction applications. The approach relies on crowd workers' feedback to determine whether an adaptation of a rule produces a positive impact on extracting information or not. 

Alternatively, in recent years some solutions focused on offloading analysts from adapting rules~\cite{tabebordbar2018adaptive,tabebordbar2019adaptive}. For example, These solutions, consider a rule a set of features and determines the performance of rules by adding or removing features~\cite{tabebordbar2019adaptive,tabebordbar2018adaptive}. GC et al.~\cite{gc2015big}, relied on a relevance feedback algorithm~\cite{rocchio1971relevance} to determine the performance of features for adapting a rule. The algorithm based on the analyst feedback proposes an adaptation to make the rule applicable and precise. 

\section{Sensemaking of the Curation Environment}
\label{conceptmap_2}
This section explains techniques focused on enhancing users' comprehension of curation environments. In particular, we discuss solutions proposed for assisting users in the sensemaking of data and formulating their preferences. 

In a large curation environment, users' information needs can range from relatively simple tasks, e.g., looking up disputed facts or finding weather information, to rich and complex ones, e.g., job seeking and planning vacations. Typically, user interaction with a curation environment may vary based on the amount of time and effort the user can invest in the curation task and the level of her expertise~\cite{hearst2009search}.

The most common interface for interacting with a curation environment is search engines, e.g., google and bing. These interfaces are more appropriate for information lookup tasks, finding information relevant to websites or answers to questions. However, as Marchionini~\cite{marchionini2006exploratory} explained, search engine interfaces are inherently limited for many of the user's curation tasks, especially when a user needs to retrieve a varied and comprehensive list of information across a large amount of data. Marchionini~\cite{marchionini2006exploratory}  makes a distinction between information lookup and exploratory search. Lookup tasks are suitable for retrieval of discrete data, question answering, numbers, dates as well as names of files and Web sites. Standard Web search interactions work well for these retrieval tasks.

On the other hand, exploratory search considers much broader information-seeking tasks, which requires \emph{learning} and \emph{investigation}. During learning, users need to issue queries, retrieve, scan, and incorporate a large amount of data. Investigating refers to a much longer search activity that requires a continuous reformulation of queries and assessing the results. The investigation may take place over a an extended period, and results may need to be analyzed before being integrated into users' knowledge sources~\cite{marchionini2006exploratory}. In the investigation, a user mostly focuses on recall rather than precision. Examples of investigative search are litigation research or academic research. 

More broadly, an exploratory search can be seen as part of a more significant task, known as \emph{sensemaking} ~\cite{pirolli2005sensemaking,russell1993cost,russell2006being}. Sensemaking is an iterative process and is defined as activities a user undertakes to frame the curation environment in a logical schema~\cite{pirolli2005sensemaking}. Search and information seeking plays a crucial role in the curation of data. Search allows users to grasp the curation environment by retrieving information relevant to their information needs. However, to make sense of data, a user needs to scan and read a large amount of information and continuously reformulates her queries, which is proven to be painstakingly difficult and time-consuming. 
Examples of sensemaking tasks include the legal discovery process, epidemiology (disease tracking), studying customer complaints to improve service, and obtaining business intelligence. Pirolli et al.~\cite{pirolli2005sensemaking}, framed the sensemaking process into four steps:

\begin{center}
$Information \longrightarrow Schema \longrightarrow Insight \longrightarrow Product$
\end{center}

Figure~\ref{fig:sensemaking-overview}, shows the stages within each step, and followings, we explain each of them in detail: 

 \begin{figure}[t]
\hspace*{-0.2cm}
\includegraphics[width=5.8in, height= 5.1in]{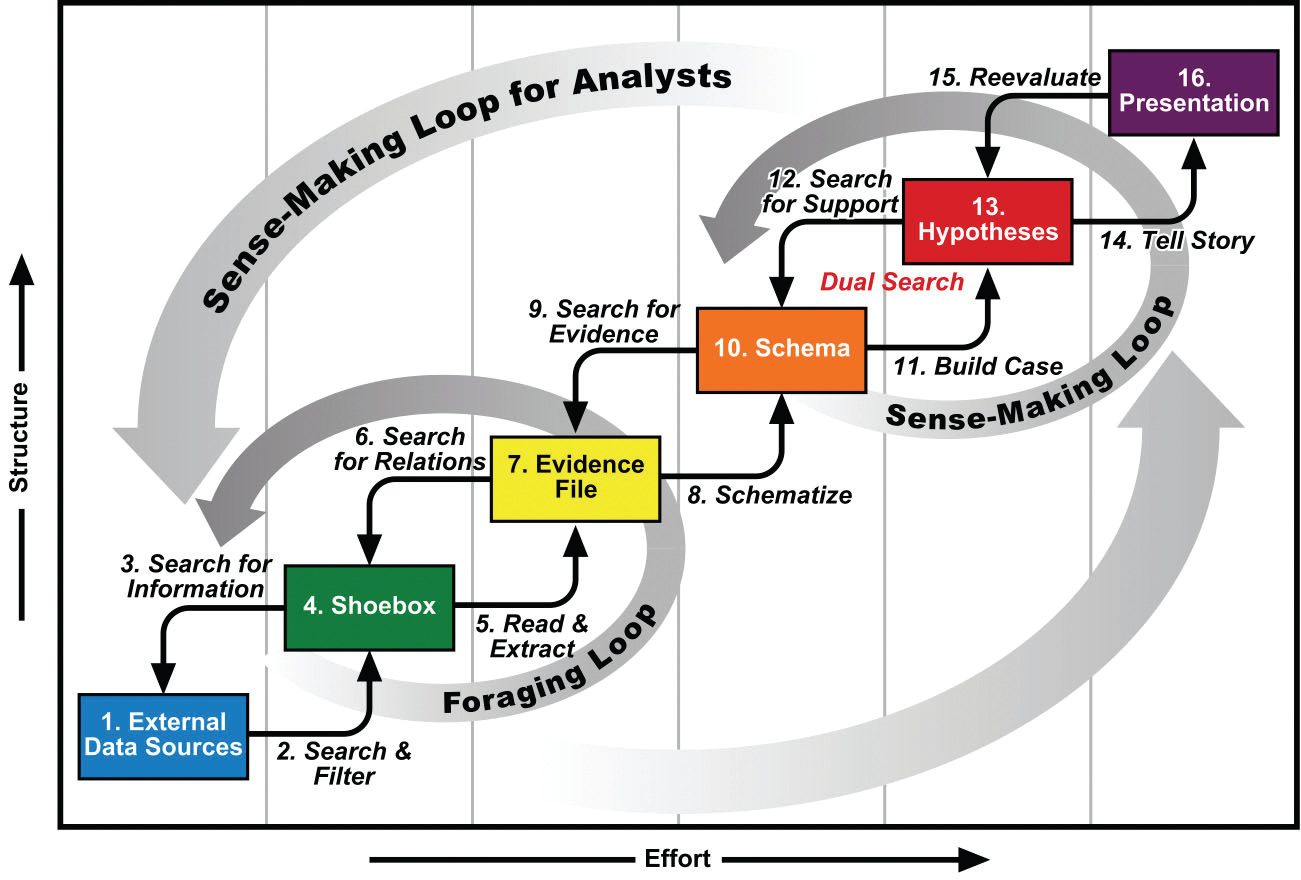}
\caption{Overview of sensemaking loop (Source:~\cite{pirolli2005sensemaking}).}
\label{fig:sensemaking-overview}
\end{figure}

\subsubsection{1. Information}
\begin{figure}[t]
\includegraphics[width=\linewidth, height=3.7in]{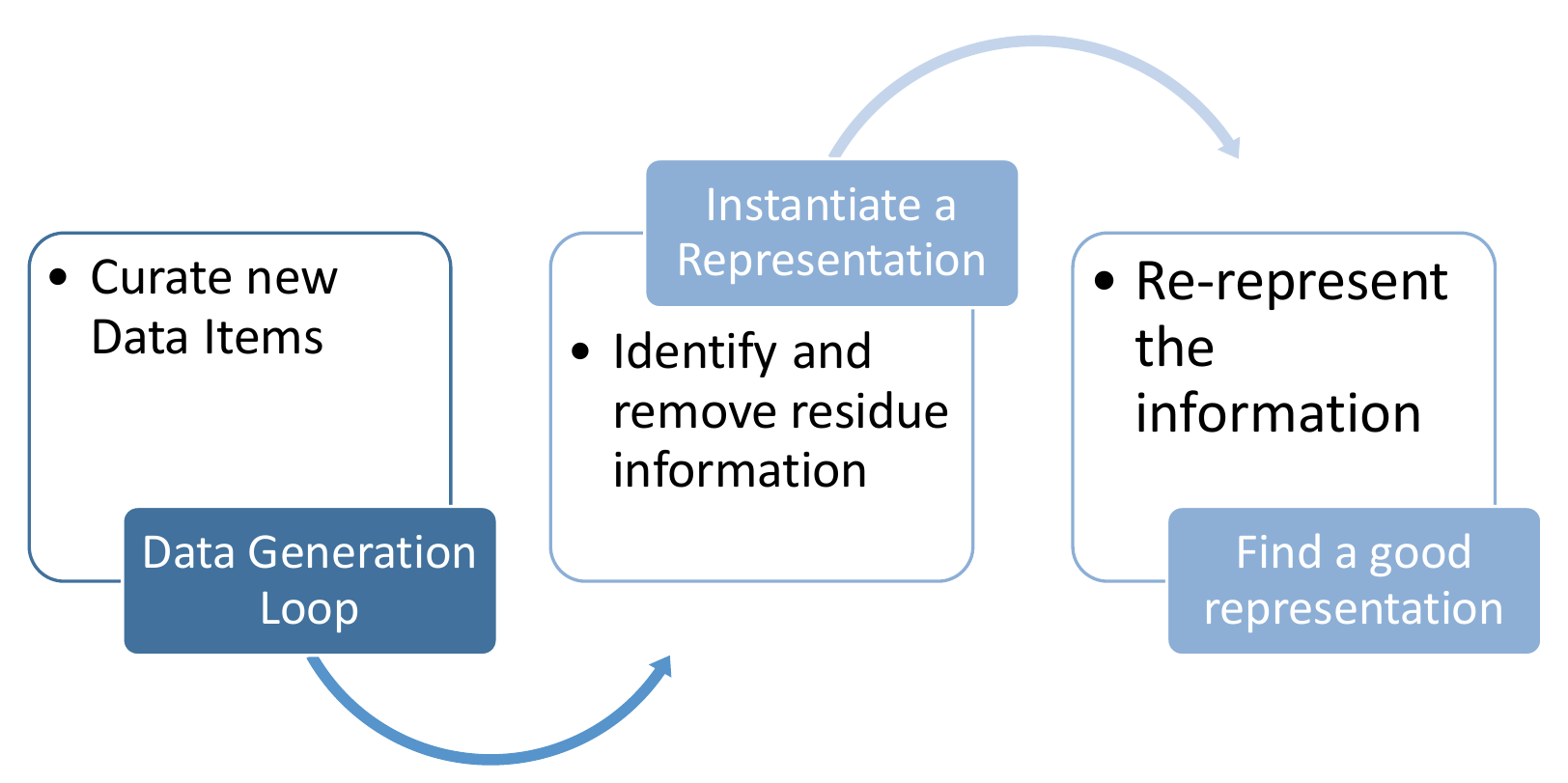}
\caption{Learning Loop Complex.}
\label{fig:sensemaking-learningloop}
\end{figure}

The first step in the sensemaking of a curation environment is retrieving information to support users to understand the data (stage 1 to 7 in Figure~\ref{fig:sensemaking-overview}). This stage is also known as the `foraging or learning loop complex' (Figure~\ref{fig:sensemaking-learningloop}), where a  user investigates the curation environment to identify a good representation of data. In this stage, the user's information needs may evolve as the user learns about the curation environment by analyzing the retrieved information~\cite{bates1989design}. 

Typically, to retrieve information, a user starts with a set of imprecise queries to approximately fetch the relevant part of data. Then, the user reformulates (see Section~\ref{insight_relwork_2} for the explanation on how reformulation happens) her queries by examining retrieved information. In the past years, many curation systems~\cite{dennis2002web,huang2003relevant,fonseca2003discovering, baeza2004query,ma2008learning,blanco20118th,song2012query,vahabi2013orthogonal} have attempted to support users in retrieving their information needs through elaborating their vague queries and recommending better ones. Many of these systems relied on logs accumulated from previous searches. An example of such systems, DirectHit~\cite{culliss2004personalized}, reformulates user's preferences by suggesting new query terms to narrow down their information retrieval tasks. 

Another technique that users relying on to investigate a curation environment is \emph{Boolean Operators}. Boolean operators have been supported by a large number of data curation systems. However, Boolean operators are difficult to use, and a user hardly could apply these operators to curating their information needs~\cite{dinet2004searching,hargittai2004classifying,hertzum1996browsing,hildreth1989general}. For example,  an examination of the search engine log over 1.5M queries revealed that only 9.7\% of queries were contained boolean operators ~\cite{jansen2007web}. Another study in 2006 over nearly 600,000 users queries revealed that only 1.1\% of the queries were contained boolean operators (double quotes, +, -, and site:) and only 8.7\% of the users used an operator at any time ~\cite{white2007investigating,hearst2009search}. 

\subsubsection{2. Schema}
\begin{figure}[t]
\includegraphics[width=\linewidth, height=2.5in]{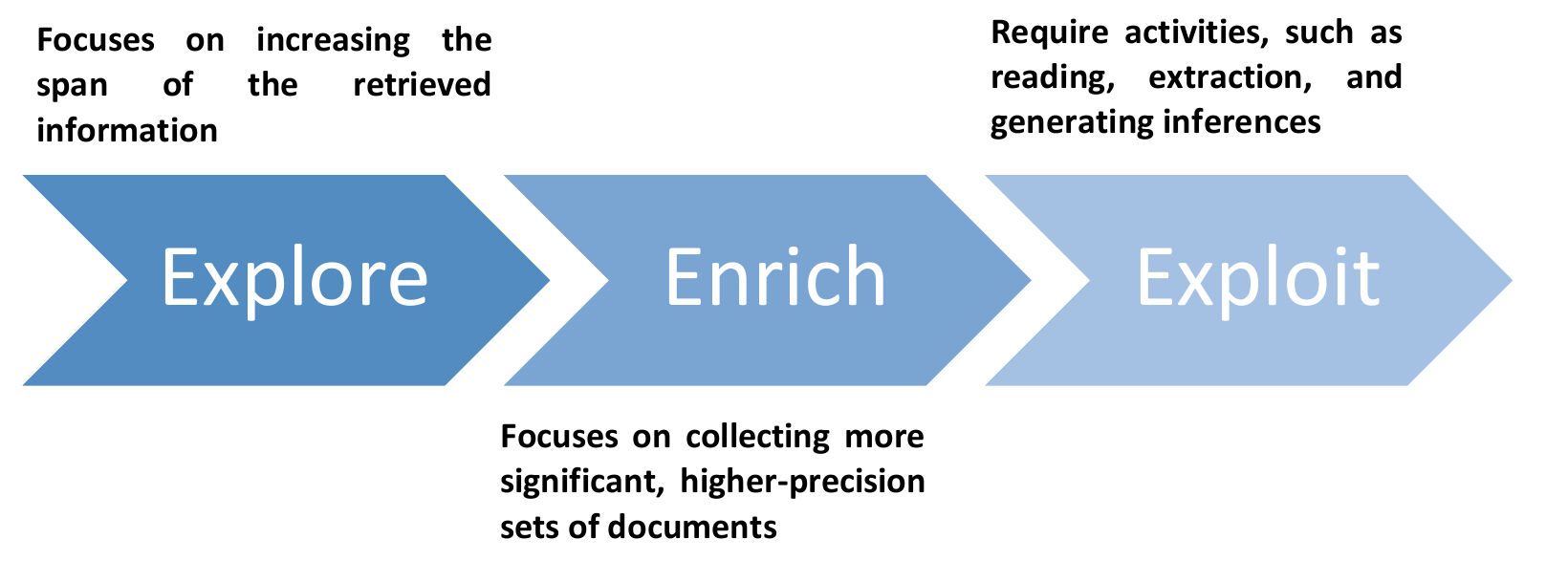}
\caption{Exploration Cycle in a Curation Environment.}
\label{fig:explore-exploit}
\end{figure}

The second step of the sensemaking process is to create a mental structure of the curation environment by analyzing results retrieved from users' queries (stage  8 to 10 in Figure~\ref{fig:sensemaking-overview}). In this stage, a user attempts to encode the curation environment in a new representation, which is more compact and better describes her information needs. The re-representation may informally occur in the mind of the user, aided through pen and pencils, or even computers. Often, in re-representation, the user tries to discard the residue information to identify information relevant to her information needs~\cite{russell2006being}. 
Figure~\ref{fig:explore-exploit} shows how users explore the curation environment to create a mental schema of a curation environment. 
Initially, to create a mental structure of the curation environment, the user begins with a broad set of documents and then narrows down that set into successively smaller rings. Patterson et al.~\cite{patterson2001predicting}, discusses that as a trade-off between \emph{Exploring, Enriching, and Exploiting} of data. Followings explain each of these steps in detail:
\begin{enumerate}
    \item \textbf{Exploring}: Focuses on increasing the span of the retrieved information and corresponds to improving the recall of the information search.
    \item \textbf{Enriching}: Focuses on collecting more significant, higher-precision sets of documents by removing the residue data.
    \item \textbf{Exploiting}: In this process, a user more involves in activities, such as reading, extraction, and generating inferences.
\end{enumerate}

Over the past years, many data curation systems (e.g.,~\cite{di2016rank,gc2015big,sun2014chimera,tabebordbar2019conceptmap}) attempted to support users in creating a mental structure of data. Many of these solutions rely on augmenting user comprehension of data through different visualization elements. Ranked lists of items, is one of the most common techniques that have been used to aid users in creating a mental schema of data. A ranked list ranks documents based on their relevance to the user query (e.g.,~\cite{freyne2004experiment,coyle2004searchguide}). The advantage of ranked lists is that users are familiar with the presentation arrangement and know where to start their scan for documents that seem relevant to their information needs~\cite{di2016rank}. A study by Shani et al.~\cite{shani2013displaying}, notes that augmenting ranked lists with bars lowers the users' cognitive load in grasping the curation environment. On the other side, ranked lists limit the number of items a user can examine within the curation environment as they imply a sequential search, and only a small subset of items are visible to users~\cite{di2016rank}. In addition to ranked lists, other techniques for supporting users to create a mental structure of curation environments are:

\begin{enumerate}
\item \textbf{$Focus + Context$ ~\cite{card1999readings}:} Enables users to examine objects relevant to their information needs in full detail, while users can get an overview impression of all other available information at the same time. Focus+Context systems allow having the information of interest in the foreground and the rest of the information in the background. It is made up of three components:~(1) It provides both overview and detail information together,~(2) Information provided in the overview can be different from those presented in detail, and~(3) The context and the overview information can be combined within a single (dynamic) display.
 \item \textbf{$Overview + Detail$~\cite{burigat2013effectiveness}:} Focuses on simultaneously displaying both an overview and a detailed view of a curation environment. This design shows each overview and detail in a distinct presentation area. For example, consider two images that are used for presentation. In an Overview + detail interface, the first image shows an overview of the whole curation environment, while the second image shows a small portion of the curation environment and visualizes details.
\end{enumerate}

\subsubsection{3. Insight}
\label{insight_relwork_2}
The third step of the sensemaking is developing insight through manipulating the representation created in the previous step (stage 11 to 13 in Figure~\ref{fig:sensemaking-overview}). In this step, a user examines the curation environment to extract evidence relevant to her information needs. The user examines different hypotheses and concludes the relevancy of evidence to her information needs by analyzing the relationship between documents. Pirolli et al.~\cite{pirolli2005sensemaking}, provides below guidelines for verifying hypotheses and evidence: 

\begin{enumerate}
    \item \textbf{Span of attention between hypotheses and evidence:}
    Humans have a limited memory capacity in absorbing the information, which limits the number of hypotheses, evidence, and the relation between hypotheses and evidence can heed. This problem exacerbated while users require to conduct reasoning of the extracted evidence and hypotheses as it has an exponential cost structure. 
    
    \item \textbf{Generating alternative hypotheses:}
    Typically, humans comprehension is biased towards the interpretation of information into some prejudged expectations. Human reasoning is also biased to some heuristics that deviate from `normative rationality'~\cite{pirolli2005sensemaking}. This problem limits the ability of people to generate new hypotheses. Besides, factors such as time pressures and data overload decrease humans' ability to produce, manage, and evaluate their hypotheses effectively. 

    \item \textbf{Confirmation bias:} 
    People typically fail to consider the diagnosticity of evidence and the disconfirmation of hypotheses. A solution would be to understand users need to distribute their attention to profoundly suggestive evidence and also search for disconfirming relations within the information space.
\end{enumerate}

Over the past years, data curation systems have been focused on augmenting users in deriving insight and verifying their curation hypotheses. These solutions mainly focused on enhancing users whist they reformulating their preferences. An early study~\cite{jansen2005temporal} of search engine logs, showed that during a curation task, least $50\%$ of users are modifying their queries to discover their information. Query recommendation is one of the conventional techniques that have been employed by Web search systems to aid users in deriving insight. A query recommendation system helps users to better verify their hypotheses by showing terms related to their queries. An example of such systems is spelling corrections or suggestion systems~\cite{kukich1992techniques,li2006exploring,cucerzan2004spelling}.

Additionally, query expansion is another technique for supporting users to formulate their preferences. Query expansion focuses on formulating users' information needs based on previous users' searches. A study by Jansen et al.~\cite{jansen2007web} suggests that at least 6\% of users who were exposed to query suggestion systems chose to click on them ~\cite{anick2008longitudinal}.

Relevance feedback~\cite{rocchio1971relevance} is another method and proposed to help users to derive insight through reformulating their queries. The main idea of relevance feedback is to determine the relevancy of documents and queries. In some variations of relevance feedback, users specify the terms within documents that are relevant to their queries~\cite{koenemann1996case}. Then the system computes a new query using the feedback received from the user. Although relevance feedback successfully integrated with non-interactive systems, they were not successful from the usability aspect and couldn't incorporate with data curation interfaces~\cite{ruthven2003survey,allan2004hard,kelly2005loquacious}

~\subsubsection{4. Product}
This step focused on aiding the user to organize curation results to understand the data and makes the decision (stage 14 to 16 in Figure~\ref{fig:sensemaking-overview}). Two common systems for organizing the curation results are: \emph{category systems} and \emph{clustering}.
\begin{enumerate}
    \item \textbf{Category System}:
    
A category system is made up of a set of labels that formed in a way to represent concepts related to a domain~\cite{hearst2009search}. A category system needs to be consistent and impeccable with predictable and consistent structure across a curation environment. Examples of category systems are faceted, flat, and hierarchical categories~\cite{sebastiani2002machine}.

\begin{enumerate}
    \item \textbf{Flat Categories:} 
    A flat category is a list of topics or subjects that are grouped to help a user in organizing her information. Flat categories also can be used for filtering or classifying documents. The early studies on the usability of flat categories have shown that these systems are not useful for organizing a large amount of information with an extended number of topical subspaces. Instead, flat categories showed positive feedback for more focused information-gathering tasks~\cite{dumais2001optimizing,kules2008users}.
    
    \item \textbf{Hierarchical Categories:} 
    Hierarchical categories first used for file system browsing, e.g., explorer window. In Web search interfaces, hierarchical categories have been introduced by Yahoo to organize popular sites into a browsable fashion.
    
    \item \textbf{Faceted Categories:}
    Faceted categories utilize both flat and hierarchical categories and are suitable for organizing curation environments with a large number of documents.
    
\end{enumerate}
\item \textbf{Clustering:}

 Clustering refers to the grouping a set of items that share some measure of similarity. In document clustering, the analogy is computed using the commonality among features, where a feature can be a keyword or a phrase~\cite{cutting2017scatter}. Clustering presents a fully automated strategy for representing the information within a curation environment. An example of clustering is to group documents by the language they have written, e.g. English, German, and Japanese. However,  Also, clustering algorithms require high computational power and is difficult for use in real-time Web search or information retrieval tasks.
\end{enumerate}
\subsection{Sensemaking Challenges}
Sensemaking of the information space is the quintessential part of every data curation system~\cite{marchionini2006exploratory}. Sensemaking is known as a challenging and time-consuming task, especially when a user needs to extract information across a large number of data~\cite{peltonen2017topic}. As we discussed in the previous sections, over the past years, different solutions have been proposed to aid users in sensemaking of a curation environment. These solutions focus on augmenting users comprehension of the information space through different visualization elements, e.g., ranked lists, tables, keyword expansion, clustering and categorization. Although, relying on such solutions lower user's cognitive load in understanding the curation environment. Still, users need to scan and examine a large amount of data to identify the relationship between different attributes in the curation environment. Besides, in large curation environments, many of these relations remain invisible either due to users limited memory capacity in absorbing information or visual clutter~\cite{sultanum2019doccurate}. To alleviate this problem in Chapter~\ref{Chapter4}, we discussed our solution for enhancing the user's comprehension and sensemaking of a curation environment. We proposed a summarization technique that generates a conceptual summary of data without the need to scan or investigate the curation environment. Our approach automatically discovers associations among different attributes. It boosts the user's comprehension of data by formulating their preferences as a set of high-level concepts such as topic, category, and locations.

\section{Conclusion and Discussions}
\label{conclusion_2}
In this chapter, we reviewed state of the art on data curation systems. We started the chapter by defining data curation and frameworks proposed for framing the curation tasks. Then, we continued our discussion on techniques proposed for transforming the raw data. We discussed their strengths and weaknesses and explained that current solutions require analysts to conduct various time-consuming and tedious curation tasks to curate the data. We also demonstrated that analysts need to spend an extended period to scan curation environments to identify and extract features that best describe their curation needs. In the next chapter, we accentuate our proposed feature-based solution for curating the raw data. We introduce different types of features that can be extracted from data and tools to automate many of curation tasks. We present the notion of Knowledge Lake and compared it with a data lake in deriving insight and extracting knowledge.\newline
    
Next, we discussed data curation rules. We explained different techniques for curating data in dynamic and changing environments. We discussed why relying on algorithmic approaches fail to curate data in dynamic environments. We noted how rules could complement curation algorithms for curating data in dynamic and constantly changing environments. We introduced different rule languages and enrichment techniques for enriching data curation rules. Finally, we wrapped up the section by reviewing methods proposed for adapting data curation rules. In Chapter 4, we discuss our proposed solution for adapting curation rules. We explain how learning algorithms can be utilized to offload analysts from adapting rules in dynamic and changing environments. Besides, we explain how rules can be boosted to curate data at the conceptual level to annotate a larger number of items.\newline

Finally, we reviewed state of the art on augmenting user's comprehension of curation environments. We discussed the importance of the sensemaking in formulating users preferences and curating data. Then, we explained the sensemaking process, including information, schema, insight, and product, and how these steps may impact the user's comprehension of a curation environment. We wrapped up the section, with challenges in the sensemaking process and possible solutions to augment user's understanding of data. In Chapter~\ref{Chapter4}, we discuss our proposed solution for the sensemaking of data. We discuss how summarization enhances users to comprehend the data without the need to scan or investigate the curation environment.

 \chapter{Feature Based and Automated Data Curation Foundry}\label{chapter5}
In this chapter, we present a feature-based data curation foundry for extracting value. We introduce a set of API's that made available publicly to automate curation tasks. We discuss, an algorithm for creating a Knowledge Lake (e.g., a contextualized data lake), to facilitate the transformation of the raw data (e.g., a Tweet in Twitter) into a curated item. 

The rest of this chapter is organized as follows: We introduce the research problem in Section~\ref{intorduction_3}. In Section~\ref{relatedwork_3}, we provide background
and related works. We present our solution in Section~\ref{solution_3}. 
In Section~\ref{result_3}, we present the implementation and the evaluation results of our approach. Finally, we conclude the chapter with remarks for future directions in Section~\ref{Conclusion_35}.

The content of this chapter is derived from the following paper(s):
\begin{itemize}
    \item A Beheshti, B Benatallah, A Tabebordbar, H R Motahari-Nezhad, M C Barukh, and R Nouri, \textbf{Datasynapse: A social data curation foundry}, Distributed and Parallel Databases Journal (2018), 1–34 (ERA Rank A).
    \item   A Beheshti, A Tabebordbar, B Benatallah, R Nouri, \textbf {On automating basic data curation tasks.} In Proceedings of the 26th International Conference on World Wide Web Companion 2017 Apr 3 (pp. 165-169).
\end{itemize}

\newpage

\section{Introduction}
\label{intorduction_3}
By the expansion of various data generation platforms, e.g., social media, Web, sensors, and big data processing systems have become the quintessential method for extracting knowledge and deriving insights from vastly growing data~\cite{rebele2016yago}. This increase in the volume of data, made opportunities for organizations and governments~\cite{ProcessAnalytics,russom2011big} to extract knowledge and generate value. For example, over the last few years, several companies started mining social media contents to personalize the advertisements in elections
~\cite{tene2012big}, analyse citizens' opinions on urban issues~\cite{alizadeh2019capturing}, improve government services~\cite{chen2012business}, predict intelligence activities~\cite{lohr2012age}, unravel human trafficking activities~\cite{de2016deepdive}, as well as to improve national security and public health ~\cite{tene2012big}. 

Social media, e.g., Twitter, LinkedIn, and Facebook, have provided an unprecedented opportunity for data generation platforms to propagate people opinions in real-time. In this context, a fundamental principle is to provide an efficient technique to transform the raw data generated by users into curated data, e.g., contextualized data and knowledge that is maintained and made available for use by end-users and applications. This process significantly enhances business operations, especially when it comes to decision-making processes and analysis. Data curation involves various curation tasks, including identifying relevant data sources, extracting data and knowledge, cleaning, maintaining, merging, enriching and linking data and knowledge (see Chapter~\ref{Chapter1} for more detail). For example, a government can analyze citizens' opinions regarding urban issues~\cite{alizadeh2019capturing}, by curating their Tweets, Posts and comments on social media platforms, or an organization may target an advertisement for a group of users based on the contents posted on their social media page. Thus, data curation acts as the glue between the raw data and analytics, providing an abstraction layer that relieves analysts from time-consuming, tedious and error-prone curation tasks.

Despite wide-spread efforts for data analytics, big data systems are still in their preliminary stages, with several unsolved theoretical and technical challenges stemming from the lack of adequate support for complex data curation tasks~\cite{stonebraker2013data}. At present, current approaches mostly rely on:
(1)~Purely algorithmic approaches~\cite{gc2015big}, while these approaches are dominant in a predefined context, they cannot be easily adapted for a large number of curation tasks that suffers from lack of enough training data,
(2)~Scripting languages, while offering increased flexibility, they demand sophisticated programming and mastery over the associated low-level libraries to create and maintain complex curation.

\begin{figure} [t]
\centering
\includegraphics[width=1.1\linewidth,height=2in]{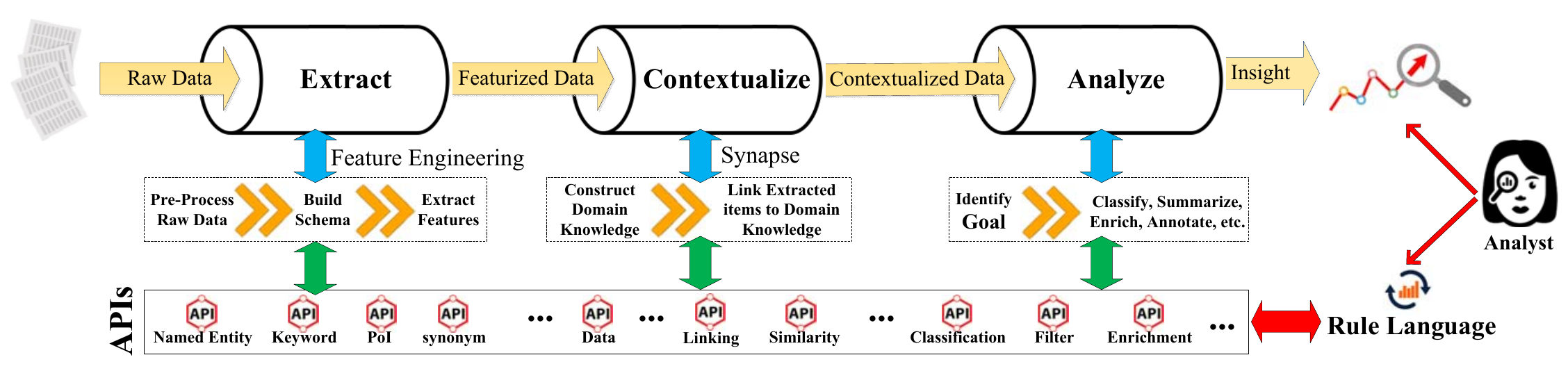}
\caption{Overview of Proposed Data curation Foundry~\cite{beheshti2018datasynapse}.}\label{fig:data-curation-pipeline}
\end{figure}

To facilitate the curation process, in this chapter, we present the notion of Knowledge Lake, (e.g., a contextualized data lake) ~\cite{beheshti2017coredb,terrizzano2015data}, which provides a foundation for data analytics by automatically curating the raw data into actionable insights. We leverage a Cross Document Co-reference Resolution (CDCR)  algorithm to assist analysts in linking the raw data to domain knowledge and derive insight. The algorithm facilitates the transformation of data by offering a customizable feature extraction to harness the desired features from the data. The unique \textbf{contributions} of this chapter can be summarised as:

\begin{enumerate}

\item We present the notion of Knowledge Lake, to facilitate data analytics by automatically curating the raw data and preparing them for deriving insights. The term \emph{Knowledge} here refers to a set of facts, information, and insights extracted from the raw data using data curation techniques such as extraction, linking, summarization, annotation, enrichment, and classification.

\item We proposed an approach for transforming the raw data, using a `Feature-based' data extraction technique. We also define a set of service-based APIs to facilitate data curation tasks, e.g., ingesting, extracting, cleaning, Summarizing and classifying data, as well as extracting features. Examples of an API in the category of `extraction' include: named entities\footnote{A named entity is a phrase that clearly identifies one item from a set of other things that have similar attributes, such as people, organization and places}, keywords~\footnote{a word or concept of great significance.}, synonyms~\footnote{a word or phrase that means exactly or nearly the same as another word or phrase in the same language.}, stem and part-of-speech\footnote{part-of-speech is a category to which a word is assigned in accordance with its syntactic functions, such as noun, adjective and verb.}.

\item We offer an algorithm for linking the extracted data to the domain knowledge by producing a summary of data and developing its contextualization. To do so, we leverage a CDCR technique~\cite{Beheshti2016} to identify coreferent entities within the data. For example, considering an analyst who is interested in gaining an accurate and deep understanding of cyberbullying, a keyword-based summary can enhance her comprehension of the threats exists within the data.

\item We provide a simple rule language to assist analysts in querying the Knowledge Lake to facilitate \emph{analytical} tasks.

We have implemented our approach as a set of reusable APIs, which are available publicly on Github\footnote{https://github.com/unsw-cse-soc/Data-curation-API.git}. We adopt a typical scenario for analyzing urban issues from Twitter. We demonstrate how our approach improves the quality of extracted data compared to the classical curation pipeline (in the absence of feature extraction and domain-linking contextualization). Figure~\ref{fig:data-curation-pipeline} illustrates the proposed data curation foundry.
\end{enumerate}

\section{Related Works and Background}
\label{relatedwork_3}
Data curation have been studied extensively in the past years. One of the main domain of data curation is to transform social data into actionable insights. Data curation has been defined as the active and ongoing management of data through its lifecycle of interest and usefulness~\cite{cavanillas2016new}. 
In this chapter, we primarily aim at data creation and value generation, rather than maintenance and management of data over time. More specifically, we focus on curation tasks that transform the raw social data (e.g., a Tweet in Twitter) into contextualized data and knowledge include extracting, enriching, linking, annotating and summarizing social data.

The contributions of this chapter aimed at breathing meaning into the raw data generated on social media and transforming it into contextualized knowledge, for effective consumption in social analytics and insight discovery. For example, information extracted from Tweets is often enriched with metadata on geolocation. 
Current approaches in data curation rely mostly on data processing and analysis algorithms
including machine learning-based algorithms for information extraction, item classification,
record-linkage, clustering, and sampling. Snorkel~\cite{ratner2017snorkel}, is an example of an algorithmic curation system for the rapid generation and annotation of the raw data. The system relies on weak supervision and a set of user-defined functions to train the learning algorithms and labelling the data. DeepDive~\cite{de2016deepdive}, is an algorithmic curation system for knowledge base construction. The system relies on statistical inference and machine learning for extraction, cleaning, and integration of data into a knowledge base. For example, consider a system, which extracts named entities from Tweets (e.g., `ISIS' and `Palmyra' in `There are 1800 ISIS terrorists in Palmyra, only 300 are Syrians'). The system may link the entities to
a knowledge base (e.g., Wikidata~\footnote{https://www.wikidata.org}, Google Knowledge Graph~\footnote{https://developers.google.com/knowledge-graph/}), to annotate and classify the Tweets into a set of predefined topics
(e.g., using naive Bayes classifier). Learning algorithms are undoubtedly the core components of data-curation platforms, where high-level curation tasks may require a non-trivial combination of several algorithms\cite{brainwash}, e.g., IBM Watson question-answering system uses hundreds of various algorithms for producing an answer~\cite{ferrucci2012introduction}.

Another set of related works ~\cite{beheshti2016galaxy,pu2015topic,sellam2015semi,karlgren2014semantic,wang2016linked} focuses on the semantical analysis of social media contents to breathe meaning into information extracted from the raw
data. Many of these approaches directed at creating, enriching or reusing Knowledge Graphs (KGs). KGs are large knowledge-bases that contain a wealth of information about entities (e.g., millions of people, organizations, places, topics, events) and their relationships (for a list of existing knowledge bases and their description, please refer to Chapter~\ref{Chapter2}). 
A knowledge base can be curated manually or automatically. Many of knowledge bases are interlinked at the entity level, (i.e., a Web of linked data) to provide more insights and knowledge which in turn would be an excellent asset for facilitating data curation pipelines. For example, cognitive applications, knowledge-centric services, deep question answering, and semantic search and analytics can benefit from the knowledge bases.

Alternatively, many data analytics platforms rely on scripting languages ( rule and query-based languages) for curating data~\cite{ShareInsights}. Examples of these languages in academia, include DEL~\footnote{https://w3c.org} (Data Extraction Language), as well as AQL~\cite{chiticariu2010systemt} (for more detail on curation languages and their description, please refer to Chapter~\ref{Chapter2}).  Typically, scripting languages use regular expressions, dictionaries and taxonomies to curate user-defined information needs~\cite{gc2015big}. Overall, even sophisticated and professional data scientists tools force analysts to use scripting languages to retrieve and curate their information~\cite{ShareInsights}.

Finally, there has been considerable  amount of works on curating open data. These works provide domain-specific solutions for different curation tasks, including leveraging crowdsourcing techniques to extract keywords from Tweets in Twitter~\cite{tene2012big,beheshti2018crowdcorrect}, Named entity recognition in tweets~\cite{ritter2011named}, linking entities for enriching and structuring social media content~\cite{troncy2016linking}, and sentiment analysis and identifying mental health cases on Facebook~\cite{ruder2011suicide}. However, to the best of our knowledge, there has been no work proposed a general-purpose approach for curating open data. Our proposed solution enables analysts to automatically link the data and knowledge generated on different social networks, uncover hidden patterns and generate insight.

\textbf{Motivating Scenario.}
Consider an analytic task related to \emph{ `understanding a government budget in the context of urban issues'}:
A typical government budget denotes how policy objectives are reconciled and implemented in various categories and programs. In particular, budget categories (e.g., `health', `social services', `transport', and `employment') defines a hierarchical set of programs (e.g., `medicare benefits' in health, and `aged care' in social-services). These programs refer to a set of activities or services that meet specific policy objectives of the government\cite{BudgetMap}. Using traditionally adopted budget systems would be challenging to evaluate the government services requirements and performance. For example, it is paramount to stabilize the economy through timely and dynamic adjustment in expenditure plans by considering related \emph{social issues}. For instance, a problem or conflict raised by a society ranging from local to national issues such as health, social security, public safety, welfare support, and domestic violence~\cite{BudgetMap}. Therefore the opportunity to link ongoing social problems to budget categories provide the public with increased transparency, and government agencies with real-time insight about how to make decisions.

\section{Solution Overview}
\label{solution_3}
Social media allows people of different walk of life to share their ideas and views by tagging, commenting or retweeting each other Posts. Examples of social media networks include Twitter~\footnote{www.twitter.com}, Facebook~\footnote{www.facebook.com}, and LinkedIn~\footnote{www.linkedin.com}. Analyzing social media posts allowing companies and business owners to promote brands, connect to new customers and foster their business. However, this requires businesses to transform the raw social media data into meaningful insights. In this context, we propose an automated and feature-based data curation foundry. We augment users in curating data by suggesting a set of curation services that offloads analysts from many tedious and time-consuming curation tasks. We propose a set of features to enhance analysts in curation tasks to grasp the salient aspect of data. An example of a feature is mentions of a person in Tweets or other social media Posts. Followings, we describe different types of features we extract from social media data.

\subsection{Feature Extraction}

 In this section, we introduce different types of features we extract from social media contents. Today, we extract two types of features: \emph{surface level features} and \emph{semantic level features}. Surface level feature refers to the type of features that can be extracted from social media by analyzing their syntactical characteristics. In contrast, semantic level feature refers to the type of features that describe social media contents semantically.
 \begin{enumerate}
    \item \textbf{Surface Level Features}
     \begin{enumerate}
         \item \textbf{Schema-based features:} This feature is related to the information we extract from the properties of a social item. For example, according to the Twitter schema~\footnote{https://developer.twitter.com/en/docs/tweets/data-dictionary/overview/tweet-object}, a Tweet may have attributes such as text, source and language, and a user may have attributes such as username, description and timezone. 
         \item \textbf{Lexical-based features:} This feature extracts information from social media texts, e.g., keywords, topic, phrase, abbreviation, special characters (e.g., a quotation in the text of a Tweet), slangs, informal language and spelling errors. 
         \item \textbf{Natural language-based features:} This feature is related to entities that can be extracted by the analysis and synthesis of natural language (NL) and speech, such as part-of-speech (e.g., verb, noun), named entity type (e.g., person, organization, product). For example, an instance of an entity type such as `Malcolm Turnbull' is an instance of an entity type `person'. 
         \item \textbf{Time-based features:} This feature is related to the information that can be extracted from the time in the schema of social media contents (e.g., `Tweet.Timestamp' and `user.TimeZone'). For example, a Tweet in Twitter or a Post in Facebook may contain a date, e.g., `3 May 2017'. 
         \item \textbf{Location-based features:} This feature is related to the mentions of locations in the schema of items. For example, in Twitter `Tweet.GEO' and `user.Location', or the content of a Tweet may contain mentions of locations, e.g., `Sydney', a city in Australia. 
        \item \textbf{Meta data-based features:} This feature is related to a set of data that describes and gives information about the social items. For example, it is important to know the number of followers (followers count) and friends (friends count), the number of times a social item has been visited (view count), liked (like count), or the sentiment of the content posted on a social media. 
     \end{enumerate}
     \item \textbf{Semantic Level Features}
     \begin{enumerate}
         \item \textbf{Schema-based semantics:} We use knowledge services such as Google Cloud Platform~\footnote{  https://cloud.google.com/}, Alchemyapi~\footnote{ https://www.ibm.com/watson/alchemy-api.html.}, Microsoft Computer Vision API~\footnote{ https://azure.microsoft.com/en-gb/services/. } and Apache Prediction IO~\footnote{ https://github.com/PredictionIO/.} to extract various features from the social media properties. For example, if a Tweet contains an Image, it is possible to extract objects (e.g., people) from the image.
         
         \item \textbf{Lexical-based semantics:} We leverage knowledge sources such as dictionaries and WordNet~\footnote{ https://wordnet.princeton.edu/. } to enrich lexical-based features with their synonyms, stems, hypernyms~\footnote{ A hypernym is a word with abroad meaning constituting a category into which words with more specific meanings fall. For example, colour is a hypernym of red. } hyponyms~\footnote{ A hyponym is a word of more specific meaning than a general or superordinate term applicable to it. For example, a spoon is a hyponym of cutlery.} and more.
         
         \item \textbf{Natural language-based semantics:} We leverage knowledge sources such as WikiData, GoogleKG and DBPedia to enrich Natural language based features with similar and related entities. For example, $Malcolm Turnbull$ is similar to $Tony Abbott$ \textemdash they both acted as the prime minister of Australia but $Malcolm Turnbull$ is related to $University of Sydney$. 
         
         \item \textbf{Temporal-based semantics:} We leverage different knowledge sources and services (such as events and storyline mining) to enrich time-based features. For example, a Tweet posted from Australia might be enriched with all events within that time frame. For instance, if a Tweet posted on `3 May 2017', we enrich the Tweet to be related to `Australian Budget' as we know from knowledge bases that the Australian Treasurer is handing the budget on 3 May every year.
         
         \item \textbf{Metadata-based semantics:} We use metadata-based features (such as followers count and share count) to calculate semantics such as the influence of a user. These semantics will enable analysts to get more insight from the social media posts and analyze the capacity to affect the character, development, or behaviour of other social users. 
     \end{enumerate}
 \end{enumerate}

\begin{figure} 
\centering
\includegraphics[width=1.1\textwidth,height=7in]{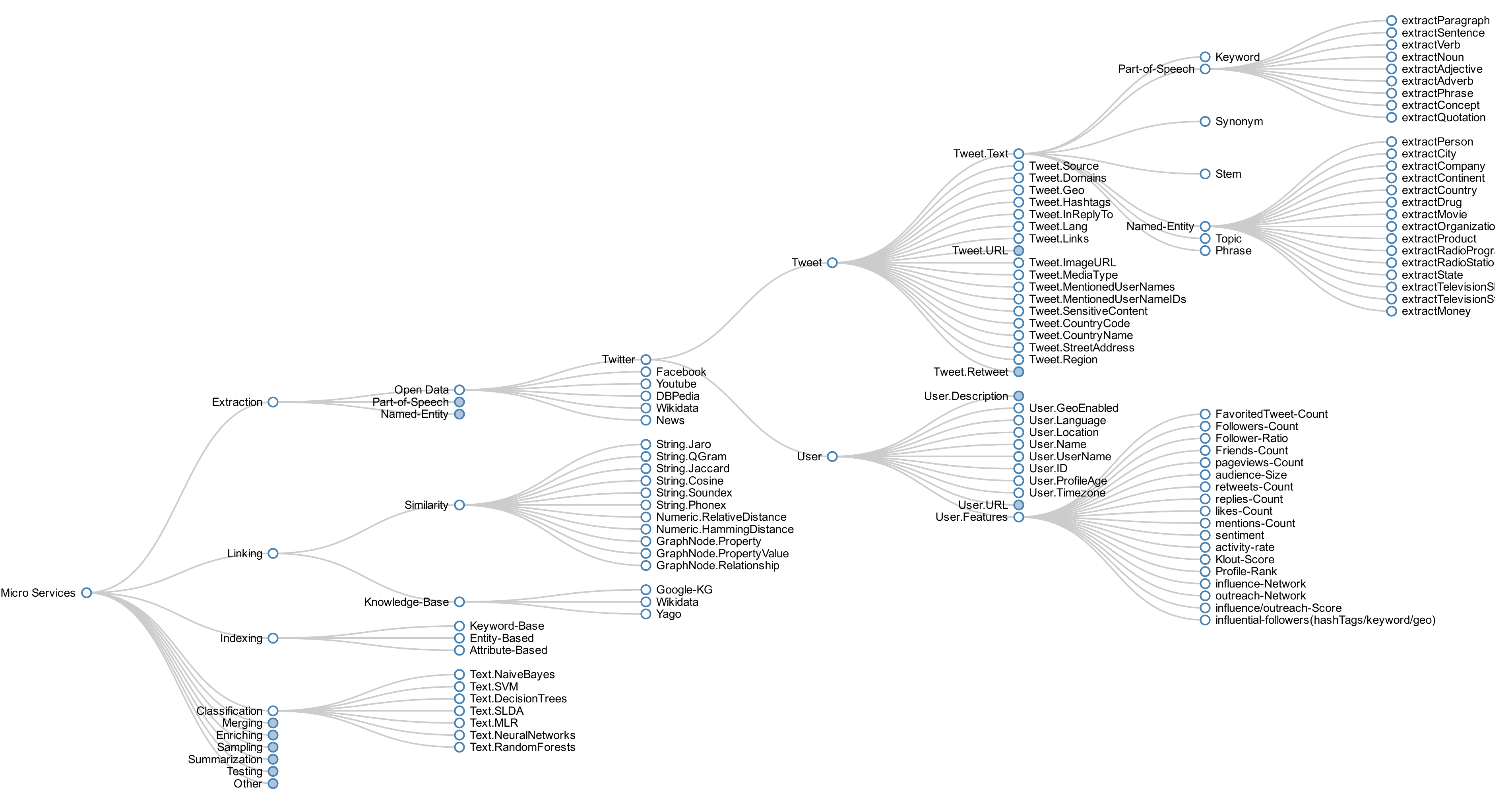}
\caption{Syntactical and semantical features that can be extracted from a Tweet~\cite{Beheshti2016}.}
\label{fig:service}
\end{figure}

Identifying and writing features is an extremely time-consuming and tedious task, especially when an analyst needs to extract features from a very large dataset~\cite{brainwash}. We have designed a set of curation APIs to assist analysts in curating data and extracting features. We implemented a set of uniformly accessible micro-services, which can be cascaded to produce analysts desired features. For example, to identify Tweets of a positive sentiment that relates to the $29^{th}$ prime minister of Australia (Malcolm Turnbull), we may craft a high-level feature that combines sentiment analysis (Metadata-based feature) with named entities (Natural-language-based feature). Figure~\ref{fig:service} illustrates examples of features that can be extracted from a Tweet.

\subsection{Data Curation Services}
To augment users in extracting features, we proposed a set of curation APIs. The APIs are implemented as micro-services and provide services such as extraction, classification, linking, and indexing. The curation services use natural language processing technology and machine learning algorithms for curating the raw data. For example, by extracting semantic meta-data from social media contents, such as information on people, places, and companies and link them to knowledge graphs such as WikiData and Google Knowledge Graph - using similarity techniques - or classify the extracted entities using classification services. 

We provide curation services for performing content analysis on internet-accessible Web pages, HTML or text content. The full description of the services is available in Chapter~\ref{Chapter6}. Also, a technical report ~\cite{Beheshti2016} is available on arXiv, which further guides analysts on utilizing the services for their curation tasks. In the following, we present an overview of the services.

Overall, the curation APIs provide four essential services for automating the curation tasks: \emph{Extraction Service, Linking Service, Classification Service, and Indexing Service}.

\begin{enumerate}
\item \textbf{Extraction Service:} This service extracts syntactical features from social media contents. The Extraction service can obtain features from both structured and unstructured data. The Service provides a wide range of APIs, including named entity recognition, part of speech, synonym, stem, and URL extraction.

\item \textbf{Linking Service:} We rely on Linking Service to automate the extraction of semantic level features. This service can be used for enriching social media contents, summarization, and computing the similarity between objects.  The enrichment service utilizes knowledge bases such as Google Knowledge Graph and Wikidata. The summarization service identifies and groups the semantically related keywords. 

\item \textbf{Classification Service:} This service facilitates utilizing the machine learning algorithms for classifying social media contents. The Classification Service assigns social media contents to a set of pre-defined target categories or classes.  This service facilitates the usage of algorithms, such as naive Bayes, Support Vector Machine (SVM), decision tree, random forest, linear regression, logistic regression, and neural network.

\item \textbf{Indexing Service:} This service enables analysts to scan and retrieve a curation environment quickly without the operational burden of managing it. For Indexing the social media contents, we utilized ElasticSearch~\footnote{https://www.elastic.co/}, which speeds up querying the data and deriving insight.

\end{enumerate}

\begin{figure}
\centering
\includegraphics[width=1.1\textwidth]{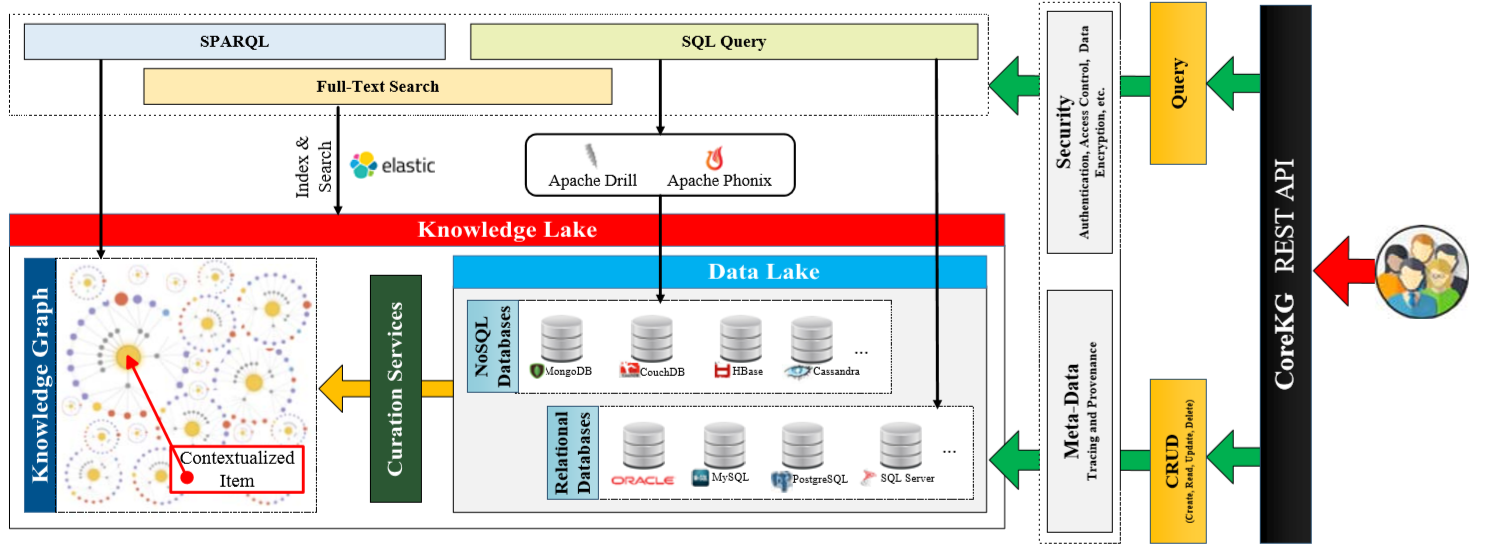}
\caption{Overview of A Knowledge Lake~\cite{beheshti2018corekg}}
\label{fig:coreKG}
\end{figure}

\section{Knowledge Lake}
In the previous section, we discussed different features that we can extract from the social media contents. However, to transform the raw data from a large number of independently-managed datasets such as Twitter, Facebook, and LinkedIn, into actionable knowledge there is a need to organize and facilitate the way users deal with these datasets. In recent years, data lake has been introduced as a centralized repository containing limitless amounts of the raw data ingested from different data sources. The rationale behind the data lake is to store the raw data and let the data analyst decide how to curate it later. Instead, we introduce the notion of \emph{Knowledge Lake} \textemdash a contextualized data lake. The term \emph{Knowledge} in a Knowledge Lake, refers to a set of facts, information, and insights to create a contextualization layer for transforming the raw data into knowledge.
A Knowledge Lake provides the foundation for big data analytics by automatically curating the raw data in a data lake and prepare it for deriving insights.

On top of the Knowledge Lake, we provide a rule language to enable analysts to query and retrieve the data. Figure~\ref{fig:coreKG} illustrates the architecture and the main components of Knowledge Lake. Technical details of Knowledge Lake and how it organizes the information can be found in~\cite{beheshti2018corekg}. In the rest of this section, we discuss how we can automatically link information extracted from social media contents to a Knowledge Lake. 

\subsection{Building Knowledge Lake}
Data extracted from social media may be interpreted in many different ways. To make sense of extracted data and to augment user's comprehension, it is beneficial to enrich the data through different features to produce \emph{contextualized knowledge}. We do this by building a \emph{Knowledge Lake} that implements a rich structure of relevant entities, their semantics, and relationships. We then utilize a CDCR technique to link the information extracted from social media contents to entities in Knowledge Lake. In this manner, we can discover hidden relationships and knowledge amongst extracted entities, or to group related entities (text or non-text), or to find paths describing the relationships among entities.

\begin{figure} [t]
\centering
\includegraphics[width=1.1\textwidth,height=3.1in]{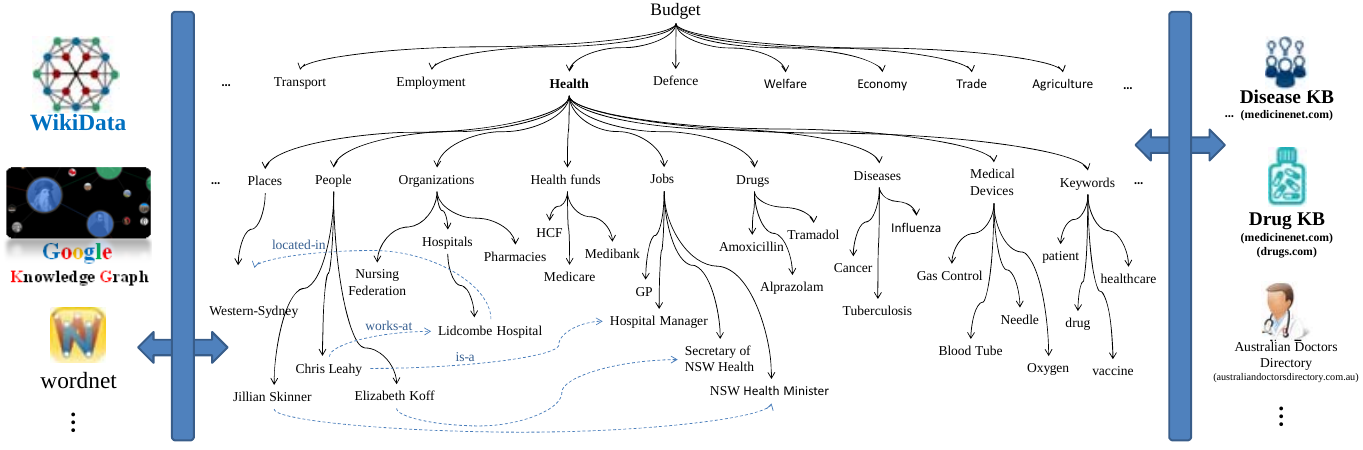}
\caption{A sample fragment of the Budget-KB.~\cite{beheshti2018datasynapse}}
\label{fig:BudgetKG}
\end{figure}

\textbf{Step 1: Constructing Budget-KB.}
This section, we explain the technique to build \emph{Budget-KB}, a domain specific knowledge base that represents entities relevant to the Australian Government's Budget. The Budget-KB is made up of a set of concepts related to the Australian budget organized into a taxonomy, instances for each concept, and relationships among these concepts. Figure~\ref{fig:BudgetKG} illustrates a sample fragment of the Budget-KB. 
To build the knowledge base, we first identified the list of budget categories and their related programs provided by Australian government data services \footnote{http://data.gov.au/}. Then, we filtered out the irrelevant categories and selected popular ones. For example, we have identified:
\begin{enumerate}
\item~\textit{people}, from GPs and nurses to health ministers and hospital managers,
\item~\textit{organizations}, such as hospitals, pharmacies and nursing Federation,
\item~\textit{locations}, states, cities and suburbs in Australia,
\item~\textit{health funds}, such as Medibank, Bupa and HCF,
\item~\textit{drugs}, such as amoxicillin, tramadol and alprazolam,
\item~\textit{diseases}, such as cancer, influenza and tuberculosis,
\item~\textit{medical devices}, such as gas control, blood tube and needle,
\item~\textit{job titles}, such as GP, nurse, hospital manager, secretary of NSW Health and NSW health minister, and
\item~\textit{keywords}, such as healthcare, patient, virus, vaccine and drug.

\end{enumerate}

\begin{figure}
\centering
\includegraphics[width=1.1\textwidth]{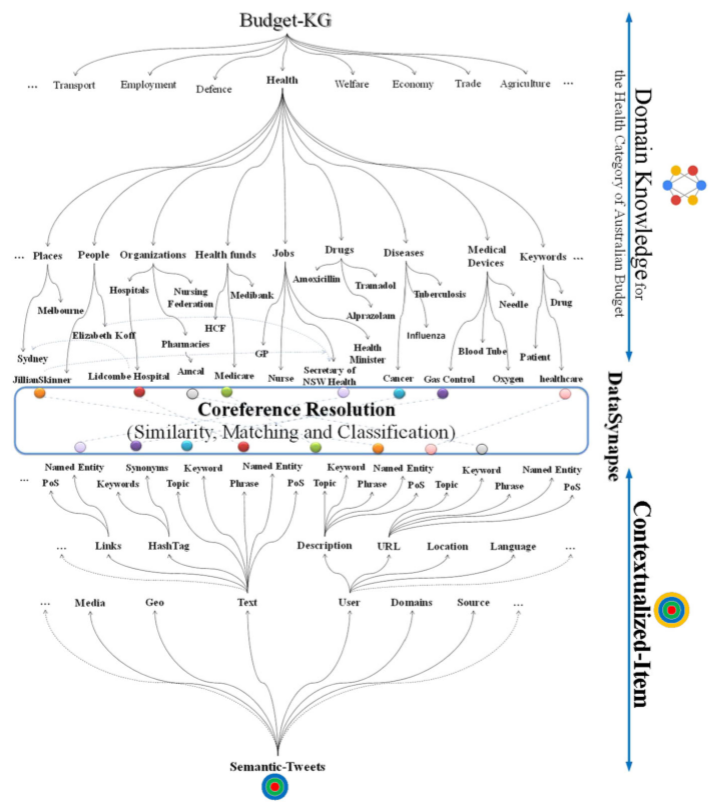}
\caption{ A typical scenario for analyzing urban social issues from Twitter as it relates to the government budget, to highlight how our solution transforms the raw data into contextualized data by leveraging a domain knowledge.~\cite{beheshti2018datasynapse}}
\label{fig:cdcr}
\end{figure}

We also extracted a set of concepts for each category using the introduced curation APIs~\cite{beheshti2017automating}:
locations from auspost\footnote{http://auspost.com.au/postcode/},
doctors from Australian doctors directory\footnote{https://www.ahpra.gov.au/} (including GPs, specialists and nurses),
hospitals from myHospitals\footnote{https://www.myhospitals.gov.au/browse-hospitals/},
health funds from health-services\footnote{http://www.privatehealth.gov.au/},
drugs from drug-index\footnote{http://www.rxlist.com/},
diseases from medicine-net\footnote{http://www.medicinenet.com/},
medical devices from FDA\footnote{http://www.fda.gov/},
job titles from compdata\footnote{http://compdatasurveys.com/compensation/healthcare}, and
keywords from Australia national health and medical research council\footnote{https://www.nhmrc.gov.au/}. The concepts work as the seed data for the categories, which enriched using several readily available knowledge bases such as
Wikidata\footnote{https://www.wikidata.org/},
Google Knowledge Graph\footnote{https://developers.google.com/knowledge-graph/} and WordNet\footnote{https://wordnet.princeton.edu}. For example, we extract relationships from Wikidata to form a relationship graph, e.g., `Bankstown Lidcombe Hospital' \emph{located-in} `Bankstown, Sydney, NSW, Australia', and we have used Google Knowledge Graph API to link entities to Wikipedia, e.g., by using `Jillian Skinner' as an input we have learned that `Jillian Skinner' \emph{is-a} `person', \emph{linked-to} `$https://en.wikipedia.org/wiki/Jillian\ Skinner$', \emph{is a} `member-of `New South Wales Legislative Assembly', and \emph{is-a} `New South Wales Minister for Health' for Australia. Figure~\ref{fig:BudgetKG} shows a small snippet of the created domain knowledge, which illustrates the above notions. As presented, `Jillian Skinner' \emph{is a} `person' and is the `Health Minister for New South Wales in Australia
' (see the link between this person and the job title in the Figure~\ref{fig:cdcr}). As another example `Lidcombe Hospital' is an instance of a hospital and is located in Western Sydney (a location, suburb, in NSW Australia).

\textbf{Step 2: Linking Features and the Budget-KB.} 
So far, we have presented how we construct the Budget-KB, and how we leverage the curation APIs to curate the data. In this section, we explain a method to link the curated data to Budget-KB categories. 
Identifying and linking entities across various information sources can be considered as the basis of knowledge acquisition and at the heart of analytics. We achieve this by identifying coreferences between entities extracted from data (e.g, Tweets, Posts) and those exist in the Budget-KB~\cite{Beheshti2016}. More clearly, we find the similarity among the data objects in Tweets (e.g., named entities that have been extracted from the text of the Tweet) and the entities in the Budget-KB (e.g., keywords and named entities - such as hospitals, GPs and drugs - related to health). 

We have designed a similarity API to find similarity not only among strings, numbers and dates, but also among entities (e.g., finding similarity among the attributes and their values), using a wide range of similarity techniques such as dice, cosine, TF-IDF, jaccard, euclidean, city block and levenshtein similarity techniques. For example `Bankstown-Lidcombe Hospital' is related to an item in our Knowledge Lake (Budget-KB) or an external Konwledge Graphs (e.g., Google-KG or a Webpage in Wikipedia) ~\footnote{https://en.wikipedia.org/wiki/Bankstown\_Lidcombe\_Hospital}.

\begin{figure}
\centering
\includegraphics[width=1.1\textwidth, height=8in]{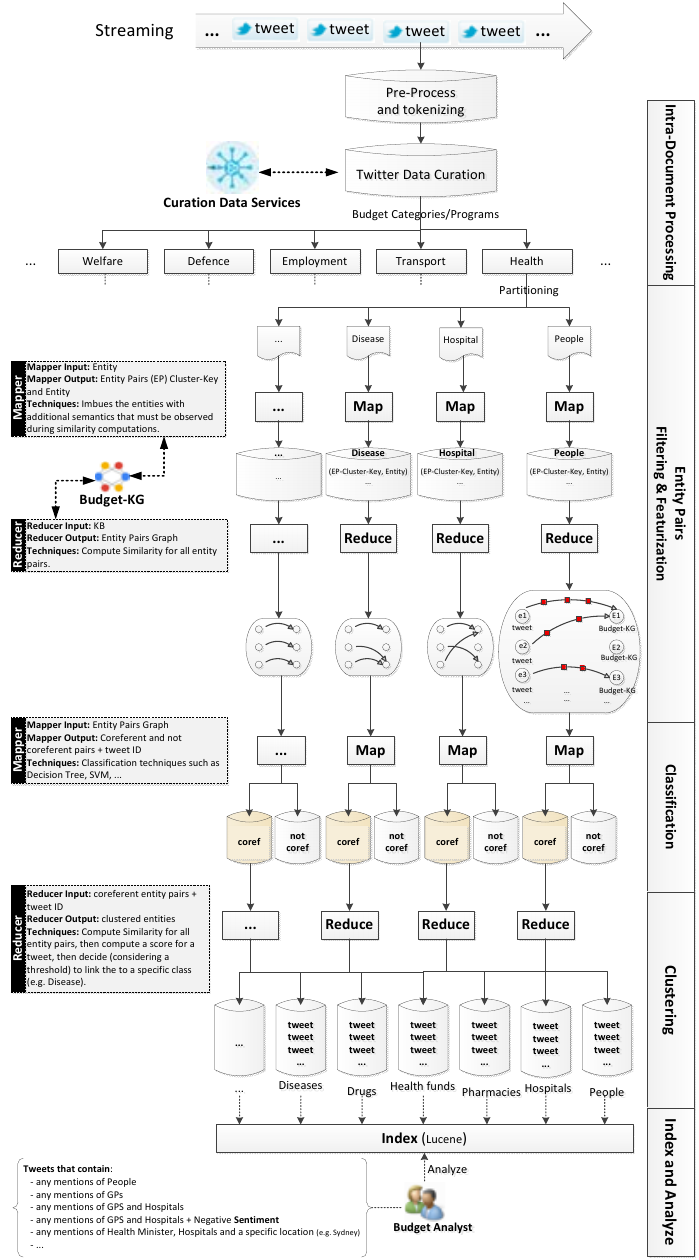}
\caption{ Scalable CDCR-similarity process.~\cite{beheshti2018datasynapse}}
\label{fig:scalability}
\end{figure}

\textbf{Scalability:} To provide a scalable approach, we divide the CDCR-Similarity process into several stages and assign each stage into a specific MapReduce (MR) job. In the first MR job, we pre-process the information item based on the social network schema. After this phase, we use the curation micro-services to generate the surface-level and semantic-level features. In the final MR job, we generate the (cross-document) co-reference entities and classify them into related summaries to assist analysts in deriving insights from the contextualized knowledge (Figure~\ref{fig:scalability}).  For example, consider an analyst who is interested in identifying Tweets on Twitter discussing a social issue related to `health'. Identifying such Tweets is largely subjective: an analyst may consider a Tweet relevant to `health' social issue if it only contains the `Health' keyword. While another analyst may consider a Tweet as relevant to `health' social issue if it contains mentions of current Australia's health minister `Hon Greg Hunt' and a negative opinion (i.e., a negative sentiment). To respond to the analyst needs in extracting the information, we generate the following summaries using the adopted CDCR process:

\begin{enumerate}
    \item Keyword-based summaries (in the category of Lexical-based features): for example, the feature keyword (`health') can be used to identify Tweets that contain mentions of the keyword `health'. 
    \item Named-entity summaries (in the category of Natural language-based features): for example, the feature named-entity (`Hon Greg Hunt', person) can be used to identify tweets that contain mentions of Australia's health minister Mr Hunt. 
    \item Negative-sentiment summaries (in the category of Metadata-based features): for example, the feature Sentiment(‘Negative') can be used to identify Tweets that express a negative opinion. 
\end{enumerate}

\textbf{Step 3: User-Guided Insight Discovery.}
The final step is to assist the analyst in extracting information relevant to her information needs through querying the Knowledge Lake. The preceding steps focused on the extraction of raw data, followed by contextualizing the data. However, even when an analyst has a clear goal of her information needs, it is important to pinpoint her required insight effectively. For example, the analyst may assume a Tweet as relevant to Australian budget if the Tweet has mentions of Australia (an instance of type Location) and Tweeted on or around 3 May 2017 (in Australia, the Treasurer handed down the budget each year on 3 May). 



To assist analysts to formulate their information needs through our proposed summaries, we relied on curation rules (See Chapter~\ref{Chapter4} for detailed description on data curation rules). We define a curation rule as a set of features in forms of:

\begin{center}
$< Feature > ::= < Dataset >.<Function>.<Operator>(<string|integer|boolean>) $\\
$< Rule > ::= < Feature_1 > [AND|OR|NOT < Feature_2 >]$
\end{center}
Where $Dataset$ represents the source a rule operates for curating the data. $Function$ performs the curation task, and $Operator$ represents the condition for a rule to curate an item. For example, for curating Tweets contains `health' and `Hon Greg Hunt' the curation rule will be inform of:

\begin{center}
    $Rule_1\ = Tweet.Keyword.Contains(`Health')\ AND\ Tweet.Entity.Person(`Hon Greg Hunt')$
\end{center}
or to extract a Tweet that contains health with negative sentiment, the analyst may write a rule as:
\begin{center}
    $Rule_1\ = Tweet.Keyword.Contains(`Health')\ AND\ Tweet.Sentiment.Negative(`true')$
\end{center}


\section{Implementation and Experiment}
\label{result_3}

\subsection{Implementation}
We identify and implement a set of APIs and made them available (on GitHub~\footnote{https://github.com/unsw-cse-soc/Data-curation-APIs.git}) to researchers and developers to assist them in adding features easily
-- such as extracting keyword, part-of-speech, and named-entities (e.g., persons, locations, organizations, companies, products, diseases, drugs, etc.) providing synonyms and stems for extracted information items leveraging lexical knowledge bases for the English language (e.g., WordNet),
linking extracted entities to external knowledge bases (e.g., Google Knowledge Graph and Wikidata),
discovering similarity among the extracted information items, (e.g., calculating the similarity between string, number, date and time data), classifying, sorting and categorizing data into various types, and indexing structured and unstructured data - into their applications. The technical implementation of these APIs can be found in Chapter~\ref{Chapter6}.

\subsection{Dataset}
The Australian government budget sets out the economic and fiscal outlook for Australia and shows the government's social and political priorities. The Treasurer handed down the budget 2016-17 at 7.30 pm on Tuesday 3 May 2016. To properly analyze the proposed budget, we have collected all Tweets from one month before and two months after this date. In particular, for these three months, we have selected 15 million Tweets, persisted and indexed them in MongoDB~\footnote{mongodb.com}. We analyzed the performance of our approach by examining its accuracy using precision and recall. Besides, we study the efficiency of our approach over 1 million Tweets, of which 409,364 were identified as relevant to the `health' category.

\subsection{System Setup}
All the experiments were performed on Amazon EC2 machines (aws.amazon.com/ec2), Sydney Australia region, using instances running Ubuntu Server 14.04. To demonstrate the usability of our approach, we experiment its \emph{Accuracy}, using metrics such as recall and precision. We also, demonstrate the \emph{efficiency} of our approach in pairing entities in term of execution time. For the efficiency, we have scaled the experiment over three different configurations on Amazon EC2: single machine, four machines and eight machines.

\subsection{Evaluation}
\begin{figure}
\centering
\includegraphics[width=1.1\textwidth]{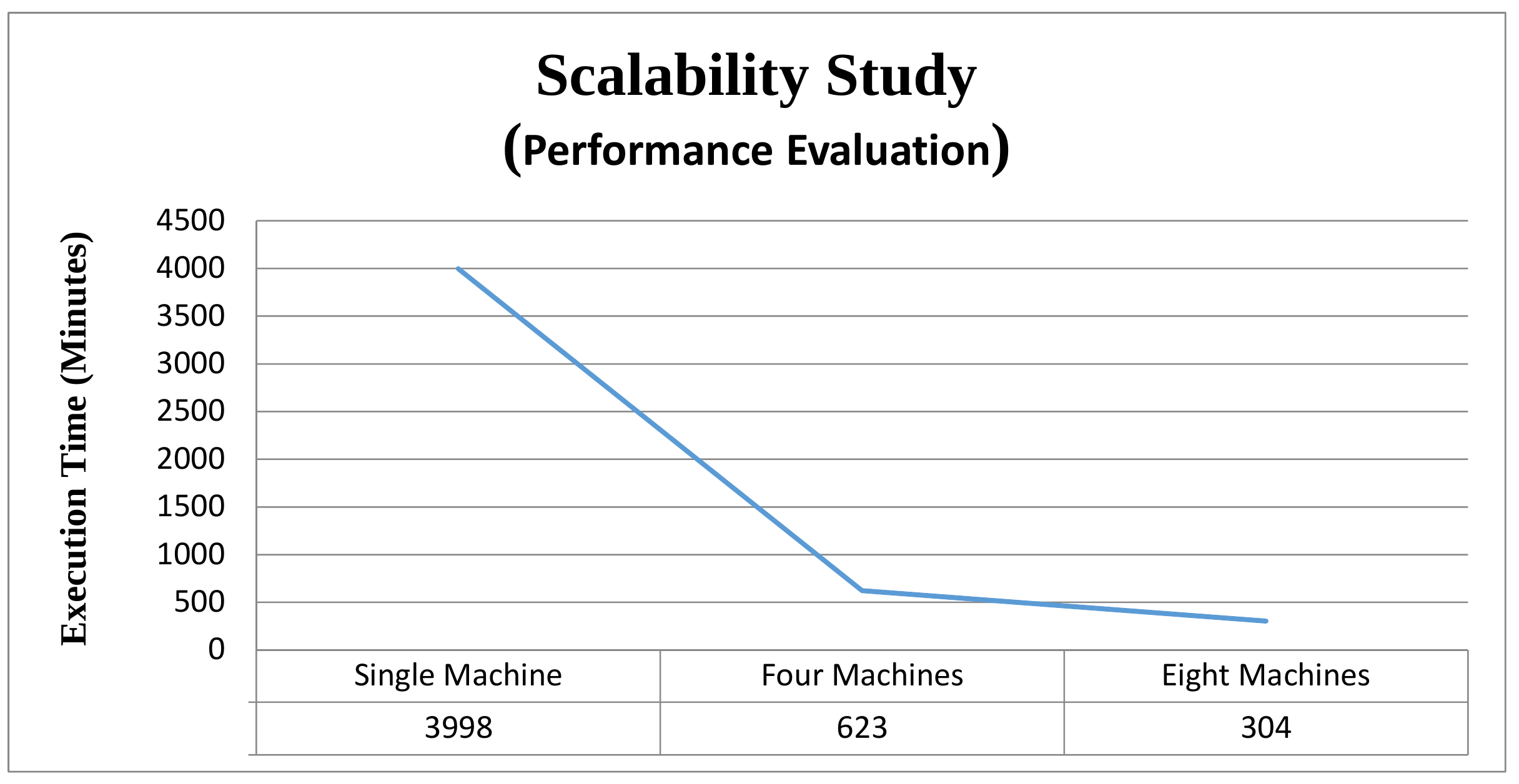}
\caption{ Execution Time: one machine, four machines, and eight machines.}
\label{fig:performance}
\end{figure}

\begin{figure}
\centering
\includegraphics[width=1.1\textwidth]{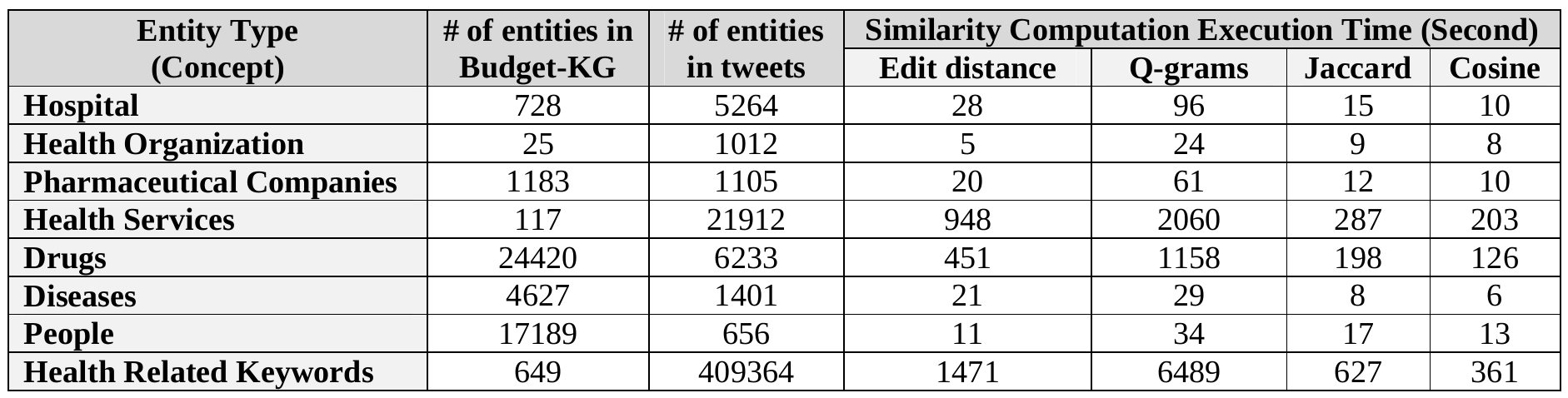}
\caption{ Execution time of linking entities and categories using different similarity metrics. }
\label{fig:similaritypair}
\end{figure}

For the initial evaluations, we focus on efficiency of our approach in terms of paring entities (e.g., hospitals, health organizations, pharmaceutical companies, health services, drugs, diseases and people) with the categories in the Budget-KB. We examined the efficiency using different similarity metrics including: edit distance, Q-grams, jaccard, and cosine. 
We calculated the \emph{average} similarity score and use it for linking the entities and categories. Here, edit distance and Q-grams are character-based functions, while jaccard and cosine functions are token-based functions. Figures~\ref{fig:similaritypair} and~\ref{fig:performance} shows the execution times taken in making coreference decisions by comparing entities. In particular, generating entity pairs and computing similarity among them is a time-consuming task and requires high-performance computing resources on very large datasets such as Twitter. For example, for around 20k entities, the algorithm generated about 9 million pairs which highlight that pairwise entity comparison will become exponential across Tweets (for further detail on efficiency, please see~\cite{beheshti2018datasynapse}).

\subsection{Analysing Budget-KB Accuracy}
In this section, we demonstrate the accuracy of our approach in curating data using metrics, such as precision and recall. Precision is the number of correctly identified coreferent pairs divided by the total number of identified coreferent pairs, recall is the number of correctly identified coreferent pairs divided by the total number of true coreferent pairs, and F-measure is the harmonic mean of precision and recall. Let us assume that TP is the number of true positives, FP the number of false positives (wrong results), TN the number of true negatives, and FN the number of false negatives (missing results). Then, Precision= $\frac{TP}{TP+FP}$, Recall=  $\frac{TP}{TP+FN}$, and F-measure=  $\frac{2*Precision*Recall}{Precision+Recall}$.

We created a set of classifiers using machine learning algorithms to classify Tweets relevant to `health'. For creating classifiers, we have used two different approaches. First, we used the classic Keyword Matching (KEYM) approach~\cite{lee2013real}, which uses a Bag of Words (BoW) to identify health-related Tweets. Next, we have used our proposed featurization technique, which uses a variety of features for annotating and extracting Tweets. The goal of this experiment is to identify the performance of the proposed featurization method in boosting the performance of classifiers in training machine learning models. The approach that better improves the precision and recall would have considered as the successful one. For the proposed featurization technique, we have used three different features: (i) We used Budget-KB Entity Matching to link an entity in a Tweet to the entities in the KB, (ii) We have used Google Knowledge Graph API to indicates the existence of a health-related entity in the Google KG with an entity in a Tweet, and (iii) We have used URL Entity Matching to analyze the content of the URLs provided in the tweet and to identify the health-related entities and keywords. Then, we created a set of binary classifiers to classify the extracted Tweets. A binary classifier receives a collection of input data as the training set and creates a model to identify an item is relevant or not. For example, in our scenario, a classifier predicts a Tweet is related to `health' or not. Next section explains how we created the training set and classifiers in detail. 

To train binary classifiers, we created two different training sets. The first training set was created through KEYM approach, and the second training set was created through the proposed featurization technique. Considering that we have around 15 million tweets, using the KEYM approach, we identified 50 thousand Tweets as relevant to health. Next, we applied some preprocessing on tweets: for example, we eliminated Tweets containing less than four keywords, Tweets that contain non English words, and the URLs. We also removed the duplicate Tweets (e.g., retweeted tweets). Finally, we have generated around 20 thousand preprocessed Tweets. We labelled the extracted Tweets as relevant and fed them as an input to the machine learning algorithm (naiveBayes, KNN and SVM classifiers). In addition, we feed the classifier with a dataset of irrelevant Tweets from our previous works~\cite{beheshti2018crowdcorrect} which manually labelled through crowds. For the test set, we have manually labelled 600 Tweets which contain 322 health-related and 278 unrelated Tweets. We consider each Tweet as a document, and process it by stemming, removing stop words, punctuations and numbers, and lower casing the entire Tweet. We followed the same procedure to create the second dataset for evaluating the performance of the featurization technique. 

\begin{figure}
\centering
\includegraphics[width=1.1\textwidth]{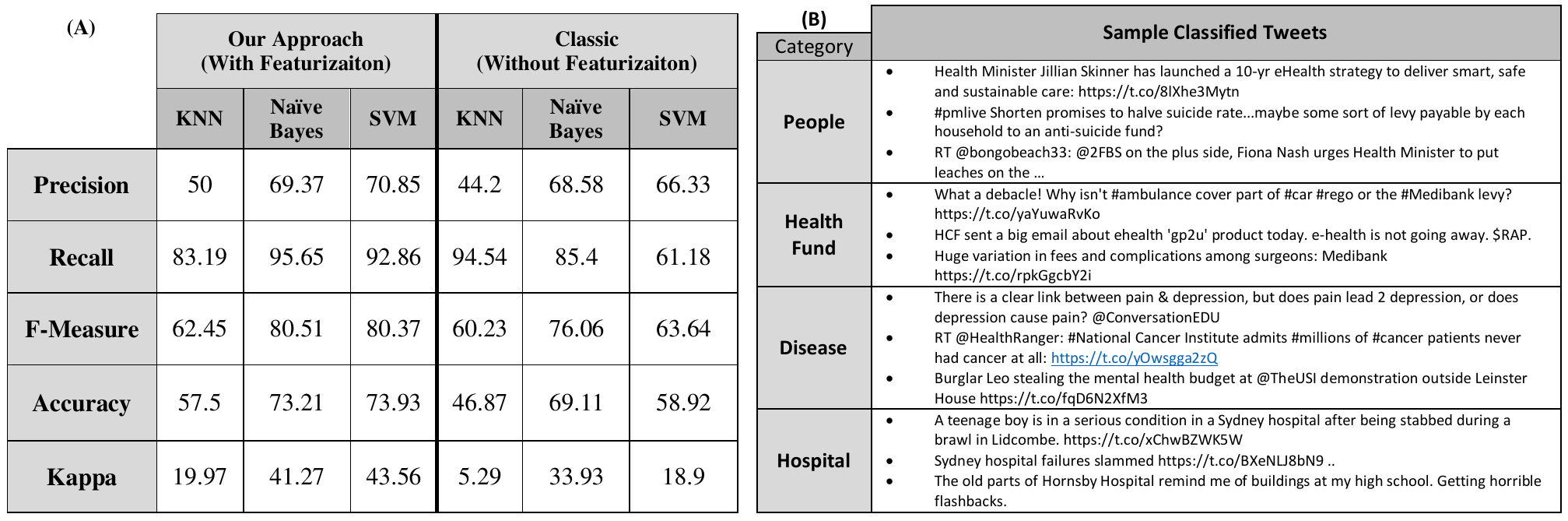}
\caption{(A) Comparison between featurized and classical classification, (B) Sample of classified Tweets Using the proposed solution}
\label{fig:classifyresult}
\end{figure}

As illustrated in Figure~\ref{fig:classifyresult}, our proposed approach significantly improves the quality of extracted knowledge compared to the classical curation pipeline (in the absence of feature extraction and domain-linking contextualization). The proposed technique could identify many relevant Tweets (that should be contained in the returned results) and accordingly, the accuracy of the result can be improved. Notice that, accuracy is the proximity of measurement results to the true value, and calculated as 
$ accuracy = (TP+TN)/(TP+TN+FP+FN))$, where $TP$ is True Positive, $TN$ is True Negative, $FP$ is False Positive, and $FN$ is False Negative. As an ongoing work, to improve the precision and recall:~(i) We are going to use rules in combination with the machine learning approach for further filtering results,~(ii) We will use some refinement techniques, e.g., to merge the results obtained from KEYM approach with the Budget-KB,~(iii) We will add more feature to our model, and~(iv) We are enhancing the model, currently supporting unigram, to support n-grams and leverage multiple machine learning techniques for further filtering the results.

\textbf{Social Issues.}
Identifying social issues is challenging as it requires the budget analyst to understand the candidate Tweets properly. To provide the candidate Tweets, we identified the Tweets having negative sentiments.
To achieve this goal, we have used classified Tweets. For example, our proposed approach classified
5823 Tweets linked to anxiety,
2934 Tweets related to diabetes,
22430 Tweets related to cancer, and
16931 Tweets associated with Mental Health.
We have reused the sentiment classifier implemented in the Apache PredictionIO (http://prediction.io) to identify the tweets with negative sentiment.
For example, out of 2934 diabetes-related Tweets, the algorithm identified 615 tweets with negative sentiment.
As another example, we have identified 1549 tweets with negative sentiment in the mental health category.
Later on, the analyst is able to use the proposed declarative language to analyze the candidate tweets based on a specific goal, e.g., identify Tweets discussing a social issue related to health and specifically about the medicare~\footnote{Australian federal health insurance program} or an issue related to public hospital services.

\section{Conclusion and Future Work}
\label{Conclusion_35}

Big data analytics have become the quintessential engine for extracting knowledge and deriving insights from the vastly growing amounts of local, external and open data.
With the advent of widely available data capture and management technologies, coupled with intensifying global competition, fluid business and social requirements, organizations are rapidly shifting to data-fication of their processes~\cite{ProcessAnalytics}.
For example, understanding and analyzing open data now is recognized as a strategic priority for governments.
In this context, the \emph{data curation} process becomes a vital analytics asset for understanding the data.
To address this need, we have introduced a general-purpose data curation pipeline.
The goal is to facilitate analytical tasks through transforming raw data into featurized (through the proposed feature engineering approach) and after that contextualized (requires contracting the domain knowledge and linking extracted data to that) data~\cite{tabebordbar2020feature}.
We have designed and implemented a set of reusable APIs to assist analyst through the curation process. 
As an ongoing and future work, we are extending the Budget-KB by identifying further relevant concepts and their instances in other budget categories program.

In the next chapters, we are extending our declarative rule language. We explain techniques to adapt curation rules to enable analysts to query and analyze the data more conveniently. We also discuss how a Knowledge Lake enhances users understanding of data to better formulate their preferences.

 \chapter{Feature-Based Rule Adaptation in Dynamic and Constantly Changing Environment}

In this chapter, we present an adaptive technique for adapting data curation rules in a dynamic and changing environment. Curation rules have been used increasingly to augment learning algorithms, in cases where algorithms are not working well or lack from enough training data. However, in dynamic curation environments, there is a need for an analyst to adapt rules to keep them applicable and precise. Rule adaptation has been proven to be painstakingly difficult, error-prone, and time-consuming. We proposed an adaptive approach for adapting curation rules. Our approach utilizes an online learning algorithm to learn the optimal modifications for a rule based on the feedback collects from the curation environment. We also proposed a summarization technique to boost rules to curate a larger number of items.

In Section~\ref{intro_4}, we present an overview of data curation rules. We discuss the related works in Section~\ref{related_work_4}. Then, in Section~\ref{problemstatement_4}, we explain the research problem. We discuss our solution in Section~\ref{adaptive_rule_adaptation_4}. Next, we present the performance of our approach on three different curation domains: mental health, domestic violence, and budget. The experiment results showed our approach could significantly improve the precision of rules in annotating data (by as much as 29\% precision compared to the initial results). Finally, we discuss future works and conclude the chapter in Section~\ref{conclusion_4}.

The content of this chapter is derived from the following papers:
\begin{itemize}
    \item  A Tabebordbar, A Beheshti, B Benatallah, and M C Barukh, \textbf{Adaptive rule adaptation in unstructured and dynamic environments}, International Conference on Web Information Systems Engineering, Springer, 2019, pp. 326–340. (ERA Rank A).
    
    \item Tabebordbar A, Beheshti A, Benatallah B, Barukh MC. \textbf{Feature-Based and Adaptive Rule Adaptation in Dynamic Environments.} Data Science and Engineering. 2020 Jun 25:1-7.
    
    \item A Tabebordbar and A Beheshti, \textbf{Adaptive rule monitoring system}, 2018 IEEE/ACM 1st International Workshop on Software Engineering for Cognitive Services (SE4COG), IEEE, 2018, pp. 45–51. (Best paper award).

\end{itemize}

\section{Introduction}
\label{intro_4}

Data curation indicates processes and activities related to the integration, annotation, publication and presentation of data throughout its lifecycle~\cite{beheshti2018datasynapse}.
One category of data curation is data annotation, which aims at labelling the raw data to generate value and increase productivity.
Data annotation has been used extensively in various computational machine learning algorithms for information extraction, item classification, record-linkage~\cite{ratner2017snorkel,ratner2017snorkel2,beheshti2017coredb}.
However, in dynamic environments, e.g., Twitter and Facebook, where data is continuously changing, relying on pure algorithmic approaches does not scale to the need of businesses that need to annotate data over an extended period. Because algorithms make a prediction based on the historical data only. While, in dynamic environments, the distribution of data is changing and algorithms need to be updated to capture the changes, which is expensive and time-consuming.

In recent years, several pioneering solutions (e.g.,~\cite{gc2015big,bak2014rule,milo2018interactive,liu2010refining,sun2014chimera,xie2017automatic}), have been proposed to augment algorithms with rule-based techniques. 
Rules can alleviate many of the shortcomings inherent in pure algorithmic approaches. Rules can be written by non-technician analysts, which is less expensive than training algorithms through experts~\cite{gc2015big}. Updating rules is faster than training algorithms and can supplement algorithms in cases they are not working well~\cite{sun2014chimera}.

To keep a rule applicable and precise~\cite{gc2015big,bak2014rule,milo2018interactive,sun2014chimera,xie2017automatic} there is a need for an analyst to adapt~\footnote{The process of modifying a rule to become better suited to the curation environment.} the rule based on changes in the curation environment. 
Rule adaptation has been proven to be painstakingly difficult as the analyst needs to understand the context of data and the impact of modifications she applies to the rule ~\cite{milo2016rudolf}. In many cases, the analyst needs to apply different changes to identify the optimal one. This problem exacerbated in dynamic curation environments as adaptation is not a single rule modification task, and the rule needs to be updated over its life-time.

In this chapter, we take the first step toward creating adaptive rule adaptation model in dynamic and constantly changing environments. While previous approaches rely on analysts to identify the optimal modifications for rules, we propose a different learning task. We focus on incrementally adapting a rule based on the changes in the curation environment. Hence, we focus on offloading analysts and updating a rule autonomously. We do this by utilizing a Bayesian multi-armed-bandit algorithm, which learns the optimal modification by observing rules performance over time. Besides, previous systems adapt rules at the syntactic level, e.g., keyword, regular expression. Syntactic level adaptation limits rules ability in annotating data as rules skip a large number of semantically related items. However, our work focus on coupling syntactic level features with conceptual features to boost rules to annotate a larger number of items. 

Overall, our solution is made up of the following stages: (1) Each time a rule annotates a set of items, we extract a set of candidate features (e.g., syntactic and conceptual features) as the potential modifications. (2) Then, a Bayesian multi-armed-bandit algorithm determines the optimal modification for the rule by estimating a probability distribution for candidate features. (3) Over time, by annotating more items, the algorithm learns the performance of candidate features better and modifies the rule to keep the rule applicable and precise.

\begin{figure}[t]
\includegraphics[width=\linewidth, height=3.2in]{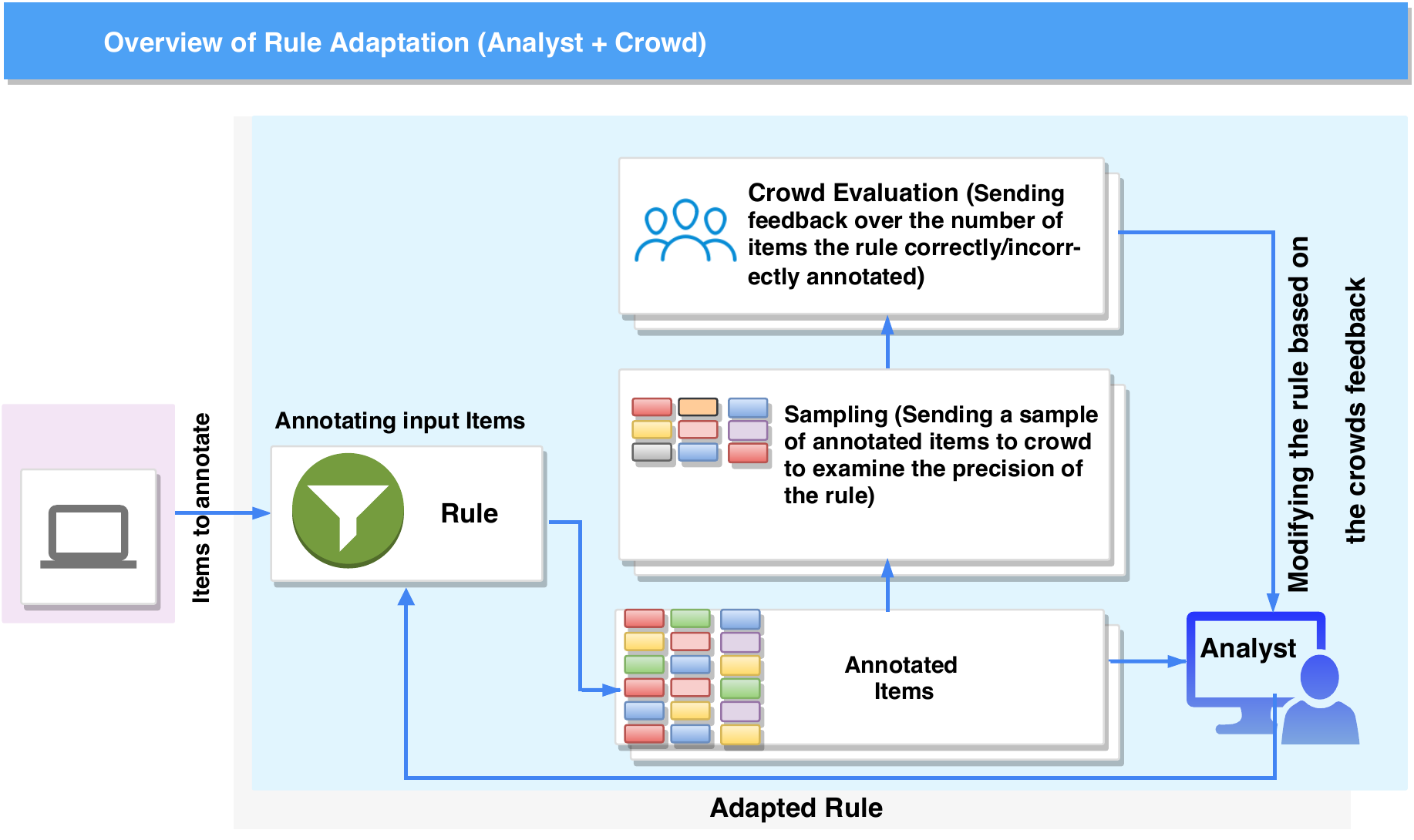}
\caption{The overview of adapting rules through analysts and crowd workers.}
\label{fig:current_approach}
\end{figure}

\noindent\textbf{Example: Rule Adaptation Through analyst.}
Consider a government that intends to analyze the quality of their social services, e.g., mental health services, domestic violence services, aged care services, based on citizens' opinions. Social media is one of the sources that decision-makers may rely on to understand public satisfaction levels about their services. However, this is particularly challenging to take a representative sample of data to train learning algorithms to analyze citizens' opinions. Because social media users are contributing a million pieces of data every day trickling in through Tweets and Posts.
 Alternatively, rules can be used to augment learning algorithms for annotating data in dynamic environments. However, as public use different keywords and hashtags for expressing their opinions, rules need to be updated to remain applicable and precise. For example, consider $Rule_1$, which annotates Tweets relevant to `Mental Health', if a Tweet contains both `Mental' and `Service' keywords:
 \begin{center}
 $Rule_1 =$ \emph{\textbf{IF} tweet contains \emph{(`Mental')} keyword\ AND\ tweet contains \emph{(`Service')} keyword \textbf{THEN} tag with `Mental Health'}.    
 \end{center}
 
  To keep $Rule_1$ applicable, the analyst needs to modify the rule based on changes in the curation environment.
  This task is particularly challenging as the social media data is `ever changing and never ending'~\cite{gc2015big}. Besides, the analyst does not know the universe of data, or may need to consider too many complex conditions, which might be difficult to integrate with the rule. Thus, the analyst may not craft the perfect rule that adequately annotates data. Figure~\ref{fig:current_approach}, shows a typical workflow of adapting rules through the analyst.

\textbf{Contributions.} This chapter makes the following contributions.
\begin{enumerate}
\item Rule adaptation is error-prone and challenging as analysts need to examine different rule modifications to identify the optimal one. We propose an autonomic approach that adapts rules without relying on analysts. We utilize a Bayesian multi-armed-bandit algorithm that learns to modify a rule-based on changes in the curation environment. 

\item To frame rule adaptation as a Bayesian multi-armed-bandit problem, we propose a reward and demote schema. The schema assigns a reward if the algorithm identifies a rule correctly annotated an item and demotes a rule if it annotates an irrelevant item. Over time, the algorithm (by observing rewards and demotes) learns a better adaptation for the rule.

\item We proposed a summarization technique to boost rules to annotate a larger number of items. The approach identifies the semantical relationship between keywords to annotate data at the conceptual level.
\end{enumerate}

\section{Related Works}
\label{related_work_4}
In this section, we discuss prior works related to rule adaptation (Section~\ref{adapt_rule}), and online learning algorithms (Section~\ref{mab}). In particular, we discuss the usage of a Bayesian multi-armed-bandit algorithm in unstructured and constantly changing environments. Besides, we consider it appropriate to discuss approaches proposed for feature extraction and how they differ from our proposed summarization technique (Section~\ref{feature_eng}). \newline

\subsection{Rule Adaptation}
\label{adapt_rule}
Rule adaptation is a continuous process focused on modifying a rule to fit the rule to the curation environment better. However, rule adaptation is a challenging and error-prone task. Thus many solutions~\cite{milo2018interactive,liu2010refining,volkovs2014continuous,he2016interactive,gc2015big,xie2017automatic} have been proposed to assist analysts in adapting rules. Several solutions ~\cite{milo2018interactive,liu2010refining,volkovs2014continuous,he2016interactive,milo2016rudolf} focused on interactively adapting rules. In these solutions, a system proposes possible adaptations for a rule, and an analyst adapts the rule through interacting with the system. For example, Milo et al.~\cite{milo2018interactive}, proposed a cost-benefit approach for generalizing or specializing fraud detection rules. The approach developed a heuristic algorithm to interactively adapt rules with domain experts until the desired set of rules is obtained. Volks et al.~\cite{volkovs2014continuous}, proposed a cost function to adapt the integrity constraint (IC) rules. The approach relies on the analyst feedback to update the cost function and resolve the inconsistencies in IC rules.
Liu et al.~\cite{liu2010refining}, proposed an interactive approach for refining a rule using a set of positive and negative results. The method uses a provenance graph to identify candidate changes that can eliminate negative results. However, these solutions focus on adapting rules that operate on structured data, where a rule may adapted with a limited number of features. Besides, many of these solutions assume the analyst can access to a ground truth, e.g., a dataset of items tagged with the correct label, to verify the effectiveness of an adaptation.

Alternatively, to adapt rules in both unstructured and dynamic environments, some solutions ~\cite{gc2015big,xie2017automatic,sun2014chimera} focused on augmenting interactive rule adaptation systems by coupling crowds with analysts. These solutions rely on crowd workers to determine the precision of rules. For example, Xie et al.~\cite{xie2017automatic}, proposed an approach for validating rules for information extraction purposes. The approach relies on a voting technique to identify whether an adaptation of a rule produces a positive impact in extracting information or not.
GC et al.~\cite{gc2015big}, designed an interactive system by coupling analysts and crowds for adapting rules. The system verifies items annotated with a rule using crowd workers, and assists analysts in identifying the optimal modification using a relevance feedback algorithm (Rocchio). Sun et al.~\cite{sun2014chimera}, proposed a rule-based technique (Chimera) for large scale data classification systems. First, the approach identifies the misclassified items in cooperation with crowd workers. Then, forwards the items to analysts to write rules and address the errors. Bak et al.~\cite{bak2014rule}, relies on visualization by showing the result of applying a rule on a set of data records. The system requires crowd workers to verify the outcome of applying the rule on the data record, indicating the optimal adjustment for the rule.

Although coupling crowd workers with interactive systems provide more flexibility in adapting rules in dynamic environments, these systems still rely on analysts for identifying the optimal modification of rules. In contrast, our approach not only offloads analysts but also it autonomically modifies a rule regarding changes in the curation environment. 

\subsection{Multi Armed Bandit Algorithm}
\label{mab}
In this section, we discuss how a Bayesian multi-armed-bandit algorithm has been used in dynamic and constantly changing environments. This algorithm increasingly used in large scale randomized A/B experimentation by technology companies~\cite{kohavi2009controlled}. One area of work that used a Bayesian multi-armed-bandit algorithm is educational learning to facilitate the learners' learning rate. For example, Williams et al.~\cite{williams2016axis}, proposed a system (AXIS) to improve explanation generation for online learning materials by employing a combination of crowds and a Bayesian multi-armed-bandit algorithm. Clement et al.~\cite{clement2014online}, used a multi-armed bandit algorithm in intelligent tutoring systems to choose activities that provide better learning for students. Other areas that relied on a Bayesian multi-armed-bandit algorithm are feature engineering~\cite{anderson2014integrated}, gaming~\cite{liu2014trading}, and online marketing~\cite{burtini2015improving}. In this context, we follow a similar trend by employing a Bayesian multi-armed-bandit algorithm with the crowd workers. Over time, the algorithm based on the collected feedback determines an adaptation for the rule to keep the rule applicable and precise. 

\subsection{Feature Extraction}
\label{feature_eng}
In addition to interactive systems for helping analysts in adapting rules, we consider it appropriate to include approaches in feature extraction to position our proposed summarization technique. Feature extraction is the process of identifying a set of variables that best describe the data~\cite{domingos2012few}. Feature extraction is an ongoing task and requires to iteratively explore the curation environment to identify features that capture the salient aspect of data. Several approaches have been proposed to aid analysts in feature extraction (e.g., ~\cite{veeramachaneni2014towards,brooks2015featureinsight,wiatowski2018mathematical,lai2018robust,chen2016deep,brooks2015featureinsight}.  For example, Anderson et al.~\cite{brainwash}, proposed BrainWash, a system that provides a pipeline to ease the process of feature extraction in large datasets. The system focused on helping a user to explore, extract, and evaluate features faster.  Cheng et al.~\cite{cheng2015flock}, relied on crowd workers for feature extraction. The approach refines the performance of machine learning algorithms based on the feedback receive from crowds. Veeramachaneni et al.~\cite{veeramachaneni2014towards}, proposed an approach to engage crowd workers in extracting features and predicting students stopout on Massive Open Online Courses (MOOC) systems. The approach provides a pipeline for evaluating and examining the relevancy features through crowd workers. 

Another type of works focused on easing feature extraction through visualization techniques. For example, Patel et al.~\cite{patel2011using}, relied on visualizing the confused region of machine learning classifiers to help analysts in extracting features. Brooks et al.~\cite{brooks2015featureinsight}, provides a visual summary of the data to aid a user to create a dictionary of features. Stoffel et al.~\cite{stoffel2015feature}, relied on visualization for examining machine learning features error. The system iteratively interacts with a user to remove ineffective features.

In contrast, we propose a summarization technique that identifies the semantical relationship among keywords and extracts features at the conceptual level. Each conceptual feature represents a group of semantically related keywords, which boosts rules to annotate a larger number of items.

\section{Preliminaries and Problem Statement}
\label{problemstatement_4}
\noindent We first introduce the component of rules used in this chapter (Section~\ref{prelim}). We then describe the problem in Section~\ref{problem}. Finally, we provide an overview of our solution in Section~\ref{solution}.
\subsection{Preliminaries}
\label{prelim}
\noindent\textbf{Feature. }We express a rule $R$ in forms of features, where each feature $f \in R$ corresponds to a function in forms of

\begin{center}
$\langle Dataset.Function.Operator \rangle\ \rightarrow Value$
\end{center}

where~$Dataset$ is the data source such as Twitter and Facebook,~$Function$ performs the curation task (e.g., feature extraction),~$Operator$ represents the condition for a feature to curate the data, and~$Value$ is the output of a feature. Examples of a feature are extraction functions, e.g., named-entities, or similarity extraction. Expressing a feature as a function allowing us to leverage the standard data-types as the feature's operator. For example, if a feature operates over textual data, the operator for the feature will include string operators, such as \textit{contains} and \textit{exact}. Similarly, if a feature curates integer data the feature will include integer operators, such as \textit{equals} and \textit{less-than}. As an example, consider the feature~$f_1 =\ \langle Tweet.Keyword.Contains\ (`Mental')\ \rangle$, which curates Tweets that contain `Mental' keyword. In this example, $Tweet$ represents the dataset the feature operates for curating the data,~$Keyword$ represents the function of the feature, and~$Contains\ ('Mental')$ is the operator and represents the condition for curating a Tweet.

\textbf{Rule. } We represent a rule $R$ as a tree of features, where each feature $f \in R$ can have $K$ children. We denote a path $p$ in the tree as a sequence of features $f_1,...,f_m$, where $f_1$ represents the root feature and $f_m$ represents the last feature in the path. More precisely, a path $p$ is a conjunction of features in the form $f_1\ \land\ ... \land f_m$. To curate an item with a rule, the item should be annotated with all features within a path. Notice that, we do not require inventing our rule language. Rather, the benefit of rules being expressed as features, we can adopt any suitable functional or rule-expression language for our purpose.

\begin{figure}[t]
\includegraphics[width=\linewidth, height=2.4in]{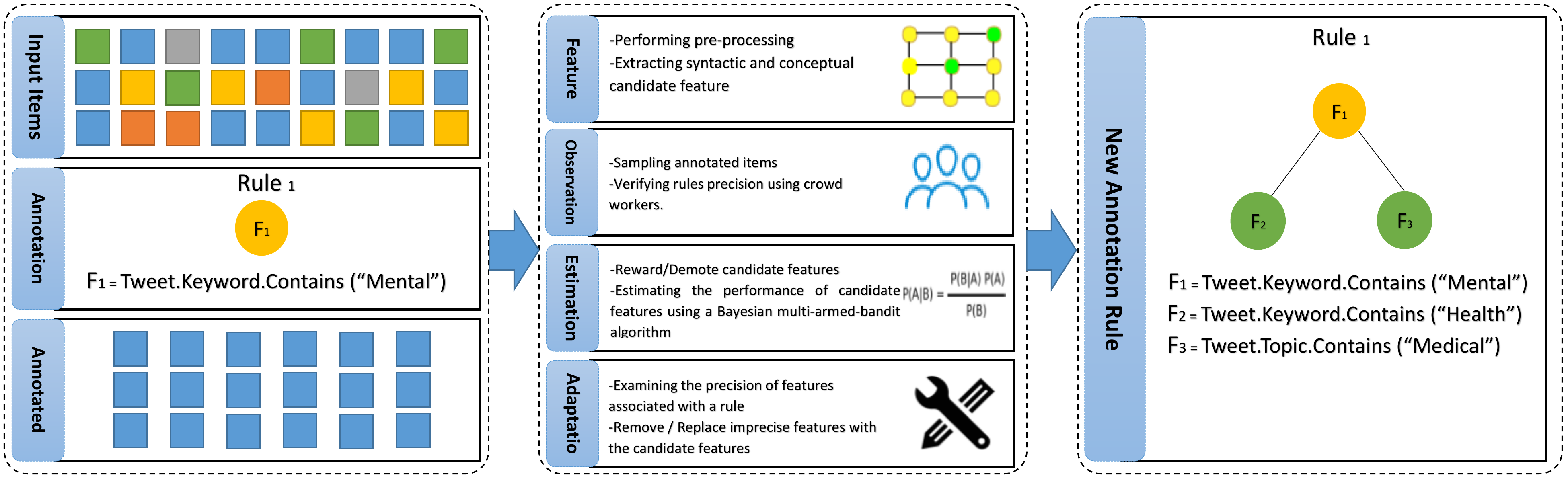}
\caption{The overview of the proposed approach for adapting rules~\cite{tabebordbar2019conceptmap}.}
\label{fig:adapt}
\end{figure}

\textbf{Tag. } A Tag is the label, e.g., `Mental Health', a rule assigns to a curated item, e.g., Tweet, to describe the item. In this chapter, we use the term \emph{tag} and \emph{annotate} interchangeably. As an example, consider the rule presented in Figure~\ref{fig:adapt}. This rule is made up of three features $\{f_1, f_2, f_3\}$, and tags a Tweet with `Mental Health', if the Tweet curated with features $\ f_1\ \land\ f_2$, or $\ f_1\ \land\ f_3$. More clearly, $Rule_1$ tags a Tweet with `Mental Health', if the Tweet contains `Mental' and `Health' keywords or the Tweet contains `Mental' and a keyword related to `Medical' topic.

\subsection{Problem Statement}
\label{problem}
In the followings, we discuss two major problems in rule-based systems.\newline

\noindent\textbf{Adaptation Through Analyst. }
Typically, to adapt a rule, an analyst examines correctly/incorrectly annotated items to identify the potential modifications that make the rule precise~\cite{milo2018interactive,milo2016rudolf,gc2015big,liu2010refining}. However, rule adaptation is challenging and error-prone as the analyst needs to evaluate the impact of each modification she applies to the rule. 
 Such a problem is categorized under the category of \emph{online learning} problem, where an analyst does not have access to the entire knowledge to craft the adequate type of rule. Instead, over time she learns to better adapt a rule through examining the annotated items. 

To offload analysts from adapting rules, we formulated the problem as a Bayesian multi-armed-bandit algorithm. The algorithm is suitable when the required information for making a decision is serially provided piece-by-piece. Each time a rule annotates a set of items, the algorithm collects feedback over the number of items the rule correctly/incorrectly annotated, over time by receiving more feedback the algorithm learns to adapt the rule better. 
 
 For example, consider  a rule $R$ that operates over a dataset and must annotate data with a threshold $\jmath$. Assume that at time $\tau_i$ rule $R$ annotated a set of items $I_{\tau i}\ = \{i_1, i_2, ..., i_n\}$. We denote by $P[R_{\tau_i}]$ as the precision of the rule observed at time $\tau_i$. Our algorithm adapts rule $R$ at time $\tau_{i+1}$, where $P[R_{\tau_i + 1}]\ >\ \jmath$.\newline

\noindent\textbf{Syntactic Level Data Annotation. }
Typically, an analyst adapts a rule at the syntactic level, e.g., keywords, regular expressions. Using syntactic level features allowing the analyst to more conveniently modify a rule by replacing irrelevant keywords or phrases with new ones. However, relying on syntactic level features limits the capacity of a rule in annotating data as these features skip a large number of semantically related items. For example, consider the rule:
\begin{center}
$Rule_{11}=$ $Tweet.Keyword.Contains(`Mental')$ ${\land}$ $Tweet.Keyword.Contains(`Health')$ $:\ `Mental Health'$
\end{center}
This rule tags a Tweet if the Tweet contains `Mental' and `Health' keywords. However, there exists a large number of Tweets relevant to `Mental Health', which could not be tagged with the $Rule_{11}$ as those Tweets may not contain both `Mental' and `Health' keywords. 
\subsection{Solution Overview}
\label{solution}
The overview of our proposed solution shown in Figure~\ref{fig:adapt}. The approach consists of four steps, feature extraction, observation, estimation, and adaptation. \newline

\noindent\textbf{Feature Extraction. }
The initial step in the workflow is feature extraction, which extracts a set of candidate features $T = \{t_1, t_2, ..., t_n\}$ from annotated items. The approach extracts candidate features both at the syntactic and conceptual levels. Each syntactic level feature represents a keyword extracted from items annotated with a rule, while a conceptual feature represents a group of semantically related keywords. For extracting conceptual features, we propose a summarization technique, which is made up of two steps: (1) we map each syntactic level feature to an abstract concept using a knowledge base, and (2) we group features with the same concept and consider each group as a conceptual candidate feature. In Section~\ref{featureextraction}, we accentuate how our approach extracts candidate features for adapting a rule.\newline

\noindent\textbf{Observation.}
The second step in the workflow is observation, which gathers feedback to update a Bayesian multi-armed-bandit algorithm about changes in a curation environment. For gathering feedback, we rely on the crowd workers~\footnote{https://www.figure-eight.com/}. Each time a rule annotates a set of items $I= \{i_1, i_2, i_3, ..., i_n\}$, the algorithm receives feedback over a sample of annotated items $S =\{i^{\prime}_1, i^{\prime}_2, i^{\prime}_3, ..., i^{\prime}_n\}$, where $S \subset I$ to identify the latest rule performance in annotating the data. Crowds verify whether a rule correctly tagged an item or not. In Sections~\ref{observation} and~\ref{gatherfeedback_4}, we review how crowd workers contribute in verifying items.\newline

\noindent\textbf{Estimation. }
The third step in the workflow is estimation, where a Bayesian multi-armed-bandit algorithm determines the performance of candidate features by estimating a probability distribution $\theta$. The algorithm calculates the performance of features using workers collected feedback

To formulate workers feedback as a Bayesian multi-armed-bandit problem, we propose a reward/demote schema. Each time the rule annotates a set of items, the schema calculates a reward/demote for candidate features to update the algorithm about changes in the curation environment. In Section~\ref{estimation}, we review how the approach estimates the probability distribution for candidate features.\newline

\noindent\textbf{Adaptation. }
Given a set of candidate features $T$ along with their probability distribution $\theta$, we identify potential modifications that keeps a rule applicable and precise. We do this by removing or restricting features that deteriorate the rule performance. In Section~\ref{adaptation} we review how our approach modifies a rule.\newline

\section{Adaptive Rule Adaptation}
\label{adaptive_rule_adaptation_4}
In this section, we explain components (feature extraction, observation, estimation, and adaptation) of our proposed solution.

\subsection{Feature Extraction}
\label{featureextraction}
The first step in our workflow is feature extraction, where we extract a set of candidate features as the potential modifications for a rule. Each time a rule annotates items, we extract a set of candidate features to calculate their performance in adapting the rule. We extract two types of candidate features, syntactic and conceptual. A syntactic level feature represents a keyword within an annotated item, while the latter represents a group of semantically related keywords. Followings explain how we extract features.

\noindent\textbf{Syntactic Candidate Feature:}
\noindent For extracting syntactic candidate features, we conduct a preprocessing task on annotated items $I$. The preprocessing performs tokenization, normalization, and noise removal. In tokenization, we split each item $i \in I$ into smaller tokens. Normalization removes stop words and conduct stemming., and noise removal skips certain characters, e.g., emoji, URLs, that occur in items. We consider the remaining tokens as the candidate feature type of \emph{keyword}.

\noindent\textbf{Conceptual Candidate Feature:}
\noindent Conceptual candidate features are proposed to alleviate shortcomings exist in annotating data using syntactic features. Although syntactic features allow an analyst to modify a rule more conveniently, relying on these features cannot capture the salient aspect of data and limits rules capacity in annotating items. 
Thus, there is a need for more productive features to boost rules to annotate a larger number of items.
We propose a \emph{summarization} technique, which extracts and groups semantically related keywords and forms a concept. Summarization consists of two steps: (1) mapping, and (2) grouping. In the mapping step~\cite{pham2009time,saint2005general}, we map each syntactic feature using a knowledge base to an abstract concept and associate a descriptor to it. In the grouping step, we group features with an identical descriptor, and consider each group as a conceptual candidate feature. Following explains how our proposed technique extracts two new conceptual features using two readily available knowledge bases: WordNet~\cite{esuli2006sentiwordnet} and Empath~\cite{fast2016empath}. Algorithm~\ref{algo:featuresum} shows the pseudo-code of summarization technique.

\begin{algorithm}[t]
\SetAlgorithmName{Algorithm}{heuristic}{List of Heuristics}
\SetKwFunction{FMain}{Feature\_Summarization}
\SetKwProg{Fn}{Function}{:}{}
\Fn{\FMain{}}{
\KwIn {$T$}
\KwOut {$T^{\prime}$}
\ForEach{$t$ $\in$ $T$}{
    $set\_map.Add(Abstract[t])$\;
}
  \ForEach{$t_{map}$ $\in$ $set\_map$}{
    \ForEach{$t$ $\in$ $T$}{
    \uIf {$Abstract[t]$ == $t_{map}$}{
         $T^{\prime}[t_{map}].Add(t)$\;
         }
  }
}
\KwRet ${T^{\prime}}$\;
}
\label{algo:featuresum}
\end{algorithm}

\begin{enumerate}

\item \textbf{WordNet: } The first knowledge base, we rely on for extracting conceptual features is WordNet. WordNet~\footnote{https://wordnet.princeton.edu/} is a semantic lexicon, which grouped English words into sets of synonyms called synsets. We use WordNet to identify semantic relations between keywords using their hypernyms relation. A hypernym is a relationship between a generalized term and a specific instance of it.
For example, based on the hypernym relationship in the wordNet, we can describe the keyword `doctor' as a `$medical\_practitioner$'. Thus, as the mapping step, we mapped each keyword using its hypernym relation to a more generalized form, where the hypernym acts as the descriptor for the keywords.

Next, in the grouping step, we group features with the same descriptor and consider each group as a conceptual candidate feature.
For example, consider $Rule_{11}^{\prime}$, which tags Tweets with mental health.
\begin{center}
$Rule_{11}^{\prime}=$ $Tweet.Keyword.Contains(`Mental')$ ${\land}$ $Tweet.Topic.Contains(`Medical\_Practitioner)$' $:\ `Mental Health'$
\end{center}
 This rule tags a Tweet, if the Tweet contains `Mental' keyword, and a keyword relevant to `$Medical\_Practitioner$'. The topic `$Medical\_Practitioner$' represents a large number of semantically related keywords, including doctors, physician, dentist.

\item \textbf{Empath: } The second knowledge base we rely on for extracting conceptual features is EMPATH~\cite{fast2016empath}. EMPATH is a deep learning skip-gram network, which categorizes text over 200 built-in categories. It represents a token as a vector using a Vector Space Model (VSM)~\cite{pennington2014glove} and assigns tokens to categories based on their vector similarity.

To extract conceptual candidate features, we query the EMPATH vector space model to map each keyword to a category. We use categories to represent keywords in the abstract concept. Then, we group keywords with the same categories and consider each group as a conceptual candidate feature. For example, consider the following keywords $T\ = \{t_1:fund,\ t_2:illness,\ t_3:budget,\ t_4:disease\}$. To generate conceptual features, we query the EMPATH vector space model to map each keyword to a category. Assume, the following categories are identified $T\ = \{t_1:Economy,\ t_2:Health,\ t_3:Economy,\ t_4:Health\}$. Then, we group keywords with the identical categories and represent $\{fund,\ budget\}$ keywords as the \emph{Economy} topic, and $\{disease, illness\}$ keywords as the \emph{Health} topic.

\end{enumerate}

\subsection{Observation}
\label{observation}
The second step in our proposed approach is observation, which gathers feedback to update a Bayesian multi-armed-bandit algorithm about changes in the curation environment. For gathering feedback, we rely on crowd workers. Each time a rule annotates a set of items, we take a sample of the items  $S =\{i^{\prime}_1, i^{\prime}_2, i^{\prime}_3, ..., i^{\prime}_n\}$, where $S \subset I$ to send to the crowd. The crowd workers verify whether an item correctly tagged with the rule or not, e.g., if a rule tags an item with ‘Mental Health’. The task was to confirm whether the item is relevant to ‘Mental Health’ or not. For taking samples, we divided annotated items into subgroups~\cite{hunt2001stratified}, and represented each subgroup by a candidate feature — the population of subgroups determined by the frequency of candidate features in annotated items.
 
More clearly, consider the following candidate features $T\ = \{t_1:fund,\ t_2:illness,\ t_3:budget,\ t_4:economy\}$ that extracted from items annotated with a rule. The approach divides annotated items into four subgroups, where feature $t_1$ represents items contain $fund$, feature $t_2$ represents items contain $illness$, and so forth. 

Our sampling strategy boosts a Bayesian multi-armed-bandit algorithm (see Section~\ref{estimation}) to better learn the performance of features in adapting a rule. For example, if we used more obvious techniques, such as random sampling, then the algorithm considers all items equally likely. Thus, it takes a longer time to learn the performance of candidate features.

\subsection{Estimation}
\label{estimation}

The third step in our proposed approach is estimation, which computes a probability distribution $\theta$ for candidate features to determine their performance in adapting the rule. This step consists of two components: (i) reward/demote schema, which calculates a reward/demote for candidate features using workers feedback, and (ii) a Bayesian multi-armed-bandit algorithm, which estimates the performance of candidate features based on their collected rewards/demotes.\newline

\noindent\textbf{Reward/Demote Schema:}
\label{evaluationss}
To adapt rules, we formulated rule adaptation as a Bayesian multi-armed-bandit algorithm. This algorithm is suitable when a system needs additional improvement to their decisions over time. The algorithm based on the feedback collects from the curation environment learns more consistent patterns of changes and takes a decision that maximizes its performance. A Bayesian multi-armed-bandit algorithm is a good fit for our problem because each time a rule annotates a set of items, it gets updated by the workers' feedback. 
To frame the rule adaptation as a Bayesian multi-armed-bandit problem, we propose a reward and demote schema using the feedback collected from workers. The schema assigns a reward/demote to candidate features $t \in T$ appear in annotated items. The schema rewards $r$, a candidate feature, if it appears in an item that verified as relevant. Similarly, it demotes $d$, a candidate feature, if a feature appears in an irrelevant item.  Over time, as a rule, annotates more items the schema updates candidate features reward/demote, allowing a Bayesian multi-armed-bandit algorithm to update its estimation regarding the performance of features in adapting the rule.

As each conceptual candidate feature represents a group of keywords, we calculate the reward/demote for these features based on the rewards/demotes collected by their associated keywords.
More precisely, consider feature $t$ as a conceptual candidate feature. Suppose $t = \{t_1^{\prime}, t_2^{\prime}, ..., t_n^{\prime}\}$, where $t^{\prime}$ represents a keyword associated with $t$. We calculate reward as $r_{t}=\ \sum_{t^{\prime}=1}^{n} r_{t^{\prime}}$, and demote as $d_{t}=\ \sum_{t^{\prime}=1}^{n} d_{t^{\prime}}$. Clearly, consider the candidate feature `$Medical\_Practitioner$', that was introduced in the previous section. Suppose the following features are associated with it: doctor, dentist, and physician. We calculate reward/demote for `$Medical\_Practitioner$' by summing up the rewards/demotes collected by doctor, dentist and physician.

\begin{algorithm}
\SetAlgorithmName{Algorithm}{heuristic}{List\ of\ Heuristics}
\SetKwFunction{FMain}{Est\_Probability\_Dist}
\SetKwProg{Fn}{Function}{:}{}
\Fn{\FMain{}}{
\KwIn {$T$}
\KwOut {$\theta$}
\ForEach{$t$ $\in$ $T$}{
  \ForEach{$I^{\prime}$ $\in$ $S$}{
    \uIf {$t$ $\in$ $I^{\prime}$ AND $I^{\prime}$ \text{is verified as} $Irrelevant$}{
         $d_t += 1$\;
         }
     \ElseIf {$t$ $\in$ $I^{\prime}$ AND $I^{\prime}$ \text{is verified as} $Relevant$}{
         $r_t +=1$\;
   }
  ${\theta_t} \gets \textbf{Beta} (r_t, d_t)$
}
\KwRet ${\theta}$\;
}
}
\caption{Estimating expected performance of candidate features}
\label{algo:Beta}
\end{algorithm}

\noindent\textbf{Bayesian Multi-Armed-Bandit Algorithm:}
\label{generalization}
This section explains how a Bayesian multi-armed-bandit algorithm estimates the performance of candidate features. We utilized Thompson sampling~\cite{russo2017tutorial}, a Bayesian multi-armed-bandit algorithm that has shown the near-optimal regret~\footnote{Given a period of time the regret is the difference between the probability distribution $\theta$ the algorithm estimated for the optimal action and the action selected by the algorithm.} bound. Thompson sampling provides a dynamic policy for choosing which feature should be selected for adapting a rule, and an algorithm for incorporating new information to update this policy based on the candidate features rewards/demotes. Thompson sampling stores an estimated probability distribution $\theta$ for each candidate feature to indicate their performance in adapting the rule. The algorithm continuously observes the curation environment and gathers new feedback to update the probability distribution estimated for candidate features, reflecting their performance in adapting the rule. 
Each time, the algorithm receives a set of candidate features $T=\{t_1, t_2, ..., t_n\}$ along with their reward/demote it updates candidate features probability distributions $\theta= \{\theta_1, \theta_2, ..., \theta_n\}$, where $0 < \theta < 1$ using the Bayesian formula:
\begin{center}
$P(\theta\ |\ t)\ =\ \frac{P(t\ |\ \theta) \times P(\theta)}{P(t)} \propto P(t\ |\ \theta) \times P(\theta)$
\end{center}
$P(t\ |\ \theta)$, represents the likelihood and $P(\theta)$, is the prior. The likelihood is a Bernulli distribution and the prior is a Beta distribution.
\begin{center}
$P(t\ |\ \theta) =\ \theta^r(1\ -\ \theta)^{n-r}, r = \sum_{r=0}^{n} t$
\end{center}

\begin{center}
$P(\theta) = \frac{\theta_n^{\alpha_n-1}(1-\ \theta_n)^{\beta_n-1}}{\beta(\alpha_n,\beta_n)}$
\end{center}
$\alpha$ and $\beta$, are the prior parameters. The initial value of $\alpha$ and $\beta$, indicates our initial belief about the performance of candidate features. We have chosen $\alpha\ =\ \beta=\ 1$,  which means initially, we considered all features to have the same performance in adapting the rule. The prior is updated continuously based on the likelihood of feedback we gather from the curation environment. 
The posterior is proportional to the product of the prior and the likelihood, with the likelihood updated continuously after receiving workers feedback. This update is easy to implement because the Beta and Bernoulli distributions are conjugate. 
 Algorithm~\ref{algo:Beta} shows how the approach estimates the value of $\theta$ for candidate features. As an example consider the following rule.
\begin{center}
$Rule_1 = Tweet.Keyword.Contains(`Mental')$ $: `Mental Health'$
\end{center}

Which tags Tweets with `Mental Health'. Assume, the following candidate features $T = \{t_1:\ medical, t_2:\ health, t_3:\ wellbeing, t_4:\ care, t_5:\ qanda\}$ are extracted from annotated items as the potential modifications for the rule. First, to identify the performance of candidate features, the algorithm calculates their reward/demote using workers feedback. Then, a Bayesian multi-armed-bandit algorithm estimates a probability distribution $\theta$ for candidate features. Each time the rule annotates a set of items, the algorithm updates the value of $\theta$ based on the feedback gathers from workers to better reflect features performance in adapting rules.

\subsection{Adaptation}
\label{adaptation}
In this section, we explain how our approach modifies a rule.
Recall from Section~\ref{prelim}, that we introduced a rule $R$ as a tree of features, where each feature $f \in R$ can have $K$ children. We also defined a path~$p$ in a rule as a conjunction of a set of features in forms of~$p\ = f_1 \land\ f_2\land\ ...\ \land\ f_n$. First, to adapt a rule, we identify imprecise paths that annotate data with a precision below a threshold~$\jmath$. The threshold represents the minimum precision a path should have to be considered as precise. We determine the precision of paths by calculating the number of relevant/irrelevant items their features annotated. After identifying imprecise paths, we determine whether to replace or further restrict their features. 

We would replace a feature in an imprecise path if the number of annotated items was below the average number of items annotated with its siblings, indicating the feature is imprecise and incapable of adequately annotating data. Conversely, we restrict a feature if the number of annotated items was greater than or equal to average, indicating the feature is applicable but should be restricted to be precise.
For replacing or restricting features, we select candidate features that yielded the highest probability distribution $\theta$ estimated by a Bayesian multi-armed-bandit algorithm.\newline

\begin{figure*}[t]
\includegraphics[width=\linewidth,height=2.3in]{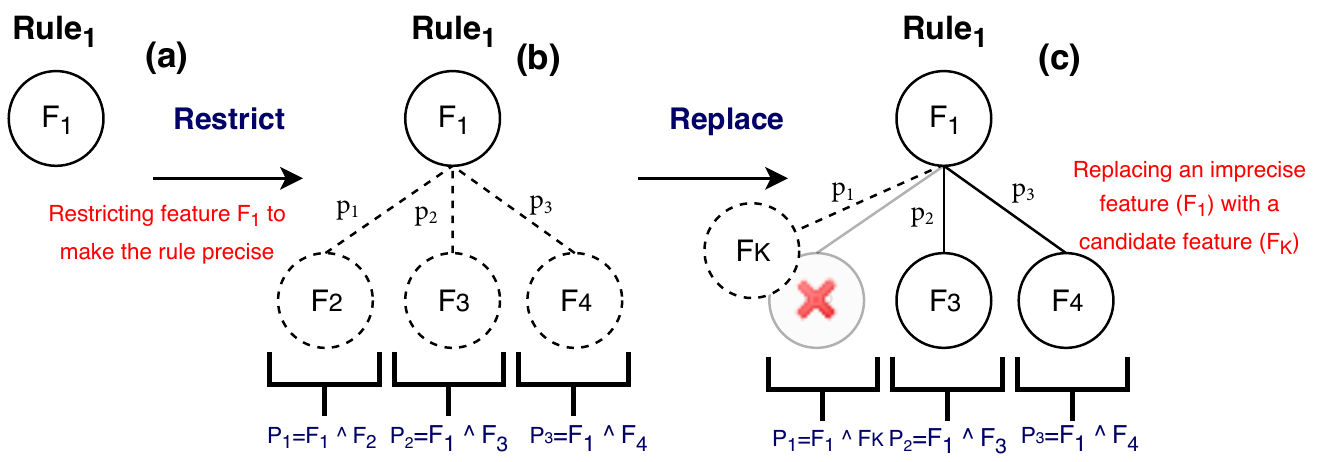}
\caption{Adapting a rule through replacing/restricting its features~\cite{tabebordbar2019adaptive}.}
\label{fig:adapt_tree}
\end{figure*}

\noindent\textbf{Example:}
\label{exexampleadapt}
Suppose, after annotating a set of items at time $\tau_i$ the algorithm identifies that the rule is imprecise~\footnote{Annotates data with a precision below $\jmath$.}. Thus, it examines the number of annotated items and adapts the rule by appending $K$ candidate features~\footnote{As feature $f_1$ is the root feature it annotates data above the average, thus satisfies the restriction condition.} that yielded the highest probability distribution (restriction) (Figure~\ref{fig:adapt_tree}b). After adaptation $Rule_1$ annotates an item if the item curated with features in paths $p_1 = \ f_1\ \land\ f_2$, or $p_2 = \ f_1\ \land\ f_3$, or $p_3 = \ f_1\ \land\ f_4$. Alternatively, the algorithm may replace a feature if it identifies the feature annotates data below the average number of items annotated with its siblings~\footnote{features $\{f_3,f_4\}$ are siblings for feature $f_2$}. For example, suppose at time $\tau_{i+n}$ feature $f_2$ is identified as imprecise and incapable of annotating data adequately~\footnote{Annotates data below the average number of items annotated with its siblings}. Thus, the algorithm removes feature $f_2$, and replaces the feature with a candidate feature that yielded the highest probability distribution value (Figure~\ref{fig:adapt_tree}c). On the other hand, to select a candidate feature, the algorithm performs a feature extraction task and estimates candidate features probability distribution $\theta$ based on the reward/demote features accumulated from time $\tau_1$ to $\tau_{i+n}$.


The proposed adaptation strategy allows to adapt rules according to changes in the curation environment. For example, by replacing an imprecise feature with a content bearing feature that obtained a high value of $\theta$ over an extended period of time, we keep the rule applicable as the new feature better captures the salient aspect of data. Similarly, by restricting an imprecise feature that annotates a large number of items, we make the rule precise by filtering out the irrelevant items.

\begin{figure}[t]
\includegraphics[width=\linewidth,height=1.7in]{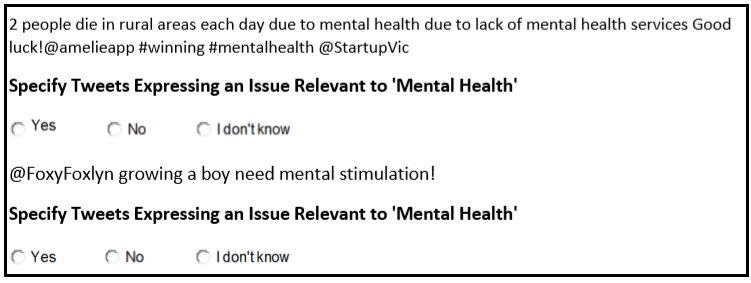}
\caption{Sample of questions to workers to verify the tag of items.}
\label{fig:crowdquestion}
\end{figure}

\section{Gathering Workers Feedback}
\label{gatherfeedback_4}
This section explains how we contribute workers to verify items annotated with rules. We created a task on Figure Eight~\footnote{https://www.figure-eight.com/} micro-tasking market. The workers' task was to confirm whether an item is relevant to the tag assigned by a rule or not. Workers could choose `Yes' if they identify the item is related to the tag, and `No', if they determine the item, is irrelevant. In cases workers could not verify an item, they could choose `I don't know'. For example, we present a Tweet to workers, which a rule tagged as relevant to 'Mental Health'. Then, workers task was to verify whether the Tweet is related to 'Mental Health' or not. Besides, we provide workers with a textual instruction to explain to them how to confirm items~\footnote{For example, in this job, we will have you to identify whether a Tweet is expressing an issue relevant to mental health or not. An issue can be a shortcoming that exists in services provided for mental health, or a threat that lack of mental health services may cause to the society, or a suggestion that helps to improve the quality of mental health services.}. We explained steps workers need to follow and provided them with three positive ~\footnote{Mental health services facing serious shortages of mental health nurses decrease of 12\% since 2010 psychiatrists.} and three negative~\footnote{if I have to hire a car and drive home from Belgium i am going to go mental stupid french air traffic control wanks on strike.} examples. For verifying each item, we paid 1 cent, and each worker verified ten items per page. At each round of the annotation task, we sent 3\% of annotated items to workers. Figure~\ref{fig:crowdquestion} shows a sample question to workers.

\subsection{Stopping Condition}
\label{stop}
In the previous section, we explained how workers verify annotated items. 
However, continuously sending items to crowds increases the cost of the adaptation task. Thus, there is a need to identify when a rule is stabilized to stop verifying more items.
To address this problem, we developed a solution using the probabilistic policy defined in Thompson sampling algorithm to determine whether a path in a rule is stabilized or not. For each path, we estimate a probability distribution $\theta$ based on the number of relevant/irrelevant items annotated. Then, we define a smoothing window $Q$ to record the value of $\theta$. We set the size of smoothing window $Q\ =\ 3$ and average as the smoothing function. We consider a path as stabilized, if the value of $Q$ increases or remains stable within $3\epsilon$, where $\epsilon\ =\ 0.01$~\footnote{we set the value of $\epsilon$ and $Q$, experimentally using simulated data.}. More clearly, consider path $p_3\ =\ f_1\ \land\ f_4$ presented in Figure~\ref{fig:adapt_tree}. Each time the rule annotates a set of items, the algorithm records the value of $\theta$ for the path. Then, the approach computes the value of $Q$, where $Q_1\ =AVG (\theta_{1}, \theta_{2}, \theta_{3})$, and $Q_2\ =AVG (\theta_2, \theta_3, \theta_4)$ and so forth. The algorithm stops sending items to workers, when the value of $Q_{i+1} +3\epsilon\ \geq \ Q_i$, indicating the path is stabilized.

\section{Experiments}
\label{experiment}
First, we discuss the dataset was used for examining the performance of our proposed approach in Section~\ref{setting}. Then, in Section~\ref{analyst}, we explain three scenarios have been defined to show the applicability of our approach. Finally, we discuss the results in Section~\ref{result}.
\subsection{Experiment Settings and Dataset }
\label{setting}
The core component of techniques described in the previous sections is implemented in Python. Three months of Twitter data (Australian region) were used as the input dataset (from May 2017 to August 2017) with $\approx\ 15\ million$ Tweets. MongoDB and ElasticSearch were used for storing and indexing the input dataset. We demonstrate the performance of our approach in three different curation domains (domestic violence, mental health, and budget). We show how our approach learns to adapt a rule to annotate data more precisely over time. As the initial rules for annotating the data, we used rules that contain only one feature. For example, the initial rule for annotating Tweets in the mental health domain was in the form of $Tweet.keyword.contai\\ns(`Mental')$ $: `Mental Health'$, which tags Tweets that contain `Mental' keyword. Then, at each timestep rules annotate a set of items, our approach adapts rules to make them more precise. We demonstrate the performance of the approach within five rounds of rule adaptation.

\begin{figure}[t]
\includegraphics[width=\linewidth, height=6.0in]{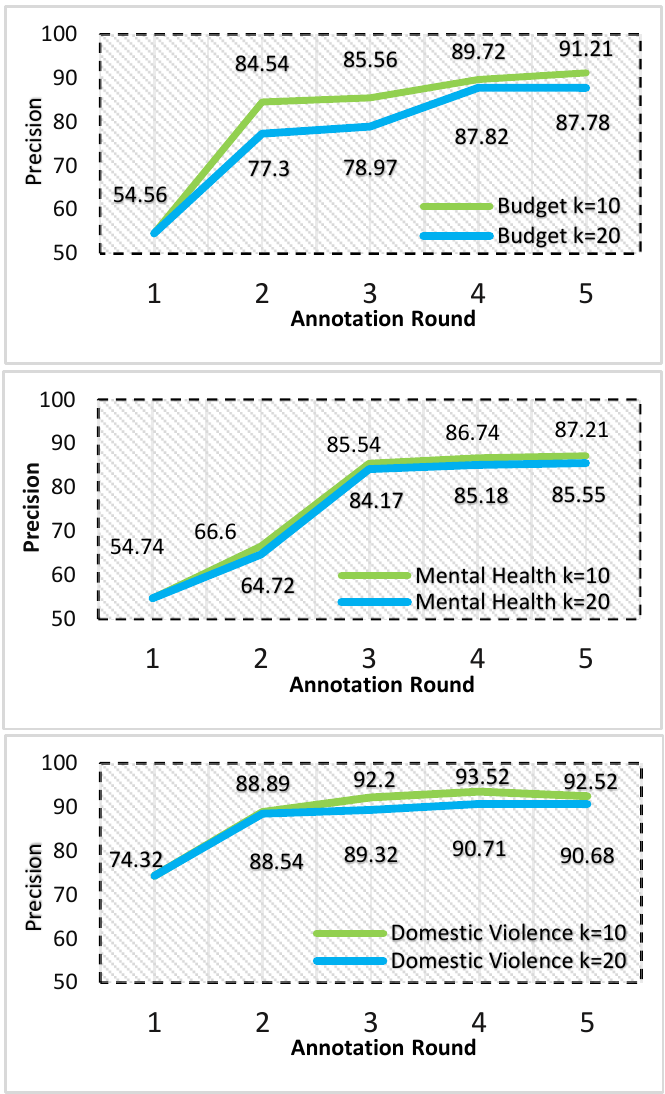}
\caption{The performance of a Bayesian multi-armed-bandit algorithm in adapting rules. As presented the algorithm could improve rules precision in all domains.}
\label{fig:curationimprove}
\end{figure}

\begin{figure}
\centering
\includegraphics[width=\linewidth,height=2.7in]{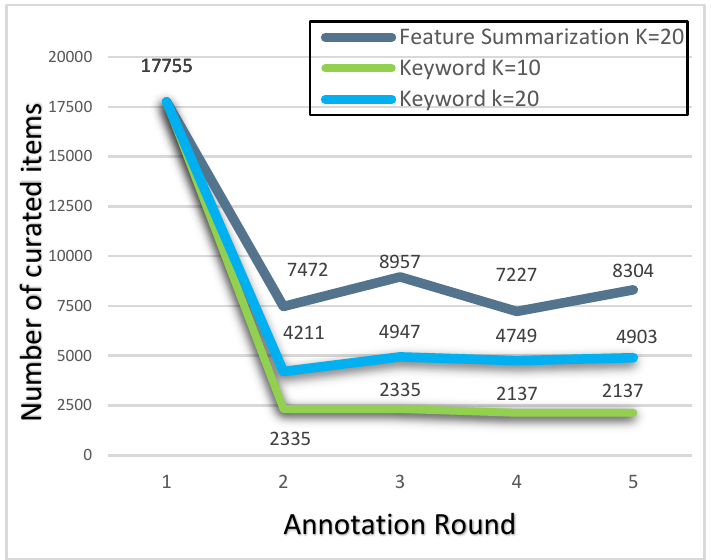}
\caption{Comparison between the number of items annotated using conceptual and syntactic level features (Budget Domain).}
\label{fig:budgetsum}
\end{figure}

\begin{figure}
\centering
\includegraphics[width=\linewidth,height=2.7in]{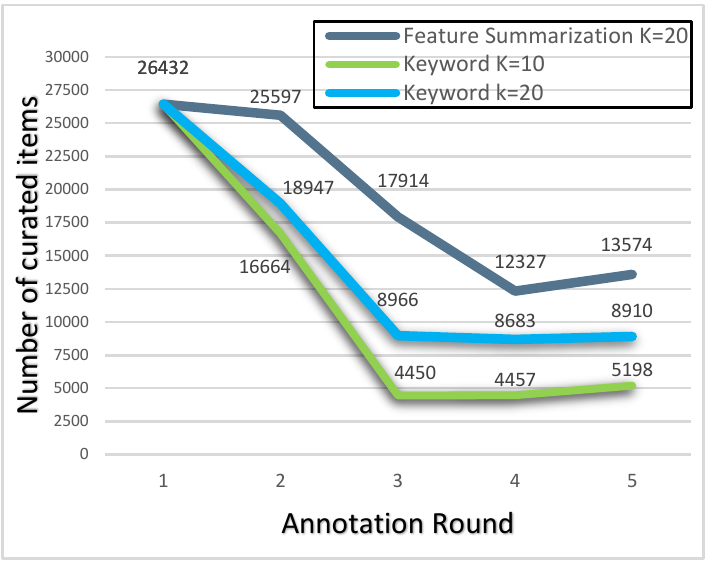}
\caption{Comparison between the number of items annotated using conceptual and syntactic level features (Mental Health Domain).}
\label{fig:mentalsum}
\end{figure}

\begin{figure}
\centering
\includegraphics[width=\linewidth,height=2.7in]{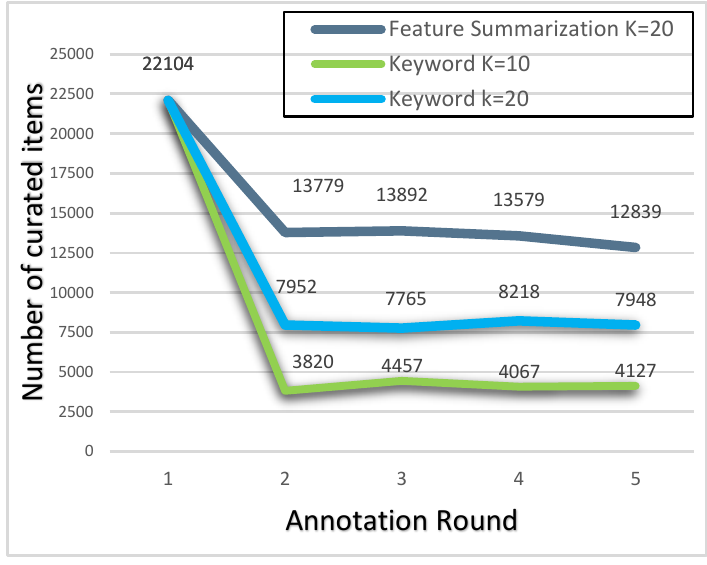}
\caption{Comparison between the number of items annotated using conceptual and syntactic level features (Domestic Violence Domain).}
\label{fig:abusesum}
\end{figure}

\subsection{Experiment scenarios}
\label{analyst}
To evaluate the performance of our solution and the applicability of the proposed algorithm, we have defined three different experiment scenarios:
\begin{enumerate}
\item \textbf{Evaluating the performance of a Bayesian multi-armed-bandit algorithm in adaptation:} We explain an experiment scenario to represent the performance of a Bayesian multi-armed-bandit algorithm in adapting rules. We demonstrate how the algorithm keeps a rule precise and applicable by adding or removing features. We adapt rules with two different choices of features ($K=\ 10,\ K=\ 20$) (see Section~\ref{prelim}). Adapting a rule with a higher number of features allows a rule to annotate a larger number of items, but with less precise ones.

\item \textbf{Evaluating the proposed feature based adaptation:} This scenario aims at demonstrating the performance of the proposed feature based technique in augmenting rules to annotate a larger number of items. We demonstrate the improvement rules make in the number of annotated items while adapting rules using both syntactical and conceptual level feature. We also, compare the obtained results with technique that adapts rules at the syntactic level only.

\item \textbf{Comparison with existing studies:} The third scenario, we conducted a controlled experiment and compared the performance of our approach with a system proposed by GC et al.~\cite{gc2015big}. The proposed system is an interactive rule adaptation system, which relies on analysts for adapting rules. Each time a rule annotates a set of items, the system sends a sample of items to crowds and receives feedback over the number of items correctly/incorrectly tagged by the rule. Then, the system tokenizes items, and weights every token using the TF-IDF weighting scheme. Subsequently, the system ranks tokens based on their TF-IDF weights and iteratively shows tokens to an analyst to adapt a rule. The system continues showing tokens until the analyst is satisfied with the resulting rule. To help the analyst to more effectively adapts the rule, the system incorporates the analyst feedback by adjusting the weight of tokens using a relevance feedback algorithm~\cite{rocchio1971relevance}. Whenever the analyst selects a token, the algorithm increases the weight of other candidate tokens that co-occurred with the selected token.
\end{enumerate}

\begin{table*}
\centering
\caption{Precision of the approach in adapting rules using \emph{ summarization} approach $K\ =\ 20$.}
\label{tbl:summaryprec}
\begin{tabular}{llllll}
\hline\noalign{\smallskip}
\textbf{Curation Domain} & \textbf{Round 1} & \textbf{Round 2} & \textbf{Round 3} & \textbf{Round 4} & \textbf{Round 5} \\
\noalign{\smallskip}\hline\noalign{\smallskip}
Budget & 54.56 & 73.12 & 78.72 & 81.11 & 84.21 \\
Mental Health & 54.74 & 57.35 & 71.40 & 80.16 & 80.61 \\
Domestic Violence & 74.32 & 83.59 & 84.60 & 85.75 & 84.43 \\
\noalign{\smallskip}\hline\hline
\end{tabular}
\end{table*}

\begin{table*}[t]
        \centering
            \footnotesize
                \caption{Precision of rules adapted through participants in \emph{Budget, Mental Health, and Domestic Violence Domains}.}
                \label{tbl:budgetin}
                \begin{tabular}{llllll}
                \hline\noalign{\smallskip}
                \textbf{Budget Domain} & \textbf{Round 1} & \textbf{Round 2} & \textbf{Round 3} & \textbf{Round 4} & \textbf{Round 5} \\
                \noalign{\smallskip}\hline\noalign{\smallskip}
                Participant 1 & 54.56 & 79.75 & 82.02 & 84.21 & 87.19 \\
                Participant 2 & 54.56 & 83.22 & 85.62 & 88.20 & 90.86 \\
                Participant 3 & 54.56 & 86.56 & 86.77 & 86.67 & 86.59 \\
                \hline\noalign{\smallskip}
                \textbf{Mental Health Domain} & \textbf{Round 1} & \textbf{Round 2} & \textbf{Round 3} & \textbf{Round 4} & \textbf{Round 5} \\
                \noalign{\smallskip}\hline\noalign{\smallskip}
                Participant 1 & 54.74 & 72.88 & 80.65 & 87.19 & 86.32 \\
                Participant 2 & 54.74 & 70.38 & 75.09 & 83.58 & 85.01 \\
                Participant 3 & 54.74 & 71.63 & 81.61 & 85.04 & 84.14 \\
                \hline\noalign{\smallskip}
                \textbf{Domestic Violence Domain} & \textbf{Round 1} & \textbf{Round 2} & \textbf{Round 3} & \textbf{Round 4} & \textbf{Round 5} \\
                \noalign{\smallskip}\hline\noalign{\smallskip}
                Participant 1 & 74.32 & 87.65 & 88.78 & 90.90 & 90.82 \\
                Participant 2 & 74.32 & 88.63 & 90.28 & 91.36 & 92.59 \\
                Participant 3 & 74.32 & 85.37 & 86.51 & 87.04 & 86.42 \\
                \noalign{\smallskip}\hline
        \end{tabular}
\end{table*}

\subsection{Result}
\label{result}
\subsubsection{Performance of a Bayesian multi-armed-bandit algorithm in adapting rules}
\label{performance}
In this section, we demonstrate the performance of a Bayesian multi-armed-bandit algorithm in adapting rules (see Section~\ref{estimation}). We show the precision of the rules adapted with two different choices of candidate features ($k\ =\ 10, k\ =\ 20$) that yielded the highest probability distribution $\theta$. As presented in Figure~\ref{fig:curationimprove} by adapting rules with 10 candidate features the algorithm could significantly improve rules precision in all curation domains. For example, in the budget domain, the algorithm could improve the precision for $36.65\%$, from $54.56\%$ to $91.21\%$. Similarly, in the domestic violence and the mental health domains, the algorithm could improve the precision for $18.20\%$ and $32.47\%$ respectively.
Also, to demonstrate the applicability of the algorithm in adapting rules, we repeated the experiment with a higher number of features ($K\ =\ 20$). This boosts rules to annotate a larger number of items, but with less precise features. Figure~\ref{fig:curationimprove} shows the obtained results for each domain. As presented, adapting rules with a higher number of features decreases the precision of rules, however, the algorithm could learn the performance of features and adapts the rule to improve its precision over time. For example, in mental health domain, the precision is improved by $30.81\%$, and in budget and domestic violence domains the precision is improved by $33.22\%$ and $16.36\%$ respectively. In this experiment, we considered features that annotate data with a precision below 75\% ($\jmath$<75\%) as imprecise. \newline

\noindent\textbf{Discussion on rules performance:} As presented in Figure~\ref{fig:curationimprove}, the initial rules added to the curation system was imprecise and annotated a large number of irrelevant items. For example, the initial precision of rules in Budget and Mental Health domains was below $55\%$. However, after collecting a set of feedback the algorithm identifies the need to restrict rules by adding a new set of features. Although the restricting rules could improve their precision, this limited rules to only annotate those items that contain the features selected by the algorithm during the adaptation. As presented in Figures~\ref{fig:budgetsum}, ~\ref{fig:mentalsum}, and ~\ref{fig:abusesum}, after adaptation rules are annotating fewer items compared to their initial states. For example, in Budget domain the number of annotated items has reduced by 15240, after two rounds of adaptation. We can see similar trends for other curation domains as well. But, the promising fact is that a Bayesian multi-armed-bandit algorithm can learn a better adaptation for rules by incrementally collecting more feedback over time. This can be seen in Figure~\ref{fig:curationimprove} that the algorithm could dramatically improve the rule precision. For example, in Budget domain the difference in precision between the adaptation that occurred at $\tau_2$ and $\tau_5$ is over $10\%$. This difference for the Mental Health domain is over $20\%$. Based on the obtained results, we concluded that a Bayesian multi-armed-bandit algorithm by collecting more feedback learns a better adaptation for rules over time, and if we can adapt rules with more robust features we can improve both precision and recall. This fact, can be approved by comparing the precision and the number of annotated items between Figures~\ref{fig:curationimprove} and Figures~\ref{fig:budgetsum}, ~\ref{fig:mentalsum}, and ~\ref{fig:abusesum}. As presented by adapting rules with a higher number of features ($K\ =\ 20$) the algorithm could annotate a larger number of items, and at the same time maintain rules precision. In the next section, we discuss how feature-based adaptation augments the performance of rules to annotate a larger number of items.

\begin{figure*}
\centering
\includegraphics[width=\textwidth,height=2.7in]{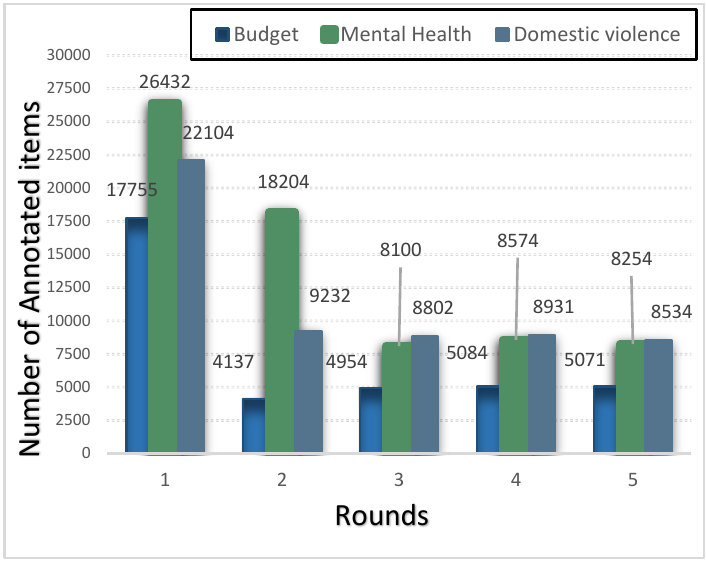}
\caption{Number of items annotated with rules adapted through participants after five rounds of annotations in three different curation domains: Budget, Mental Health, and Domestic Violence.}
\label{fig:usernumber}    
\end{figure*}

\subsubsection{Feature-Based Adaptation.}
As we discussed in the previous section~\ref{performance}, adaptation limits the ability of rules in annotating items. To alleviate this problem, we discussed that adapting rules with higher number features could boost rules to annotate a larger number of items. However, increasing the number of features has a negative correlation with precision (by increasing the number of features in adaptation the precision of rules drops). Thus, to diminish the impact of an adaptation and maintaining the performance of a rule in annotating items, we proposed feature-based adaptation. In feature-based adaptation, we hypotheses that adapting a rule with a group of semantically related features would have a similar impact on the rule precision when adapting with a single feature.
Thus, in this section, we study the impact of feature-based adaptation on rules performance. The goal is to study whether adapting a rule with a group of related features can enhance the performance of rules to annotate a larger number of items, and at the same time maintain their precision.  To test our hypotheses, we conducted two sets of experiments. First, we discuss the precision of rules adapted through our approach. Then, we compare the number of annotated items with rules adapted using syntactic level features.

Table~\ref{tbl:summaryprec} shows the precision of rules adapted using the feature-based technique. The obtained results confirm that feature-based adaptation can dramatically increase the performance of a rule in annotating items. At the same time, a Bayesian multi-armed-bandit algorithm could learn the performance of features and improves rules precision over time. This improvement for the domestic violence domain is $10.11\%$, and for mental health and budget, domains are $25.87\%$ and $28.47\%$ respectively. Although the learning rate of the algorithm using the feature-based approach is slower than syntactic level features, still the algorithm could improve rules precision in all domains. In addition, Figures~\ref{fig:budgetsum}, ~\ref{fig:mentalsum}, and ~\ref{fig:abusesum} compare the number of items annotated with rules adapted using the syntactic and feature-based approach. As presented, adapting rules with different features could boost rules to annotate a larger number of items. For example, in the domestic violence domain, the rule could annotate over 12000 items. In mental health and budget domains, rules could annotate 13574 and 8304 items respectively. These numbers are much higher than adapting rules using syntactic level features. For example, in the budget domain, the rule ($k=10$) could annotate 2137 items only. Annotating data using the syntactic level features in mental health and domestic violence domains show a similar trend and rules could only annotate 5198, and 4127 items respectively. 
\newline

\noindent\textbf{Discussion on Feature-Based Adaptation:}
An advantage of feature-based adaptation is that it allowing users to better investigate their information needs while seeking for topics that contain a large number of topical subspaces. Suppose a user intends to curate data relevant to `mental health'. There exists a large number of keywords, e.g. health, disorder, service, that are relevant to mental health but may not receive enough feedback to be considered for adapting the rule. Using, feature-based adaptation, we group all keywords that are associated with a topic, thus the rule can easily curate a varied and comprehensive list of items relevant to the user information need.

\subsubsection{Comparison with existing studies.}

In this section, we compare the performance of our approach with the state of the art technique on rule adaptation. We implemented the system proposed by GC et al~\cite{gc2015big} (see Section~\ref{analyst}), and conducted a controlled experiment. 
We asked three Ph.D. students in a lab that were familiar with the concept of learning algorithms, e.g., true positive rate, false-positive rate, to participate in the experiment. We explained to them how the system works and how they can use the system to adapt rules. Also, we allowed them to work with the system to gain the required understanding for adapting rules. To better compare the performance of our approach with the interactive system, we have asked participants to adapt rules in all domains. Then, in each curation domain, we selected the rule with the highest obtained precision and compared it with rules adapted by our approach. In this experiment, we asked participants to adapt rules with 20 features ($k=20$). Table~\ref{tbl:budgetin} shows the results. As presented, our approach has comparable performance to interactive systems. For example, in budget domain participants could adapt the rule with $90.86\%$ precision, which is $3.08\%$ higher than our proposed approach. In the domestic violence and mental health domains, participants could adapt rules with the precision of $92.59\%$ and $86.32\%$ respectively. Besides, Figure~\ref{fig:usernumber} shows the number of items annotated with the rules adapted by participants. The figure shows the most precise rules in each domain. Although our approach and participants have a shown a similar performance while using syntactic level features for adapting rule, using the proposed feature-based technique our approach could significantly annotate a larger number of items. For example, the number of annotated items in budget domain is higher by 3233 items. The difference in mental health and domestic violence domains is 5320, and 4305 respectively. The overall cost that we paid for verifying items in the mental health domain is \$35.10, and in the budget domain is \$29.92, and in the domestic violence domain is \$21.22.

By comparing the precision and the number of items annotated with our approach and participants, we believe that our adaptive approach outperforms current rule adaptation techniques. In particular, by considering the prohibitive cost of analysts for adapting rules, our proposed approach can boost companies and data enthusiasts that need to annotate data in unstructured and constantly changing environments with a limited budget.

\section{Conclusion and Future works}
\label{conclusion_4}
\noindent In this chapter, we proposed an approach for adapting data annotation rules in unstructured and changing environments. Our approach offloads analysts from adapting rules and autonomically modifies rules based on changes in the curation environment. We utilize a Bayesian multi-armed-bandit algorithm, an online learning algorithm that learns the optimal modification for rules using the feedback gathers from the curation environment. In addition, our approach adapts rules at the conceptual level, which boosts rules to annotate a larger number of items compared to current methods that rely on syntactic similarity, e.g., keywords, regular expression, for adapting rules. We evaluated the performance of our approach on three months of Twitter data in three different curation domains: domestic violence, mental health, and budget. The evaluation results showed our approach has comparable performance to systems relying on analysts for adapting rules.

There are several exciting directions for future work. In this chapter, we introduced a summarization, which boosts rules to annotate data at the conceptual level. As a part of future works, we plan to identify more features for adapting rules. Specifically, we focused on adapting rules with three other types of features, including entities, word2vec, and relation. We believe adapting rules with different kinds of conceptual features not only enhance the performance of rules to annotate a more significant number of items but also allows rules to capture the salient aspect of data better.

In the next chapter, we further expand our summarization technique and introduce and accentuate how it augments users' comprehension of curation environments. Specifically, we discuss how named entities and deep learning can be coupled with summarization technique to enable users to better formulate their preferences while seeking for a varied and comprehensive list of items.

\chapter{Enhancing Users Comprehension of the Curation Environment}\label{Chapter4}
In this chapter, we present a technique for augmenting the user's understanding and sensemaking of a curation environment. In a large curation environment, a user often conducts exploratory search for identifying and extracting information relevant to her topic of interest.
Often, however, a user needs to iteratively investigate the curation environment to formulate her preferences for Information Retrieval (IR) systems.
In recent years several visualization techniques have been proposed to help a user to better formulate her preferences.
However, using current techniques, a user needs to explicitly specify her preferences for IR systems in forms of keywords or phrases.  To address this problem, we present ConceptMap, a system that provides a conceptual summary of the curation environment and allows a user to specify her preferences implicitly as a set of concepts. ConceptMap provides a 2D Radial Map of concepts within the information space and allows a user to rank items relevant to her preferences through dragging and dropping. 
\newline
We discuss the problem of comprehending the curation environment in Section~\ref{intro_5}. In Section~\ref{related_work_5}, we discuss the related works regarding formulating users preferences and comprehending the curation environment. Section~\ref{interface_5} describes the interface of the ConceptMap. In Section~\ref{solution_5}, we discuss our proposed approach for generating a conceptual summary of the information space. Then, in Section~\ref{experiment_5}, we describe experiments and usage scenarios on retrieving citizens' opinions about issues in `Health Care System'. Section~\ref{discussion_5} details some of the challenges we encountered in implementing the ConceptMap, and Section~\ref{conclusion_5} summarizes the implications of this work.

The content of this chapter is derived from the following paper(s):
\begin{itemize}
       \item A Tabebordbar, A Beheshti, and B Benatallah, \textbf{Conceptmap: A conceptual approach for formulating user preferences in large information spaces}, International Conference on Web Information Systems Engineering, Springer, 2019, pp. 779–794 (ERA Rank A).
       \item Beheshti A, Tabebordbar A, Benatallah B. \textbf{iStory: Intelligent Storytelling with Social Data.} In Companion Proceedings of the Web Conference 2020 2020 Apr 20 (pp. 253-256). (ERA Rank A*). 
\end{itemize}

\newpage

\begin{figure}
  \centering
 \includegraphics[width=\linewidth,height=8.0cm]{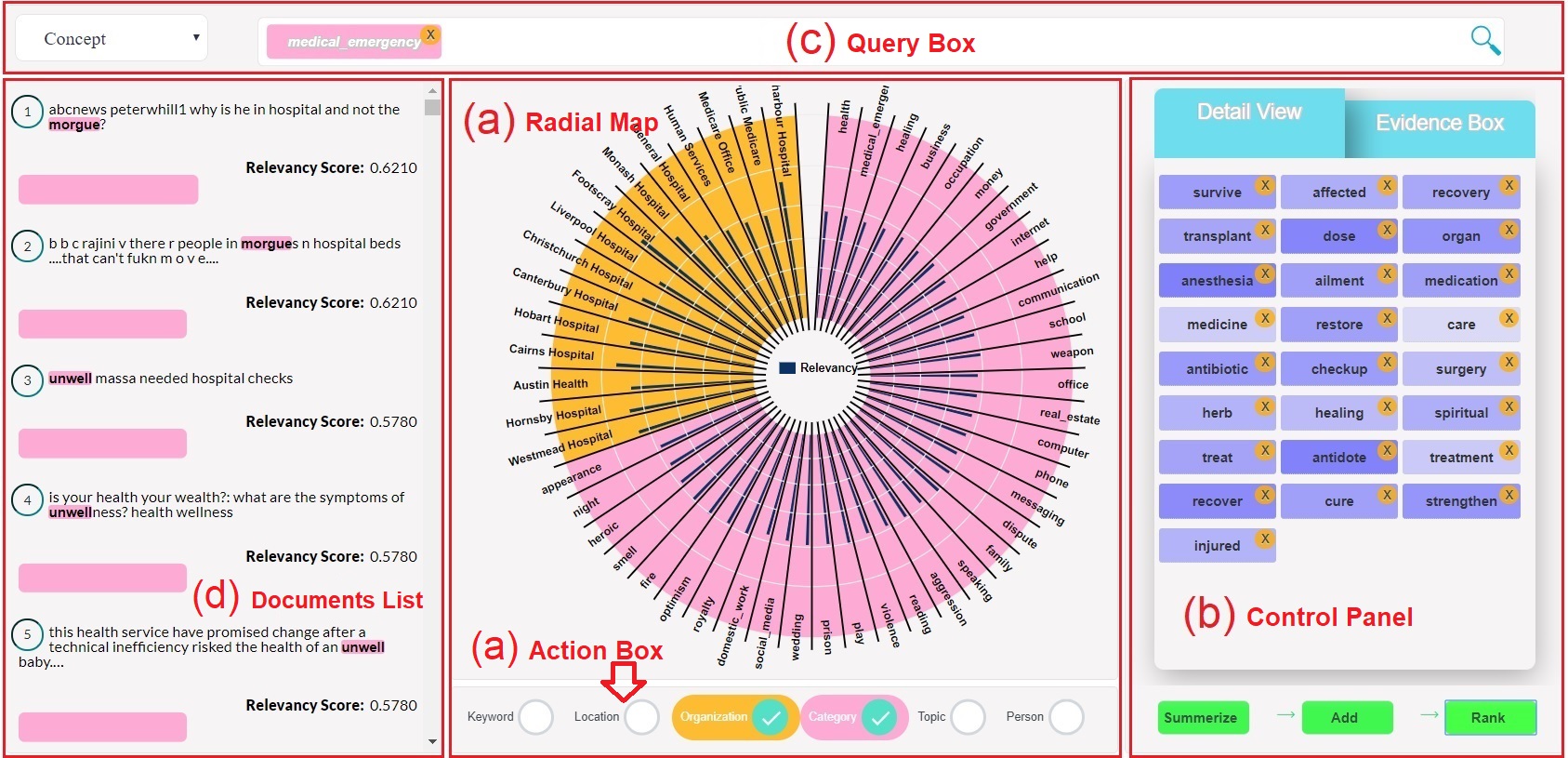}
 \caption{The ConceptMap interface is made up four components: (a) the main data view, shows a summary of concepts within the information space using a Radial Map. A user can choose summaries from the Action box. The Control Panel (b) allows a user to observe and modify attributes, e.g., keyword, Named Entities, associated to concepts. The Query Box (c) provides two interfaces (Concept and Concept + Rule) for a user to formulate her preferences. The Documents List (d) ranks documents based on their relevancy to concepts~\cite{tabebordbar2019conceptmap}.}
\label{fig:main}
\end{figure}

\section{Introduction}
\label{intro_5}
Information Retrieval (IR) systems have been extensively used to extract and locate users information.
These systems retrieve a ranked list of items ordered by their relevancy and allow a user to skim and pick items from the list.
Exploratory search is part of an information exploration process in which a user is unsure about the way to retrieve her information needs, and often becomes familiar with the information space overtime.
Usually, in an exploratory search, a user relies on text-based queries for formulating her preferences. 
Text queries are made up of a few keywords or phrases~\cite{jansen2007web,jansen2005temporal} and allow a user to explore and retrieve the information. 
However, formulating queries has been proven to be painstakingly difficult as a user needs to read and synthesize a large amount of information iteratively. This problem is exacerbated as humans have a limited memory capacity in absorbing information, which can lead to information overload or attention management~\cite{pirolli2005sensemaking}. In past years, several studies~\cite{dinet2004searching,gc2015big,sun2014chimera} have been conducted to formulate user's preferences through rules, e.g., boolean operators. However, these studies have concluded that comprehension of the information space is needed in order a user formulates her preferences accurately.

In recent years several pioneering solutions ~\cite{singh2018learn,rajaonarivo2017inline,ruotsalo2018interactive,klouche2017visual,brown2014finding,sultanum2019doccurate} have been proposed to couple Human-Computer-Interaction (HCI) techniques with IR systems to aid users to develop insight and absorb greater amounts of information. 
These solutions fuse the traditional text-based queries with various visualization elements, such as bar-graph~\cite{gratzl2013lineup}, table~\cite{wall2018podium}, and relevance map~\cite{peltonen2017topic}. Although visual encoding lowers user's cognitive load~\cite{di2016rank,gratzl2013lineup}, still the user needs to iteratively explore the information space to identify the relation between attributes (e.g., keywords, phrases, named entities), in documents to formulate her preferences. This process is challenging for several reasons: 
\begin{itemize}

\item  In many cases an exploratory search scenario contains too many topical subspaces and is difficult for a user to formulate her preferences in forms of keywords or phrases.
\item  Sensemaking of the information space is incomplete as text queries only retrieve a small part of the information space and the rest remains invisible. 
\item  Relying on text queries are time-consuming as the user is not familiar with the information space and needs to comprehend a large amount of data. 
\end{itemize}
For example, consider a user who intends to analyse citizens opinion on social media platforms, e.g., Twitter and Facebook, to identify issues in \emph{`Health Care Services`} that need improvement. Currently, the user needs to read and scan the information space to identify the query terms that properly retrieve items relevant to a large number of topical subspaces, e.g., `medical centres', `aged care services', and `mental health'. Such a search scenario needs the user to spend a long period of time to identify the content bearing terms associated with each subtopic.
Alternatively, Carterette et al.~\cite{carterette2008here} highlighted that users are more willing to express their preferences relatively, instead of precisely specifying attributes associated with it. In this context, we follow a similar trend by generating a conceptual summary of the information space and helping a user to formulate her preferences \emph{implicitly} as a set of concepts.

In this chapter, we present the \emph{ConceptMap}, a system for lowering user's cognitive load in ranking and exploring the information space. While previous systems allow a user to formulate her preferences explicitly, e.g., keywords and phrases, and observing changes in rankings to understand the data, we provide a different ranking and data presentation approach. Our work focuses on creating a conceptual summary of the information space to help a user to understand the data and relate it to her preferences. Hence, we focus on boosting a user's cognitive skill in understanding the data and formulating that understanding to extract information relevant to her topic of interest. We do this by interacting with the user to explore her preferences in a $2D$ Radial Map. A user can refine her preferences through dragging and dropping concepts into a Query Box to update document rankings, representing the relevance of concepts and documents.

ConceptMap is made up of two main technical achievements: \emph{Knowledge Lake}~\cite{beheshti2018datasynapse,beheshti2018corekg}, which is a centralized repository containing several knowledge bases, providing a contextualization layer for annotating attributes within the information space with a set of facts and information. \emph{Summarization}~\cite{fast2016empath}, takes advantage of a deep learning skip-gram embedding network~\cite{mikolov2013distributed} to learn the associations between attributes and groups the similar ones. We discuss two usage scenarios for our technique: (1) Illuminating how conceptual summary lowers the user's cognitive load in formulating her preferences, and (2) How conceptual summary and the insight developed from it can motivate a user to formulate her preferences through more advanced IR systems features, such as rules to retrieve relevant documents more precisely.

Overall, this chapter's contributions include:
\begin{itemize}
\item We introduced ConceptMap, a system that automatically generates a conceptual summary of the information space and allows a user to formulate her preferences implicitly as a set of concepts, such as topic, category and Named Entity.
\item We study how conceptual summary of data helps a user to understand the information space and formulate her preferences through rules.
\item We present two usage scenarios using Twitter data, which demonstrated how ConceptMap helps a user to explore and retrieve a varied and comprehensive list of information across a large amount of data.  
\end{itemize}

\section{Related Work}
\label{related_work_5}
In this section, we discuss prior works related to formulating user's preferences (Section~\ref{formulateuserpref}), comprehending and sensemaking of the information space (Section~\ref{comprehendsense}), and topic modeling approaches (Section~\ref{topicmodel}). 
\subsection{Formulating User Preferences}
\label{formulateuserpref}
The relevance judgement for IR systems has been made on binary scale, where a document is considered relevant to a query or not~\cite{carterette2008here}. Such judgement requires a user to precisely formulate her preferences to locate and retrieve the relevant documents.
Typically, for formulating preferences a user needs to conduct exploratory search by iteratively investigating the information space to develop insight and create its mental structure~\cite{marchionini2006exploratory}. 
Exploratory search is beyond the basic information seeking task of looking for a few relevant documents. In an exploratory search, a user has no predetermined goal or understanding of the information space and learns to formulate her preferences by investigating and learning from the context overtime~\cite{klouche2015designing}. This makes formulating user preferences challenging and time-consuming, especially in broad information spaces. Previous works have focused on augmenting users' comprehension of the information space with visual encoding to formulate their preferences more precisely~\cite{hearst1995tilebars,harrison2014ranking,kidwell2008visualizing,shi2012rankexplorer}. However, recently some solutions~\cite{carterette2008here,wall2018podium} have shown that it is easier for a user to make a relative judgement of her preferences rather than explicitly specifying attributes associated with it. For example, a user may formulate `mental health' as a `disorder', but unable to precisely determine attributes associated to it. In this context, we allow a user to formulate her preferences \emph{implicitly}, as a set of abstract concepts, such as topic, category, and Named Entity. Our approach automatically identifies the relation among attributes within the information space, without requiring to specify the exact attributes associated with it. In the next sections, we will accentuate approaches focused on augmenting users' comprehension of the information space.

\subsection{Comprehension and Sensemaking of the Information Space}
\label{comprehendsense}
Sensemaking of the information space is defined as processes and activities a user undertakes to frame the information space in an understandable schema~\cite{pirolli2005sensemaking}. 
Sensemaking has been identified as a quintessential task of information retrieval~\cite{marchionini2006exploratory}, especially when a user has varied information needs across a large number of data~\cite{peltonen2017topic}. During the past years, several solutions have been proposed to enhance user comprehension and sensemaking of the data. One category of these solutions focused on augmenting the ranked lists of search results with different visualization elements. For example, tileBar~\cite{hearst1995tilebars} represents the relevancy of ranked documents to query terms with shaded blocks. LineUp~\cite{gratzl2013lineup}, used bar charts to visualize the ranking of multi-attributes data, while other approaches highlight ranked lists with stacked bar~\cite{di2016rank}, metaphor-based layout~\cite{nguyen2006novel}, and snippet-based layout~\cite{gomez2014similarity}. However, ranked lists can only support scenarios where a user has limited information needs and is seeking for a few relevant documents. 

Another line of works have coupled visualization with HCI techniques for augmenting a user to gain a better understanding of the information space.
Comprehension of the information space allows a user to discover the aboutness of data and develop a mental structure of it~\cite{brusilovsky2018social}. For example, Wall et al.~\cite{wall2018podium}, proposed a table layout to present a holistic view of the information space for multi-attribute ranking systems. Di-Sciascio et al.~\cite{di2018study} boosts user comprehension of the information space by contributing previous users' search terms. Peltonen et al.~\cite{peltonen2017topic} provided a topical overview of the information space by interacting with the user to visualize the association between keywords on a relevance map. Di-Sciascio et al.~\cite{di2016rank} focused on transparent and controllable recommendation systems to enhance user understanding of data. However, these approaches focused on enhancing users' mental capacity to better identify the relations between attributes in documents. But, in a large information space many of these relations remain invisible to the user, either due to user inability in identifying them or visual clutter~\cite{ellis2007taxonomy}. Instead, ConceptMap provides a conceptual summary of the data and offloads user to iteratively investigate the information space to discover associations between attributes in documents.

\begin{figure}[tb]
 \centering 
 \includegraphics[width=9.0cm, height=4in]{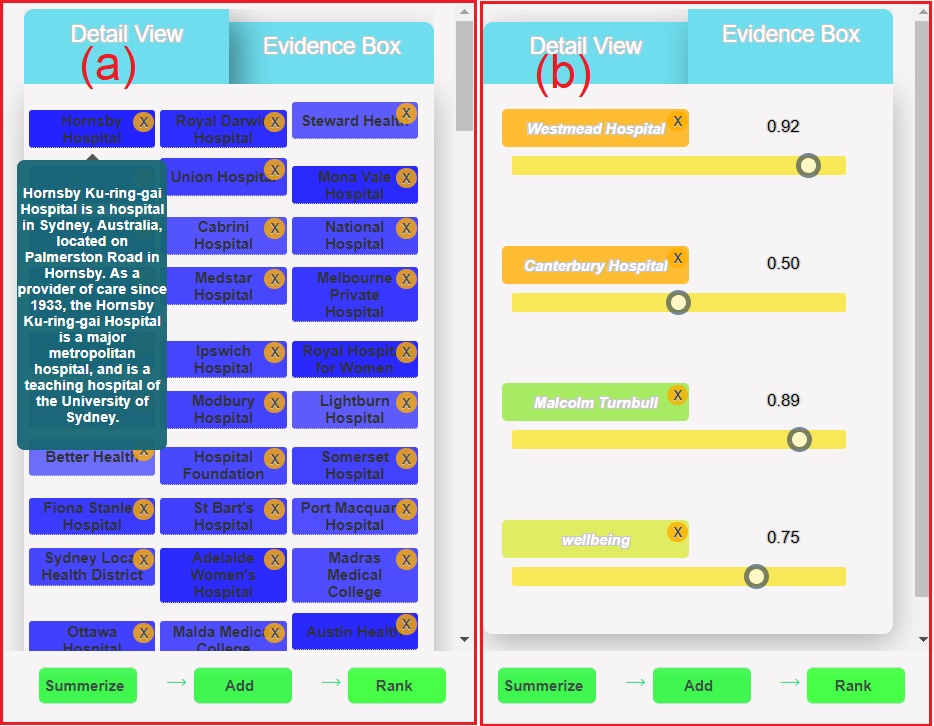}
 \caption{The Control Panel is made up of two components: (a) the Details View, allows a user to examine attributes associated to a concept, and (b) the Evidence Box, stores potential concepts relevant to a user information needs~\cite{tabebordbar2019conceptmap}.}
 \label{fig:detail_evidence}
\end{figure}
\subsection{Topic Modeling Techniques}
\label{topicmodel}
In addition to interactive methods for augmenting user comprehension, we consider it appropriate to include approaches that provide a topical overview of the information space.
Topic modelling is a generative approach, which aims at discovering groups of words that frequently co-occur in documents~\cite{blei2003latent}. Latent Dirichlet Allocation (LDA) is the most common topic modelling algorithm, which has been used extensively for providing a topical overview of the information space. For example, TIARA~\cite{liu2009interactive} is one of the early works on topic-based text summarization, which creates a visual summary of the information space through visualizing the result of LDA algorithm with a stacked graph.
Serendip~\cite{alexander2014serendip} is a topic modelling system focused on structuring exploration of information for supporting multi-level discovery. TopicNets~\cite{gretarsson2012topicnets}  is an interactive topic modelling system that visualizes documents on a graph of connected network. 
In addition, some techniques relied on hierarchical topic modelling to augment user comprehension of the information space. For example, TopicLens~\cite{kim2017topiclens} combined a lens technique with a tree-based topic modelling for exploring the data.
PolyZone~\cite{javed2012polyzoom} proposed an interactive technique, which progressively builds a hierarchy of focus regions to allow users to explore the magnification of the topic of her interest. Whereas, other researchers coupled topic modeling with different visual encoding such as panning~\cite{pietriga2008sigma}, overview $+$ detail~\cite{igarashi2000speed} or scrolling~\cite{javed2010stack}.
Although topical modelling can provide an overview of the information space, these algorithms cannot identify the semantical relation between attributes in information space~\cite{hu2014interactive}, which is required in exploratory search scenarios~\cite{di2016rank}. 
Moreover, topical modelling algorithms have a high performance requirement and are computationally expensive to rely on for dynamic and real-time search scenarios.

 On the other hand, ConceptMap discovers the semantical relation between attributes and groups them based on their similarity.
 This is different from approaches, which focused on creating topics based on the words co-occurrence.
 Let us back to our example that was introduced in the previous section, to retrieve citizens opinion about issues in \emph{`Health Care Services'}, ConceptMap may help a user by providing a conceptual summary of people who are in-charge of health care services, organizations that provide health care services, and locations associated to health care services, etc. Then, the user can rank Tweets based on provided summaries and retrieves items relevant to her information needs. Thus the user only needs to focus on her preferences rather than investigating the information space to identify the relation between attributes in documents. This speeds up the exploration of the information space, in cases where a user has varied information needs, which contains too many topical subspaces and is difficult to formulate for IR systems. 
 
\begin{figure}
 \centering 
 \includegraphics[width=\linewidth, height=0.7in]{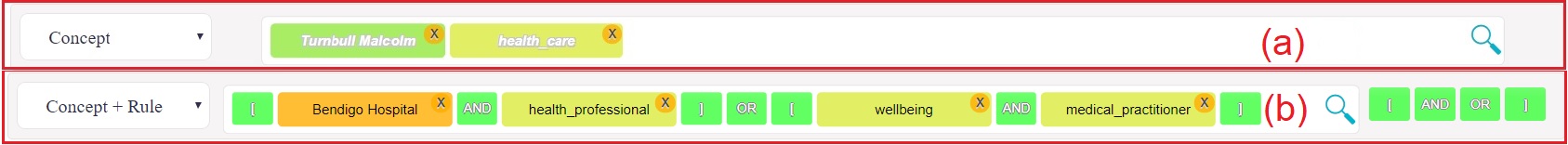}
 \caption{The Query Box provides two interfaces for formulating a user preferences: (a) the Concept Only, which allows a user to formulate her preferences as a set of concepts, (b) the Concept + Rule, which aids a user to formulate her preferences through rules~\cite{tabebordbar2019conceptmap}.}
 \label{fig:queybox}
\end{figure}

\section{ConceptMap}
\label{interface_5}
In this section, we describe ConceptMap, a system that provides a conceptual summary of the information space. ConceptMap discovers the semantical relation between attributes and enables a user to formulate her preferences as a set of abstract concepts. In this section, we discuss the character of ConceptMap interface.
\subsection{Design Components}
The ConceptMap interface is made up of four components: the Radial Map (Figure ~\ref{fig:main}a), the Control Panel (Figure ~\ref{fig:main}b), the Query Box (Figure ~\ref{fig:main}c) and the Documents List (Figure ~\ref{fig:main}d). The Radial Map is the central component of the ConceptMap. It shows concepts discovered from the information space. The Control Panel provides the controllability for a user to create her topic of interest. It contains two tabs: The Details View, which shows attributes associated with a concept and lets a user manipulate the concept by adding and removing attributes. The Evidence Box, which allows a user to develop a mental structure of data by gathering concepts relevant to her topic of interest. The third component of the ConceptMap is the Query Box, which enables a user to examine the concepts-documents relevancy. The Query box provides two interfaces: The Concept Only, which lets a user formulate her preferences implicitly as a set of concepts. The Concept + Rule, which allows a user to formulate her preferences through boolean (AND, OR) operators. The last component of the ConceptMap is the Documents List, which shows a ranked list of documents based on their relevancy to the user-selected concepts. Followings, we explain the character of each component in detail.

\subsubsection{1. Radial Map}
The Radial Map is the main data view in the ConceptMap, and shows a summary of the most frequent concepts within the information space (Figure ~\ref{fig:main}a).
A user can select her preferred summaries through interacting with the \emph{Action Box}.
It has six toggles for visual encoding of the Radial Map: \emph{Persons}, \emph{Organizations}, \emph{Locations}, \emph{Categories}, \emph{Topics}, and \emph{Keywords}.
Each toggle represents a specific summary and colors the summary if it's selected by a user.
A user can observe concepts associated with summaries through the Radial Map by pressing the \emph{Summarize} button. The coloring of concepts within the Radial Map corresponds to the Action Box, where concepts associated with a summary mirror that color. For example, ConceptMap colored the \emph{Topics} summary as red. ConceptMap also displays concepts associated with it within the Radial Map as red.

By default, ConceptMap divides the Radial Map into 50 wedges, where each wedge represents a concept. Each wedge augmented with a grid line and shows the relevancy of the concept to the information space. The value of the grid line is between 0 to 1, where zero represents the least relevancy and one represents the highest relevancy. Augmenting wedges with grid lines, enable a user to grasp an overview of the information space along with their relevancy as a whole at a glance. The following explains the types of summaries ConceptMap displays to a user. 
\begin{itemize}
 \item \textbf{Location:} Provides a summary of places within the information space based on their geographical distances. For example, location summary may represent a concept such as \emph{Suburbs in Sydney, Australia} through grouping suburbs located within it, e.g., Five Dock, Canada Bay, and Kensington.

 \item \textbf{Person:} Identifies person names within the information space and groups people based on their title. For example, person summary may create a concept like `Health Ministers' by grouping persons, such as `Greg Hunt'~\footnote{Health minister of Australia} and `Brad Hazzard'~\footnote{Minister for health and medical research in NSW state of Australia}.

 \item \textbf{Organization:} Identifies companies and organizations within the information space and groups them based on the services they provide. For example, organization summary may place financial companies into a group and consider them as a concept. 
 \item \textbf{Topic:} Provides a topical summary of the information space based on the keywords' semantical relationships. It examines the hypernym relationship of keywords and groups them based on their similarity. For example, this summary may extract the keywords pigeon, crow, eagle and seagull from the information space and groups them as \emph{bird}.
 \item \textbf{Category:} Categorizes keywords within information space into 200 pre-validated topics~\cite{fast2016empath}. This summary computes the vector similarity of keywords and categories and assign a keyword to the category that yielded the highest similarity score.

 \item \textbf{Keyword:} Presents the most frequent keywords within information space and lets a user manually examine the relation between keywords to create her topic of interest.
 \end{itemize}
Providing several summaries of the information space not only enables a user examine her preferences from a different perspective, but also enhances the user comprehension and sensemaking of data. For example, if ConceptMap only provides the topical overview of the information space, then the user's comprehension of information space may remain incomplete as the user may not be able to examine the relations between attributes in documents from other perspectives. Our studies showed that providing different summaries of the information space boosts the user's understanding to better formulate her preferences, especially when seeking varied information across a large amount of data.

\subsubsection{2. Control Panel}
The second component of the ConceptMap is the Control Panel, which allows a user to modify concepts based on her preferences. The Control Panel is made up of two components: Details View (Figure ~\ref{fig:detail_evidence}a) and Evidence Box (Figure ~\ref{fig:detail_evidence}b). We will now turn to explain the character of components in detail.

\textbf{Details View:}
The Details View provides a detailed representation of concepts to a user. It shows attributes, e.g., keywords and named entities, associated with a concept and enables a user to modify the concept based on her preferences. Attributes within the Details View are colored to represent their relevancy to a concept. The opacity of the color shows the measure of relevancy between a concept and its attributes. Darker color means higher relevancy, while lighter color shows lower relevancy. A detailed description of attributes can be seen in a tooltip by hovering over them. The tooltip provides a small textual description and lets a user better judge the relevancy of attributes to the concept (Figure ~\ref{fig:detail_evidence}a). In cases where a user identifies an attribute as irrelevant, the user may remove the attribute by pressing the \textbf{($\times$)} button located on top right side of it. A user can organize the potential concepts relevant to her preferences into the Evidence Box by pressing the Add button, located below the Control Panel.

\textbf{Evidence Box:}
\label{evidencebox}
The Evidence Box acts as a central repository and is designed to aid a user to gather potential concepts relevant to her preferences. Collecting concepts altogether in a place, allows a user to create a mental structure of the information space. It is particularly important in large information spaces as humans have a limited memory capacity in absorbing information. 
The concepts within the Evidence Box follows the same coloring scheme applied to the Radial Map (Figure ~\ref{fig:detail_evidence}b). Making it easier for a user to identify the type of summaries stored in the Evidence Box. 
A slider placed horizontally below concepts to visually encode the weight of concepts. Initially, the slider shows a pre-computed weight for each concept, which is the average TF-IDF score of attributes associated with it. A user may change the weight of a concept by moving the slider indicator to the left or right. Dragging the slider indicator to the left decreases the importance of a concept, while moving the indicator to right increases the importance of a concept.

\begin{figure}
 \centering 
 \includegraphics[width=\linewidth, height=2.5in]{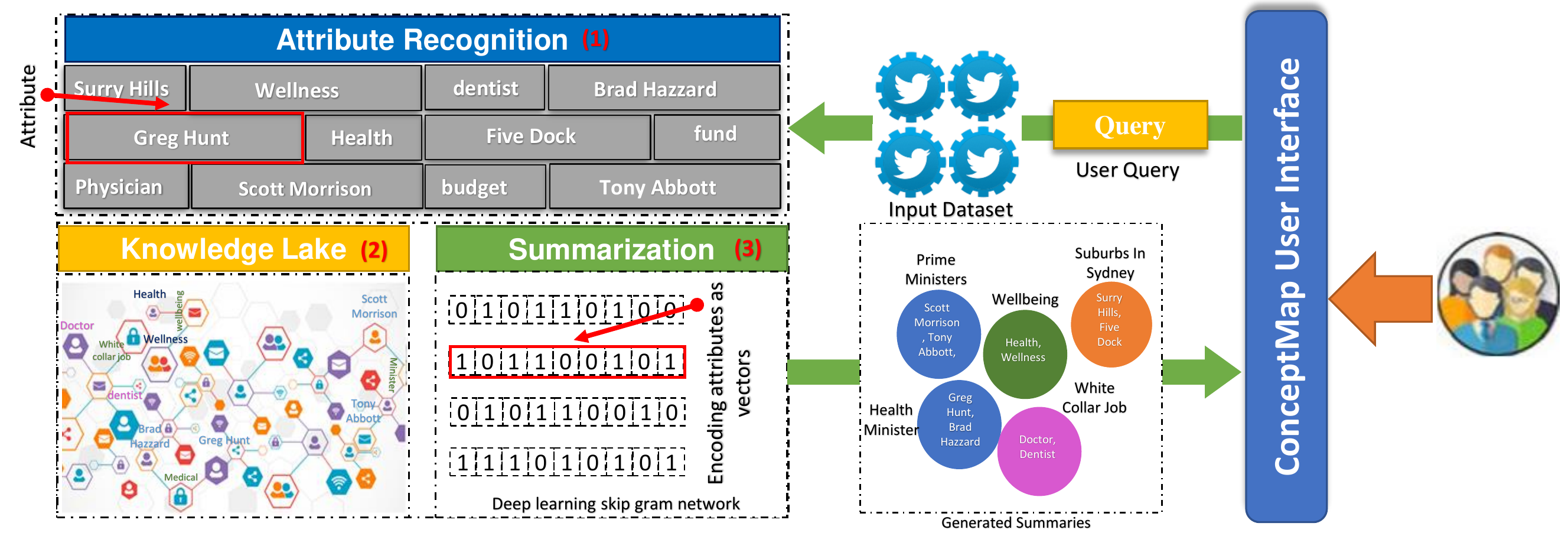}
 \caption{Overview of the proposed summarization technique: (1) Extracts the potential attributes from the information space, (2) Annotates the attribute using the Knowledge Lake, and (3) performs analogous reasoning through mapping attributes on a vector space~\cite{tabebordbar2019conceptmap}.}
 \label{fig:overview}
\end{figure}

\subsubsection{3. Query Box}
\label{query}
In this section, we explain the Query Box a component for examining a document-concept relevancy.
A user can interact with the Query Box through two interfaces: \emph{Concept Only} (Figure ~\ref{fig:queybox}a) and \emph{Concept + Rule} (Figure ~\ref{fig:queybox}b). The \emph{Concept Only} interface allows a user to specify her preferences implicitly by dragging and dropping concepts from the Evidence Box. A user can drag and drop several types of concepts, e.g., Topic, Category, Person, and Organization, into the Query Box, allowing to rank documents from different perspectives. Concepts within the Query Box colored according to their corresponding summary type, help a user to identify the type of summary used in ranking documents. A user can examine the relevancy among documents and concepts by pressing the \emph{Rank} button, where ConceptMap arranges documents accordingly in the \emph{Documents List}. 

The second interface \emph{Concept + Rule}, aids a user to formulate her preferences through rules. Rule based techniques have been coupled with IR systems for an extended period of time. Often, however, users are making mistake in using rules to formulate their preferences~\cite{dinet2004searching}. The goal of this interface is to study whether providing conceptual summary of the information space augments users ability to understand and utilize rules more effectively or not. 
The \emph{Concept + Rule} interface provides four operators for formulating user's preferences through rules `$[', `AND', `OR', `]$'. `[', indicates the start of a rule clause, `]', indicates the end of a rule clause. The `AND' operator implies a document must contain a specified concept to be ranked by the ConceptMap. The `OR' operator implies at least one of the user specified concepts must appear in a document to be ranked by the ConceptMap. For example, consider the following rule:
\begin{center}
$Q_1\ =\ [Hospital\ AND\ Health]\ OR\ [Health\ Care\ AND\ Budget ]$    
\end{center}

$Q_1$ ranks only those documents that contain both `Hospital' and `Health' concepts, or `Health Care' and `Budget' concepts. Next, we will explain how the ConceptMap formulates user's selected concepts for IR systems to rank documents.

 As each concept represents a set of attributes rather than a single keyword, we need to transform concepts into forms of text queries for ranking through IR systems. We do this by computing the cartesian product of concepts attributes and using the result set for ranking documents. More formally, we denote a concept $C= \{c_1, c_2, c_3, ...c_n\}$ as a set of attributes. We denote a preference $Q=\{C_1, C_2, C_3, ..., C_n\}$ as a set of concepts that a user dropped into the Query Box. To score documents based on the user preference, we compute the cartesian product of attributes associated with concepts $Q = C_1 \times C_2 \times C_3 \times ... \times C_n = \{(c_1, c_2, c_3, ..., c_n\ | c_n \in C_n\}$. The resulting set $Q=\{q_1, q_2, q_3, ..., q_n\}$ are the queries~\textemdash  ConceptMap computes their relevancy to documents. More clearly, suppose a user intends to analyse citizens opinion about government budget in `Health Care System' system. Assume, the user selects the `Health Ministers' concept from the \emph{Persons} summary $Health\ Ministers=\{Greg\ Hunt, Brad\ Hazzard\}$, and the `budget' concept from the \emph{Category} summary $Budget=\{fund, money, budget\}$. To generate the queries that represents the given concepts, ConceptMap computes the cartesian product of concepts as below: 
 \begin{center}
 $Q= \{(Greg\ Hunt, fund), (Greg\ Hunt, money), (Greg\ Hunt, budget), (Brad Hazzard, fund)$\\ , $(Brad\ Hazzard, money), (Brad\ Hazzard, budget)\}$
 \end{center}
 
 The following explains how ConceptMap computes the relevancy of queries and documents.
 ConceptMap arranges documents based on their relevancy to queries. To compute the relevancy of a query and a document, we implemented a Vector Space Model (VSM). The model transforms the document $d$ and query $q \in Q$ into vectors and computed their relevancy using a cosine similarity. 
\begin{center}
$ S(d,q)=\sum{\dfrac{tfidf(c,d).W_C}{||d||.||q||}}$
\end{center}
 Where $tfidf(c,d)$ is the tf-idf score for the attribute $c$ in document $d$. $W_C$ is the weight a user specified for the concept $C$ (see Section ~\ref{evidencebox}). Also, $|| d ||$ and $||q||$, are the Euclidean norms for vectors $d$ and $q$. ConceptMap arranges documents in descending order based on their cosine similarity scores to queries.

\subsubsection{4. Documents List}
The Documents List (Figure ~\ref{fig:main}d) provides a list of documents ranked based on the user's preferences. Documents List relies on stacked bar charts for visual encoding of documents. It shows a barchart below each document, illuminating the relevancy of documents to user preferences. To aid a user to better comprehend documents-concepts relevancy, Document List applies the same coloring scheme as the Radial Map. The coloring allows a user to identify the contribution of each concept to a document, and provides an explanation of why one document ranked higher than another. The ranking of documents are updated as a user modifies her preferences through adding or removing concepts from the Query Box.

\section{Solution Overview}
\label{solution_5}
In this section, we explain how ConceptMap generates a conceptual summary of the information space. To generate a summary, ConceptMap utilizes a \emph{Knowledge Lake}~\cite{beheshti2018datasynapse,beheshti2018corekg} and a deep learning skip-gram embedding network~\cite{mikolov2013distributed}. Knowledge Lake is a central repository made up of several knowledge bases and provides a contextualization layer for transforming the raw data into contextualized knowledge. It annotates attributes within the information space with a set of facts and information. Deep learning network measures the conceptual commonality existing between attributes and groups attributes with similar characteristics. Overall, our approach is made up of three stages, \emph{Attributes Recognition}, \emph{Knowledge Lake}, and \emph{Summarization}. Followings explain each stage in detail.

\subsection{Attributes Recognition}
The initial step in generating the summary is the identification of content bearing attributes exists within the information space. These attributes allow ConceptMap to discover the aboutness of data. The current version of ConceptMap extracts two types of attributes: \emph{Keyword} and \emph{Named Entity}. To extract the attribute type of keyword, ConceptMap performs a preprocessing task by removing the stopwords~\footnote{Stopwords are words, e.g., the, is, are with little meaning that commonly occur within documents.}, keeping the proper names capitalized, and filtering out of the ungrammatical and irrelevant tokens, e.g., URLs' or emoji. Also, it applies the WordNet lemmatizer over the remaining tokens to increase the probability of matching between words with the common base, e.g., `playing', `playful', `plays' all reduce to the base form `play'. The second attribute type is Named Entity. Named Entities are the span of words in a text which refer to real-world objects, such as person and company names, or gene and protein names. Examples of Named Entities include Barack Obama, New York City, Volkswagen Golf, or other proper names. The current version of ConceptMap extracts three types of named entities: persons, locations, and organizations. For recognizing named entities, we used our previous work~\cite{beheshti2017automating}. It provides a pipeline for various data curation tasks, including Named Entity Recognition, Information Linking, Similarity Computation, and Indexing. After extracting attributes, we annotate them through a Knowledge Lake to identify the concepts existing within the information space. In the next section, we explain how Knowledge Lake contributes to our work for generating the conceptual summary of the information space. 
\subsection{Knowledge Lake}
A Knowledge Lake enables a user to understand attributes within the information space and provides a foundation to measure the commonality between them. We utilize several readily available knowledge bases and taxonomies to create the Knowledge Lake:~(1) \emph{Geoname}, is a geographical database and contains information over 25 million geographical places around the world~\footnote{http://geonames.org/},~(2) \emph{Wikidata}, is a central storage of several Wikimedia data, including Wikipedia, Wikivoyage, Wikisource.~\footnote{https://www.wikidata.org},~(3) \emph{WordNet}, is a semantic lexicon and grouped English words into sets of synonyms called synsets~\footnote{https://wordnet.princeton.edu/citing-wordnet},~(4) Empath, is a deep learning skip-gram network, which categorizes text over 200 built-in categories.~\cite{fast2016empath}, and~(5) Google Knowledge Graph~\footnote{https://developers.google.com/knowledge-graph/}, is a knowledge base based on a graph database and provides information about real-world entities, including persons, locations, and business. These knowledge bases allow discovering the aboutness of data through annotating attributes with a more generalized or understandable form.

More formally, the Knowledge Lake $K$ acts as a function $K :\ c_i \longmapsto \ell({c_i})$, that receives an attribute $c_i$ as the input and returns an annotation $\ell({c_i})$ to describe the attribute. 
For example, consider the following Tweet \emph{`Malcolm Turnbull says his government will focus on growth rather than fix the budget deficit'}. To understand the Tweet, ConceptMap annotates its attributes, e.g., `Malcolm Turnbull', `government', `growth', `budget', with the Knowledge Lake from different perspectives. It may annotate the Named Entity `Malcolm Turnbull' as the `former prime minister of Australia', the keyword \emph{`budget'} as a topic for `fund', etc. ConceptMap applies the annotation for all attributes within the information space. In the following section, we explain how we employ a deep learning neural network to measure the commonality between attributes to generate a summary of the information space.

\subsection{Summarization}
\label{summary}
In this section, we explain how ConceptMap generates a conceptual summary of the information space. For example, we explain how it identifies two persons, e.g., `Greg Hunt'~\footnote{Health Minister of Australia} and `Brad Hazzard'~\footnote{New South Wales Minister for Health and the Minister for Medical Research} can be similar and forms a concept.

As we discussed, ConceptMap annotates attributes~\footnote{ConceptMap generates \emph{topic} and \emph{category} summaries from the attribute type of keyword, while \emph{person}, \emph{organization} and \emph{location} summaries from the named entities.} within the information space through the Knowledge Lake. ConceptMap uses these annotations to generate the summaries. For generating the \emph{Topic} summary, ConceptMap groups attributes based on their hypernym relations. For example, it groups attributes, such as $\{doctor, physician, dentist\}$ as `Medical\_Practitioner', while attributes like $\{health\ and\ wellness\}$ as `wellbeing'.

For generating the \emph{Category} summary, ConceptMap relies on the EMPATH, which categorizes text into 200 built-in human-validated categories. For example, it may categorize $\{doctor\ and\  physician\}$ attributes as relevant to `medical\_emergency', `occupation' and `white\_collar\_job' categories, while $\{health\}$ as `medical\_emergency', but not `occupation' and `white\_collar\_job'. ConceptMap visualizes categories with the highest frequency within the information space.

To generate the \emph{Person, Organization, and Location} summaries, ConceptMap 
takes advantage of a deep learning skip-gram network~\cite{mikolov2013distributed} to predict the semantic similarity between attributes within the information space. For example, the network may learn that the word `health' may predict `medical', but not of `happiness'.

By training the skip-gram network, it learns a representation of words within the information space, which known as neural embeddings. The neural embeddings construct a vector space model and allow to measure the similarity between attributes in an unsupervised fashion. We used word2vec neural embeddings model~\footnote{https://drive.google.com/file/d/0B7XkCwpI5KDYNlNUTTlSS21pQmM/edit} to map attributes onto a vector space.

For attributes annotated with the Knowledge Lake, ConceptMap encodes attributes as vectors by querying the vector space model trained on the word2vec. Then, it performs `analogous reasoning'~\footnote{A form of comparison to highlight respects in which two attributes can be similar} ~\cite{fast2016empath} by conducting the vector arithmetic on generated attributes vectors, e.g., the vector arithmetic for words `Women + King - Man' generates a vector similar to `Queen'. The following explains how ConceptMap performs analogous reasoning to measure the similarity between attributes.

For each attribute $c$, ConceptMap tokenizes its annotation $\ell({c})$ into a set of words $\ell({c})=\ \{\ell_1^{\prime}({c}), \ell_2^{\prime}({c}), \ell_3^{\prime}({c}),..., \ell_n^{\prime}({c})\}$. Then, for each $\ell^{\prime}({c}) \in \ell({c})$, ConceptMap queries the vector space model and extracts the vector $V(\ell^{\prime}({c}))$ corresponds to it. It performs analogous reasoning by computing the vector sum of all $V(\ell^{\prime}({c}))$. The resulting vector $V(c)$ represents the attribute $c$ in vector space.
 \begin{center}
 $V(c)\ =\ \sum_{i = 1}^{n}\ell_i^{\prime}(c)$
 \end{center}
 ConceptMap groups similar attributes based on their vectors similarity using the cosine measure. For example to compute the vector similarity of two attributes, the cosine similarity computed as:
 \begin{center}
 $cos(\theta)= \dfrac{V(c_1)\ .\ V(c_2)}{||V(c_1)||.\ ||V(c_2)||}$
 \end{center}
 Where $V(c_1)$ and $V(c_2)$ representing vectors of attributes $c_1$ and $c_2$, and $||V(c_1)||$ and $V(c_2)$ are their lengths. 
More clearly, consider the attribute $c = \{Greg\ Hunt\}$, which annotated with the Knowledge Lake as the $\ell(c) = \{ Health\ Minister\ of\ Australia\}$. To represent this attribute as a vector ConceptMap tokenizes $\ell(Greg\ Hunt)=\{Health,\ Minister,\ Australia \}$ and query a VSM to extract their corresponding vectors $\ell(Greg\ Hunt)=\{V(Health),\ V(Minister),\ V(Australia)\}$. Then, it computes the vector sum of all attributes $V(Greg\ Hunt)\ = V(Health)\ + V(Minister)\ + V(Australia)$. The resulting vector $V(Greg\ Hunt)$, represents the attribute $c = \{Greg\ Hunt\}$ in vector space, and allows ConceptMap to compute its similarity with other attributes vector using the cosine similarity. ConceptMap groups attributes that their cosine similarity is above a pre-defined threshold~\footnote{Currently, we consider attributes over 0.7 \% cosine measure as similar.}. Figure~\ref{fig:overview} illustrates an overview of our proposed summarization technique.

\section {Experiments}
\label{experiment_5}
In this section, we discuss the structure of the ConceptMap, as well as the dataset used in our experiments. Also, we discuss two usage scenarios to show how the ConceptMap helps a user to formulate her preferences.
\subsection{ConceptMap Architecture and Datasets}
The core component of techniques described in the previous sections is implemented in Python and JavaScript. We gathered over 300K Tweets (Australian region) relevant to health care and budget to create the input dataset. We used ElasticSearch~\footnote{https://www.elastic.co/start} as the search engine to index and retrieve Tweets. In our preliminary implementation, we only indexed Tweets and performed the annotation and summarization simultaneously whenever a user interacts with the ConceptMap. However, it turns out that the annotation is the most time-consuming task, which hinders the ConceptMap to effectively response to the user interactions. We alleviate the ConceptMap response time by separating the annotation and summarization tasks. First, for each summary, we indexed attributes and their annotation along with the Tweet text. Then, whenever a user interacts with the ConceptMap to fetch a summary of the information space, ConceptMap only computes the similarity between attributes (see Section~\ref{summary}) to group similar concepts. In this manner, it avoids annotating attributes constantly and could generate summaries promptly.

\subsection{Experiment Settings}
In this section, we study the performance of ConceptMap in formulating users preferences with respect to a traditional keyword-based UI. The study followed a repeated measures design ANOVA~\cite{girden1992anova} with two independent variables: \emph{tool}: ConceptMap, which consists of Concept-Only and Concept-Rule interfaces, and a traditional keyword-based UI \textemdash\  and \emph{items}. The Keyword-Based UI allows users to investigate the information space by entering their keywords and observing the resulting set ordered by their relevancy in the Documents List. Also, to aid users to better identify the relationship among keywords, we visualize the most frequent keywords co-occurred with users' selected keywords. To counterbalance the experiment, we conducted the study on two different topics relevant to social issues: \emph{Health Care} and \emph{Budget}, where topic treated as a random variable.

The study simulates an exploratory search scenario, where users need to write queries to retrieve Tweets relevant to the given topics. We divided users task into two subtasks: a focused exploratory search scenario and a broad exploratory search scenario. The focused search scenario requires users to investigate the information space to retrieve items for a limited number of topical subspaces, e.g., retrieve a list of Tweets contain information relevant to medical centres within Australia. The broad search scenario simulates cases where users need to retrieve items for a larger number of topical subspaces, e.g., retrieve citizens' opinions using the Twitter about people who involved in health care system, and Tweets about the quality of the services provided by health care centres.

We invited five post-graduate students from a research lab to take part in experiments. None of the participants was knowledgable in the topics selected for the study. For evaluating the performance of tools, participants selected a topic and performed the search scenarios. The goal of these experiments were to reflect the behaviour of users in formulating their preferences while they were investigating the information space with different information needs. For each task, participants filled a 7-point likert scale NASA TLX questionnaire. The questionnaire is a multidimensional assessment tool that rates the perceived workload in order to assess a task or system. We limited the duration of focused search scenario to five minutes and the broad search scenario to 10 minutes. During the experiment, we reminded participants when their allotted time was almost over, but we didn't force them to stop using the tools.

\begin{figure*}[t]
 \centering 
 \includegraphics[width=\linewidth,height=1.8in]{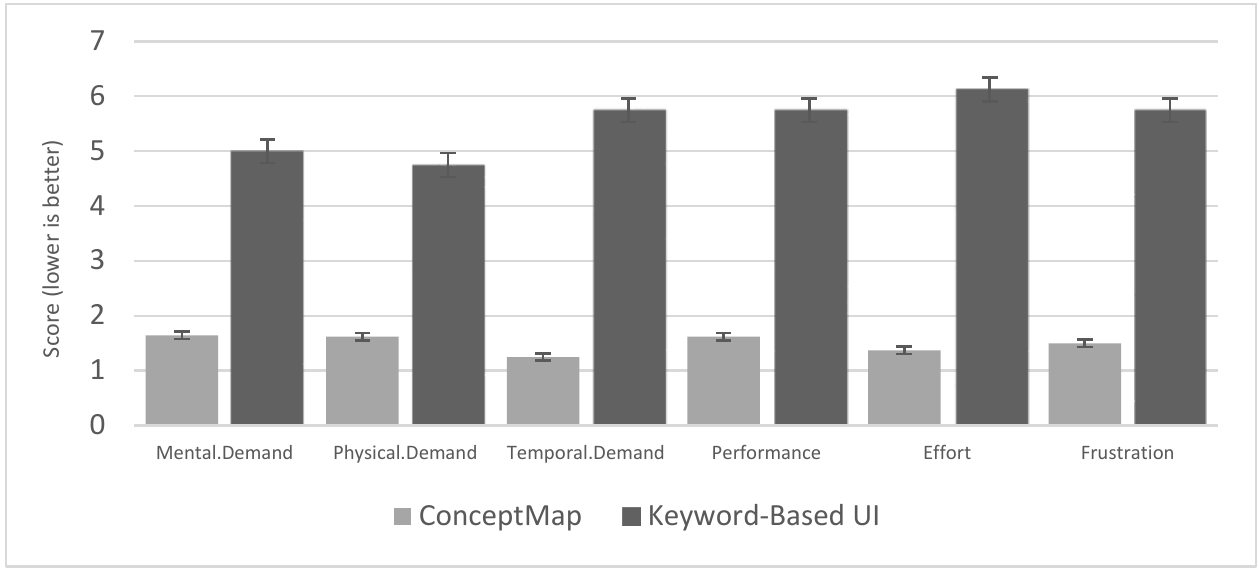}
 \caption{The bar chart shows ConceptMap imposes lower workload across different dimensions. Error bars shows the standard error.}
 \label{fig:tlx}
\end{figure*}

\textbf{Results: Workload and Performance Analysis:}
A repeated measure ANOVA revealed the impact of ConceptMap in lowering participants overall workload $F(1,5)=58.803$, $p=0.05$. This tendency can be observed in detail in Figure~\ref{fig:tlx}, which shows participants were more relaxed while interacting with ConceptMap for formulating their preferences. The results showed ConceptMap could significantly lowers participants temporal demand $F(1,5)=162.00$, $p=0.001$. We also observed a similar impact on improving participants efforts $F(1,5)=83.308, p=0.003$, and performance $F(1,5)=43.560$, $p=0.007$. Based on the obtained results, we concluded that providing several summaries of information space could boost participants comprehension and sensemaking of data: this impact is more significant on improving users performance and time.

We also analyzed the effectiveness of ConceptMap in aiding participants in formulating their preferences through rules. A repeated measure ANOVA showed that ConceptMap could lower participants overall workload $F(1,5)=12.60, p=0.05$. The results revealed that ConceptMap reduces participants effort in crafting rules $F(1,5)=18.00$, $p=0.005$. We also observed a similar impact on participants performance $F(1,5)=22.04, p=0.003$.

\begin{figure*}[t]
 \centering
 \includegraphics[width=\linewidth, height=4cm]{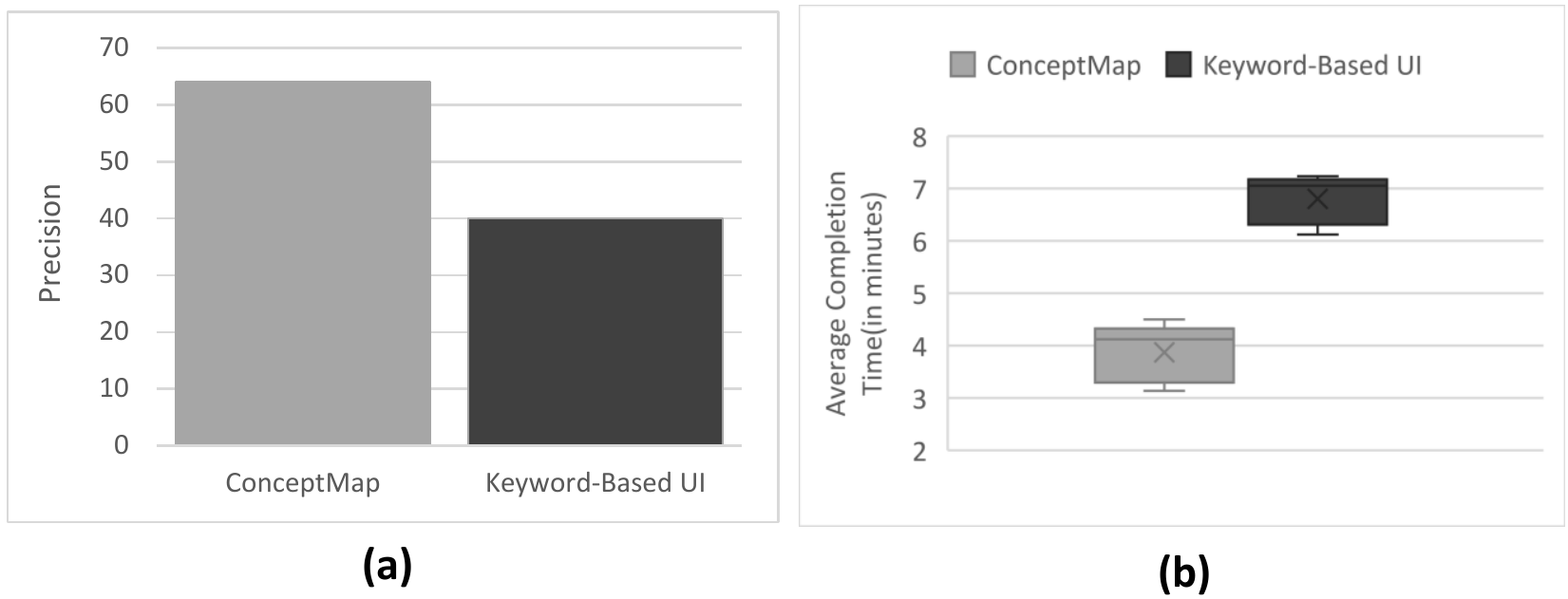}
 \caption{ (a) The average precision of participants obtained using ConceptMap and Keyword-Based UI, and (b) The time participants spent for retrieving their information.}
 \label{fig:prectime}
\end{figure*}

In addition to previous experiments, we analyzed the performance of tools by aggregating the top 20 items collected by participants and verifying their relevancy. Thus, we created two datasets from the aggregated items. The first dataset represents items collected by participants through interacting with the ConceptMap, and the second dataset contains items collected through interacting with the Keyword-Based UI. Then, we verified whether a retrieved item is relevant to the topics assigned by participants or not. The results showed that participants could retrieve items more precisely using the ConceptMap compared to the Keyword-Based UI. Figure~\ref{fig:prectime}a shows the average precisions obtained using the tools. Based on the observed results, allowing users to examine the relevancy of data from different perspectives has a positive impact on retrieving items.

\textbf{Results: Completion Time and Usability Analysis:}  In the second study, we analysed the completion time and the usability aspect of ConceptMap. We assigned a task to participants and let them accomplish the task using the tools in more natural settings, e.g., without times-up. The task was a broad exploratory search scenario for retrieving Tweets relevant to issues and people involved in budget planning. Then, we asked participants to fill a standard Software Usability Scale (SUS) questionnaire~\cite{brooke1996sus}. SUS provides subjective assessments of software usability, where a statement is made, and respondents can indicate the degree of agreement and disagreement on a five-point scale.  We averaged over all participants questions, the mean score amounted to 87.5 out of 100, which falls ConceptMap in the 90-95 percentile range in the curve grading scale interpretation of SUS scores~\cite{sauro2012standardized}.

We also calculated the time participants spent to accomplish their task. We observed using the ConceptMap participants could accomplish their task in a shorter time compared to the Keyword-Based UI (Figure~\ref{fig:prectime}b). The results confirm the impact of ConceptMap on lowering participants temporal demand.

At the end of the study, we asked participants to share their impressions on the strengths and weakness of the ConceptMap. All participants agreed that ConceptMap could enhance users comprehension and sensemaking of the information space. One of the participants noted to the Evidence Box, allowing her to store potentially relevant concepts in one place and shortening the time needs to examine the relevancy of concepts and documents. Another participant mentioned that providing several summaries of the information space could help him to formulate his preferences from different perspectives. Another participant mentioned the potential of ConceptMap to support users in exploratory search scenarios where there is no well-defined goal.

\section{Discussions}
\label{discussion_5}
In this section, we discuss our goal for designing the ConceptMap interface, and its limitations.
\subsection{ConceptMap Interface} To design the ConceptMap interface, we focused on providing the controllability and transparency a user needs to effectively formulate her preferences. ConceptMap interface, provides the controllability over the information space by allowing a user to combine concepts from different perspectives. ConceptMap also lets a user modify concepts based on her preference by removing attributes associated with them. In addition, in cases a user could not formulate her preference using the provided summaries, ConceptMap enables the user to manually create her topic of interest using the Keyword summary.

ConeptMap also provides transparency by relying on various visual encodings. ConceptMap uses a consistent coloring scheme across all components to help a user to understand the cause of retrieving an item. ConceptMap also provides an explanation of why an attribute associated with a concept in a tooltip, which can be seen by hovering the mouse over the attribute. The explanation facilitates the comparison of attributes grouped as a concept allows a user to identify inconsistencies that may occur in summarizing the attributes. 

\subsection{Limitations and Future Works} Today, the ConceptMap Query Box provides two interfaces for capturing user preferences. However, the Rule + Concept interface supports `AND' and `OR' operators only. In order to more effectively formulate user preferences through rules, ConceptMap needs to support other operators like `Not'. We are planning to add this operator as a part of future work. The second limitation is that for each concept within the information space, ConceptMap labels the concept using the attribute with the highest frequency. For example, if ConceptMap generates a concept that represents hospitals within Sydney, it labels the concept with the hospital name in the concept that has the highest frequency within the information space. However, in our experiments, we discovered that users prefer to see more descriptive names for the concepts visualized on the Radial Map.   
\section{Conclusion}
\label{conclusion_5}
In this chapter, we introduced the ConceptMap, a system that automatically identifies the relation between attributes, e.g., Keywords and named entities, within the information space. ConceptMap produces several summaries of data, e.g., Topic, Category, Person, and Organization, and enables a user to formulate her preferences implicitly as a set of abstract concepts. To generate the summaries, ConceptMap relies on a Knowledge Lake and a deep learning skip-gram network, which groups attribute based on their conceptual similarity. Our results showed that providing a conceptual summary of the information space enables a user to better formulate her preferences, especially when seeking for varied information in a large information space.

\chapter{Automating Basic Data Curation Tasks (Software Prototype)}\label{Chapter6}
In this chapter, we present a software prototype to assist analysts in curating the raw data and deriving insight. We develop a set of APIs to automate different curation tasks and creation of data curation pipelines. The APIs are available as open source on GitHub~\footnote{https://github.com/unsw-cse-soc/Data-curation-API}. Additionally, we have provided a set of rest services and a Web interface to support analysts to curate their data without writing code. 

The rest of this chapter is organized as follows: We introduce the problem of data curation in Section~\ref{introduction_6}. In Section~\ref{solution_6}, we introduce the services provided by the curation APIs. Then, in Section~\ref{scenario_6}, we discuss different usage scenarios and how it assists analysts to derive insight and extract value. Finally, we conclude the chapter in Section~\ref{conclusion_6}. 

The content of this chapter is derived from the following paper(s):
\begin{itemize}
    \item A Beheshti, A Tabebordbar, B Benatallah, R Nouri, \textbf {On automating basic data curation tasks.} In Companion  Proceedings of the 26th International Conference on World Wide Web Companion 2017 Apr 3 (pp. 165-169). (ERA Rank A*).
   \item Beheshti, A., Tabebordbar, A. and Benatallah, B., 2020, April. iStory: Intelligent Storytelling with Social Data. In Companion Proceedings of the Web Conference 2020 (pp. 253-256).
\end{itemize}
\newpage

\section{Introduction}
\label{introduction_6}
Understanding and analyzing big data is firmly recognized as a powerful and strategic priority~\cite{chen2012business}. For deeper interpretation of and better intelligence with big data, it is important to transform raw data (unstructured, semi-structured and structured data sources, e.g., text, video, image data sets) into contextualized data and knowledge that is maintained and made available for use by end-users (e.g., data scientists and researchers) and applications (e.g., data and machine learning applications). In particular, data curation acts as the glue between raw data and analytics, providing an abstraction layer that relieves users from time-consuming, tedious and error-prone curation tasks. Data curation involves identifying relevant data sources, extracting data and knowledge, cleaning, maintaining, merging, enriching and linking data and knowledge. For example, consider a tweet in Twitter [Kwak et al. 2010]: a micro-blogging service that enables users to Tweet about any topic within the
140-character limit and follow others to receive their tweets. It is possible to extract various information from a single tweet text such as keywords, part of speech, named entities, synonyms and stems~\cite{gattani2013entity}. Then it is possible to link the extracted information to external knowledge graphs to enrich and annotate the raw data. Later,
this information can be used to provide deeper interpretation of and better intelligence with the huge number of tweets in Twitter: every second, on average, around 6,000 tweets are tweeted on Twitter, which corresponds to over 350,000 tweets sent per minute, 500 million tweets per day and around 200 billion tweets per year.

In particular, the data curation process enables extracting knowledge and deriving insights from the vastly growing amounts of local, external and open data. This task is vital for recent data analytics initiatives include: improving government analytical services, personalized advertisements in elections, and predicting intelligence activities~\cite{tene2012big}. In this chapter, we identify and implement a set of basic data curation APIs and make them available to researchers and developers as services to assist them in transforming their raw data into curated data. The curation services enable developers to easily add features - such as extracting keyword, part of speech, and named entities such as Persons, Locations, Organizations, Companies, Products, Diseases, Drugs, etc., providing synonyms and stems for extracted information items leveraging lexical knowledge bases for the English language such as WordNet, linking extracted entities to external knowledge bases such as Google Knowledge Graph and Wikidata, discovering similarity among the extracted information items, such as calculating similarity between string, number, date and time data, classifying, sorting and categorizing data into various types, forms or any other distinct class, and indexing structured and unstructured data into their data applications. These services can be accessed via a REST API, and the data is returned as a JSON file an easy-to-parse structure, that can be integrated into (data and machine learning) applications. The basic data curation APIs are available as an open-source project on GitHub. The technical details for the curation APIs can be found in a technical report~\cite{CurationAPIs}. The rest of the chapter is organized as follows. In Section~\ref{solution_6}, we present an overview of the curation services, and in Section~\ref{scenario_6}, we further discuss the usage of our APIs through presenting a demonstration scenario.

\section{Curation Services Overview}
\label{solution_6}

\begin{figure} [t]
\centering
\includegraphics[width=1.1\linewidth]{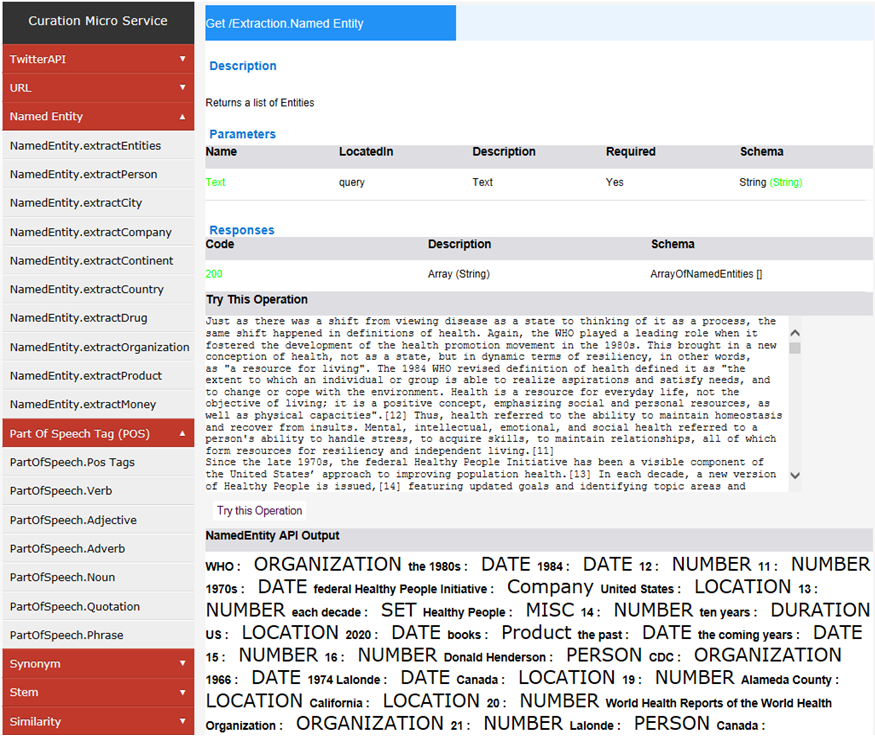}
\caption{Data Curation Services ScreenShot- Named Entity Extraction~\cite{beheshti2017automating}.}\label{fig:nameentity-curationservice}
\end{figure}

To augment users in extracting features, we proposed a set of curation APIs. The APIs are implemented as micro-services and provide services such as Extraction, Classification, Linking, and Indexing. The curation services use natural language processing technology and machine learning algorithms to transform the raw data, e.g., by extracting semantic meta-data from content, such as information on people, places, and companies and link them to knowledge graphs such as WikiData and Google KG - using similarity techniques - or classify the extracted entities using classification services. We provide curation services for performing content analysis on Internet-accessible web pages, posted HTML or text content. A technical report is also provided~\cite{Beheshti2016}, which further guides analysts on how to utilize the services for their curation tasks. In the following, we present an overview of the services.

\begin{enumerate}
\item \textbf{Extraction Services:} The majority of the digital information produced globally presented in the form of web pages, text documents, news articles, emails, and presentations expressed in natural language text. Collectively, such data is termed unstructured as opposed to structured data that is normalized and stored in a database. The domain of information extraction is concerned with identifying information in unstructured documents. In most cases, this activity concerns processing human language texts utilizing Natural Language Processing (NLP). Accordingly, analysts may need a collection of natural language processing APIs to extract entities, Part of Speech (PoS), keywords, synonym, stems and more.

\begin{enumerate}
\item \textbf{Named Entity Recognition (NER)}, can be used to locate and classify atomic elements in text into predefined categories such as the names of persons, organizations, locations, expressions of times, quantities, monetary values, and percentages~\cite{beheshti2017systematic}. In particular, named entities carry important information about the text itself, and thus are targets for extraction. Accordingly, NER is a key part of information extraction systems that supports robust handling of proper names essential for many applications, enables pre-processing for different classification levels, and facilitates information filtering and linking. State-of-the art NER systems for English produce near-human performance. Figure~\ref{fig:nameentity-curationservice} illustrates a screenshot of the basic data curation services presenting an example for the named entity extraction task. 

\item \textbf{Part-of-Speech (PoS)} is a category of words (or more generally, of lexical items)with a similar grammatical properties~\cite{jurafsky2000speech}. Words assigned to the same PoS display a similar behaviour in terms of syntax, grammatical structure of sentences, and morphology. Commonly listed English PoS are noun, verb, adjective, adverb, pronoun, preposition, conjunction, interjection, and sometimes numeral, article or determiner.

\item \textbf{Keyword} In corpus linguistics, a keyword is a word which occurs in a text more often than we would expect to occur by chance alone. Keywords are calculated by carrying out a statistical test which compares the word frequencies in a text against their expected frequencies derived in a much larger corpus, which acts as a reference for general language use. To assist analysts filtering and indexing open data, it will be important to extract keywords from unstructured data such as tweets' text. 

\item \textbf{Synonym} is a word or phrase that means exactly or nearly the same as another word or phrase in the same language. An example of synonyms is the words begin, start, and commence. Words can be synonymous when meant in certain contexts, even if they are not synonymous within all contexts. For example, if we talk about a long time or an extended time, `long' and `extended' are synonymous within that context~\cite{manning-EtAl:2014:P14-5}. While analyzing the open data, it is important to extract the synonyms for the keywords and consider them in the analysis steps. For example, sometimes two tweets can be related if we include the synonyms of the keywords in the analysis: instead of only focusing on the exact keyword match. It is important as the synonym can be a word or phrase that means exactly or nearly the same as another word or phrase in the tweets. 

\item \textbf{Stem} is a form to which affixes can be attached. To assist analysts understand and analyze the textual context, it will be important to extract derived form of the words in the text. For example, considering the keyword `health’, using the Stem service, it is possible to identify derived forms such as healthy, healthier, healthiest, healthful, healthfully, and healthfulness, and more accurately identify the information items, e.g., tweets, that are related to health. 


\item \textbf{Information Extraction from a URL:} A Uniform Resource Locator (URL), is a reference to a Web resource that specfies its location on a computer network and a mechanism for retrieving it. Considering a tweet that contains a URL link, it is possible to extract various types of information including: Web page title, paragraphs, sentences, keywords, phrases, and named entities. For example, consider a tweet which contains URL links. It is possible to extract further information from the link content and use them to analyze the Tweets. 
\end{enumerate}

\item \textbf{Linking Services}
\begin{enumerate}
 \item \textbf{Similarity} Approximate data matching usually relies on the use of a similarity function, where a similarity function $f(v1,v2) \longrightarrow s$ can be used to assign a score $s$ to a pair of data values $v1$ and $v2$. These values are considered to be representing the same real-world object if $s$ is greater than a given threshold $t$. Different similarity functions have been proposed for comparing~\cite{beheshti2017systematic}: strings (e.g., edit distance and its variations, Jaccard similarity, and TF-IDF based cosine functions), numeric values (e.g., hamming distance and relative distance), images (e.g., Earth Mover Distance) and more. Accordingly, analysts may need a collection of similarity metrics to measure the Cosine similarity of two vectors of an inner product space and compares the angle between them, the Jaccard similarity of two sets of character sequence, the length of the longest common sub-sequence between two strings using an edit distance algorithm, the hamming distance between two strings of equal length and more. 

\item \textbf{Knowledge Base} While extraction services can augment analysts to extract various features, e.g., named entities, keywords, synonyms, and stems, from a text, it is important to go one step further and link the extracted information into the entities in the existing Knowledge Graphs (e.g., Google KG and Wikidata). For example, consider, we have extracted ‘M. Turnbull’ from a Tweet text. It is possible to identify a similar entity (e.g., ‘Malcolm Turnbull’) in Wikidata. As discussed earlier, the similarity API supports several functions such as Jaro, Soundex, QGram, Jaccard and more. For this pair, the Jaro function returns 0.74, and the Soundex function returns 1. To achieve this, we have leveraged the Google Knowledge Graph, Wikidata, and ConceptNet knowledge bases to link the extracted entities from the text to the concepts and entities in these knowledge bases. 
 \end{enumerate}

\item \textbf{Classification Services}
Classification is a data mining function that assigns items in a collection to target categories or classes. The goal of classification is to accurately predict the target class for each case in the data. For example, a classification model could be used to identify loan applicants as low, medium, or high credit risks. A classification task begins with a dataset in which the class assignments are known. For example, a classification model that predicts credit risk could be developed based on observed data for many loan applicants over a period of time. In the terminology of machine learning, classification is considered as an instance of supervised learning. At the same time, the unsupervised procedure is known as clustering, and involves grouping data into categories based on some measure of inherent similarity or distance ~\cite{jajuga2012classification}. An algorithm that implements classification, especially in a concrete implementation, is known as a classifier. Examples of classification algorithms include: Linear classifiers, Support vector machines, Quadratic classifiers, Kernel estimation and Decision trees. Figure~\ref{fig:data-curation-classification} illustrates a screenshot of the basic data curation services presenting an example for the classification task. 

\item \textbf{Indexing Services}
enable analysts to scan and retrieve the curation environment quickly without the operational burden of managing it. For indexing the content of the curation environment, we utilized elastic search, to allow the user to query the data and derive insight.

\item \textbf{Converter Services}
The basic curation APIs may be applied to different data sources and file formats. To facilitate this task, the converter API can be used to convert PDF, Word, PowerPoint, XPS, and HTML documents into a text file to be fed to the basic curation APIs, where the result is returned as a JSON file, an easy-to-parse structure, and can be integrated into data applications. As an ongoing work, we are identifying various data sources and file formats to facilitate converting documents without user interaction.
\end{enumerate}

\begin{figure}
\centering
\includegraphics[width=\linewidth,height=7.0in]{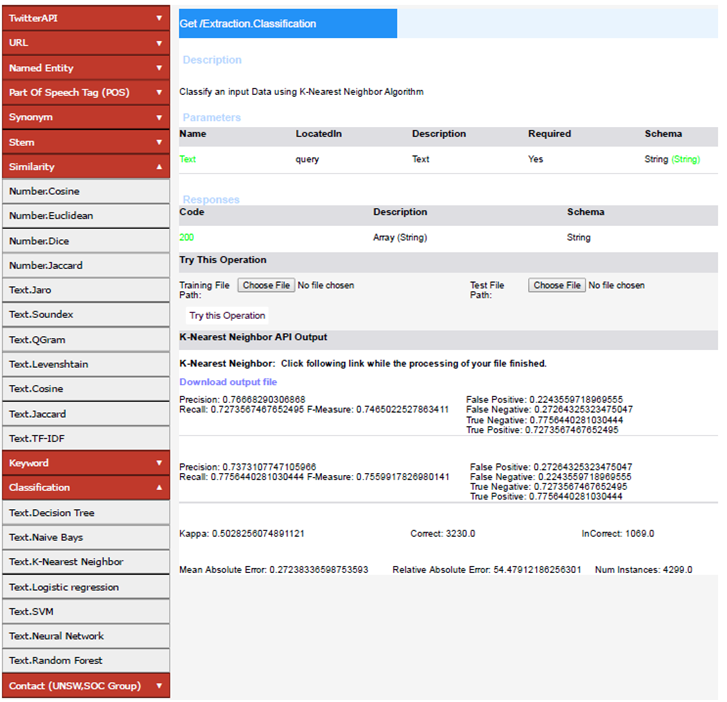}
\caption{Data Curation Services ScreenShot- Classification~\cite{beheshti2017automating}.}
\label{fig:data-curation-classification}
\end{figure}

\section{Demonstration Scenarios}
\label{scenario_6}
In this section we demonstrate a scenario to illustrate how an analyst can leverage different curation APIs to curate Tweets relevant to Australian budget.

Consider a data analyst, say Alan, that intends to analyze Tweets to understand citizens' opinions regarding the quality of services provided through the government. These services includes a large number of government projects such as light rail, highways, road maintenance, public transportation, health cares. Analyzing and understanding citizens opinions regarding these services allows decision-makers to prioritize them and better plan their budget. Using the curation APIs, Alan can leverage different curation services, including linking, extraction, classification, and indexing, to extract insight from Tweets. For example, Alan, can use the extraction service to extract named entities, keywords,
synonyms, PoS and stems from tweets. Then, he may use the linking service to link extracted information to knowledge bases such as Wikidata and Google Knowledge Graph. Next, Alan may utilize the classification service to classify Tweets that are related to and, finally index the retrieved information for fast access and further analysis.\newline

%

Additionally, the curation services can allow Alan to perform more focused analysis, such as extracting Tweets relevant to the health category of the government budget. For example, Alan, can extract named entities such as (i) People, from GPs and Nurses to health ministers and hospital managers from Australian doctors directory, (ii) Organizations, such as Hospitals, Pharmacies and Nursing Federation from myHospitals, (iii) Locations, states, cities and suburbs in Australia from auspost10, (iv) Health funds, such as Medibank, Bupa and HCF from health-services, (v) Drugs, such as Amoxicillin, Tramadol and Alprazolam from drug index, (vi) Diseases, such as Cancer, In uenza and Tuberculosis from medicine-net, (vii) Medical Devices, such as Gas Control, Blood Tube and Needle from FDA14, (viii) Job titles, such as GP, Nurse, Hospital Manager, Secretary of NSW Health and NSW Health Minister from comp data, and (ix) Keywords, such as healthcare, patient, virus, vaccine and drug from Australia national health and medical research council. Then he can enrich these named entities using Knowledge Bases such as Wikidata, Google Knowledge Graph and Wordnet. Then, we use the classification service to identify the Tweets with negative sentiment. Notice that, for the sentiment analysis, the classification API leverages the sentiment classifier implemented in the Apache PredictionIO (http://prediction.io). For example, using the Curation APIs, out of 2934 diabetes-related Tweets, we could identify 615 tweets with negative sentiment. As another example, we have identified 1549 tweets with negative sentiment in the mental health category.

\begin{figure} [t]
\centering
\includegraphics[width=\linewidth]{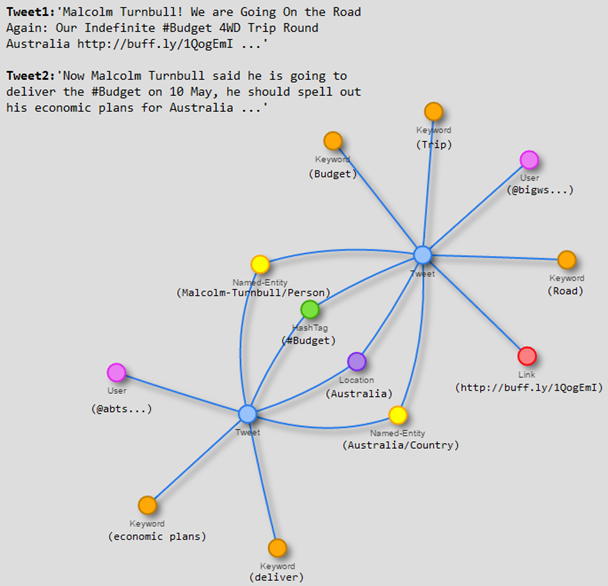}
\caption{Use extracted features from Twitter to link related tweets~\cite{beheshti2017automating}.}
\label{fig:data-curation-user-tweet-feature-extract}
\end{figure}

\textbf{Example:} Figure~\ref{fig:data-curation-user-tweet-feature-extract} shows two real Tweets and different information that have been extracted, e.g., named entities, keywords, and hashtags, to generate a graph where nodes are the main artifacts and extracted information are the relationships among them. As illustrated in this figure, Tweets are linked through
named entities and hashtags and this will generate an interesting graph which reveals the hidden information among the nodes in the graph: for example it is possible to see the path (transitive relationships among the nodes and edges) between user1 and user2 (in Twitter) which in turn reveals that these two users are interested in the same topics, and consequently may be part of some hidden micro-networks.\newline


\section{Conclusion}
\label{conclusion_6}
In this chapter, we identified and implemented a set of curation services to make them available to researchers/developers
to assist them in transforming their raw data into curated
data. We have provided the technical details for the curation APIs in a technical report~\cite{CurationAPIs}. As
an ongoing work, we are identifying and implementing more
services to support enriching, annotating, summarizing and
organizing raw data.

\chapter{Conclusion and Future Works}\label{conclusion_7}

 Data curation has become a vital asset for organizations and governments to derive insight and extract value from the raw data. For example, over the last few years, enterprises started to curate and analyze vastly growing open data to personalize the advertisements in elections, improve government services, predict intelligence activities, as well as to improve national security and public health. However, often for curating the data analysts need to perform several time-consuming and challenging curation tasks to transform the raw data into contextualized data and make available for use by end-users and applications. \newline

In this dissertation, we discussed techniques and algorithms to facilitate the transformation and representation of the raw data. We discussed how an analyst could be aided to transform the raw data into knowledge. We presented techniques for adapting data curation rules in dynamic and constantly changing environments, and how to enhance user's comprehension
of curation environments.\newline

In Chapter~\ref{chapter5}, we introduced the notion of Knowledge Lake, i.e., a contextualized Data Lake, to provide the foundation for big data analytics by automatically curating the raw social data and deriving insights. We presented a social data curation foundry to enable analysts to engage with social data to uncover hidden patterns and generate insight. Our approach offers a customizable feature extraction to harness desired features from diverse data sources. To link contextualized information items to the Knowledge Lake, we presented a technique which leverages cross-document co-reference resolution assisting analysts to derive targeted insights. Our experimental results showed that a featurized data curation solution could significantly improve the precision of curation tasks. As future work, we plan to provide a scalable approach to the CDCR process. For example, it is possible to split up a CDCR task into several stages and assign each stage into a specific MapReduce (MR)/spark job. The first MR job can be used to pre-process and frame the information into a schema. Then, the second job can utilise different curation micro-services to extract potential features from data. The final MR job, can utilise the CDCR on extracted information and classify them into a set of related summaries. Our study showed that linking information (e.g., tweets) to the objects in the domain knowledge greatly assists with interpretation of data in a given domain. Adding the scaliability to our algorithm would greatly assists analysts to further extract knowledge and value from the raw data. Also, we are planning to identify a higher number of features from social data to better capture the salient aspect of data. As an ongoing work, we are expanding the presented declarative language to assist analysts in querying and analyzing data more conveniently. We are also extending the Knowledge Lake to cover more concepts.\newline

In Chapter 4, we proposed an approach for adapting data curation rules in dynamic and changing environments. We discussed the importance of rule-based curation systems in augmenting curation algorithms for curating data in unstructured and dynamic environments. Rules can alleviate many of the shortcomings inherent in pure algorithmic approaches. However, rule adaptation is a challenging and error-prone task, and there is a need for an analyst to adapt rules to keep them applicable and precise. Besides, analysts adapt rules at the syntactic level, e.g., keywords and regular expression. Using syntactic level features limits the ability of a rule in annotating items when the rule needs to curate a varied and comprehensive list of data. 

To alleviate the problems mentioned above, we presented an adaptive approach for adapting data curation rules in unstructured and continually changing environments. Our approach offloads analysts from adapting rules and autonomically identifies the optimal modification for rules using a Bayesian multi-armed-bandit algorithm. Besides, our proposed approach adapts rules at the conceptual level, e.g., topic, to boost rules to annotate a larger number of items. We conduct experiments on different curation domains and compare the performance of our approach with systems relying on analysts. The experimental results showed that our approach achieved a comparative performance compared to analysts in adapting rules. As a part of future work, we aim at identifying more features for adapting rules. Specifically, we focused on adapting rules with three other types of features, including entities, word2vec, and relations. We believe adapting rules with higher numbers of features enhance the performance of rules to annotate a larger number of items. 
Additionally, it is possible to automate the generation of rules using Natural Language (NLP) based approaches. Supporting analysts with NLP based solutions may reduce the analyst's workload and effort in generating rules and curating the data. Hence, as a future plan, we may fuse NLP and machine learning algorithms to automatically generate and adapt rules. There are several directions in coupling machine learning and NLP based approaches for generating rules, including: (1) whether NLP based techniques can generate a higher number of rules for curating data. (2) Scalability is another issue that needs to be studied while coupling NLP based approaches for generating rules. Additionally, (3) evaluating the analyst workload in generating and adapting rules using NLP based solution is an interesting topic that can be studied as well.
\newline

In Chapter~\ref{Chapter4}, we discussed techniques for enhancing users comprehension to formulate their preferences in a large curation environment. Understanding of data allows a user to better formulate her preferences when seeking information. However, exploring the curation environment has been proven to be painstakingly time-consuming
and challenging when a user has a varied information need with a large number of topical sub-spaces. In such environments, the user continuously issues different queries to scan and read the data and make sense of it. However, using current techniques, a user needs to explicitly specify her preferences for Information Retrieval (IR) systems in forms of keywords or phrases. Text queries limit the ability of users in comprehending the curation environment as they only retrieve a small part of data, and the rest remains invisible.

To address the above problem, we proposed a system that provides a conceptual summary of curation environments and allows users to specify their preferences implicitly as a set of concepts. Our approach lowered users' cognitive load in ranking and exploring the data. The system takes advantage of deep learning and a knowledge lake to provide a conceptual summary of the information space. User can specify her preferences implicitly as a set of concepts without the need to investigate the information space iteratively. It provides a 2D Radial Map of concepts where users can rank items relevant to their preferences through dragging and dropping. Our experiment results showed that our approach could help users to formulate their preferences better when they need to retrieve a varied and comprehensive list of information across a large curation environment. As future works, we focus on enhancing the quality of summaries by contributing the other users' feedback. This, will improve the precision of the system in retrieving user's information need and reducing the time users need to spend for formulating her preferences.\newline

\bibliographystyle{amsplain}
\bibliography{ms}

\providecommand{\bysame}{\leavevmode\hbox to3em{\hrulefill}\thinspace}
\providecommand{\MR}{\relax\ifhmode\unskip\space\fi MR }
\providecommand{\MRhref}[2]{%
  \href{http://www.ams.org/mathscinet-getitem?mr=#1}{#2}
}
\providecommand{\href}[2]{#2}
\begin{thebibliography}{100}

\bibitem{astrixdb}
\emph{Astrixdb}, https://asterixdb.apache.org/.

\bibitem{datacleaning}
\emph{https://elitedatascience.com/data-cleaning}.

\bibitem{dataingestion}
\emph{https://whatis.techtarget.com/definition/data-ingestion}.

\bibitem{orchstrate}
\emph{orchstrate}, orchstrate.io/.

\bibitem{BigSocialData}
Bilal Abu{-}Salih, Pornpit Wongthongtham, Seyed{-}Mehdi{-}Reza Beheshti, and
  Behrang Zadjabbari, \emph{Towards a methodology for social business
  intelligence in the era of big social data incorporating trust and semantic
  analysis}, Proceedings of the Second International Conference on Advanced
  Data and Information Engineering, DaEng 2015, Bali, Indonesia, April 25-26,
  2015, Lecture Notes in Electrical Engineering, vol. 520, Springer, 2015,
  pp.~519--527.

\bibitem{Trustworthiness}
Bilal Abu{-}Salih, Pornpit Wongthongtham, Seyed{-}Mehdi{-}Reza Beheshti, and
  Dengya Zhu, \emph{A preliminary approach to domain-based evaluation of users'
  trustworthiness in online social networks}, 2015 {IEEE} International
  Congress on Big Data, New York City, NY, USA, June 27 - July 2, 2015 (Barbara
  Carminati and Latifur Khan, eds.), {IEEE} Computer Society, 2015,
  pp.~460--466.

\bibitem{aizawa2003information}
Akiko Aizawa, \emph{An information-theoretic perspective of tf--idf measures},
  Information Processing \& Management \textbf{39} (2003), no.~1, 45--65.

\bibitem{akers2014building}
Katherine~G Akers, Fe~C Sferdean, Natsuko~H Nicholls, and Jennifer~A Green,
  \emph{Building support for research data management: Biographies of eight
  research universities.}, IJDC \textbf{9} (2014), no.~2, 171--191.

\bibitem{alex2008automating}
Beatrice Alex, Claire Grover, Barry Haddow, Mijail Kabadjov, Ewan Klein,
  Michael Matthews, Richard Tobin, and Xinglong Wang, \emph{Automating curation
  using a natural language processing pipeline}, Genome Biology \textbf{9}
  (2008), no.~2, S10.

\bibitem{alexander2014serendip}
Eric Alexander, Joe Kohlmann, Robin Valenza, Michael Witmore, and Michael
  Gleicher, \emph{Serendip: Topic model-driven visual exploration of text
  corpora}, 2014 IEEE Conference on Visual Analytics Science and Technology
  (VAST), IEEE, 2014, pp.~173--182.

\bibitem{alizadeh2019capturing}
Tooran Alizadeh, Somwrita Sarkar, and Sandy Burgoyne, \emph{Capturing citizen
  voice online: Enabling smart participatory local government}, Cities
  \textbf{95} (2019), 102400.

\bibitem{ratingMuh}
Mohammad Allahbakhsh, Aleksandar Ignjatovic, Boualem Benatallah,
  Seyed{-}Mehdi{-}Reza Beheshti, Elisa Bertino, and Norman Foo, \emph{Collusion
  detection in online rating systems}, Web Technologies and Applications - 15th
  Asia-Pacific Web Conference, APWeb 2013, Sydney, Australia, April 4-6, 2013.
  Proceedings, Lecture Notes in Computer Science, vol. 7808, Springer, 2013,
  pp.~196--207.

\bibitem{allahbakhsh2012analytic}
Mohammad Allahbakhsh, Aleksandar Ignjatovic, Boualem Benatallah,
  Seyed-Mehdi-Reza Beheshti, Norman Foo, and Elisa Bertino, \emph{An analytic
  approach to people evaluation in crowdsourcing systems}, arXiv preprint
  arXiv:1211.3200 (2012).

\bibitem{allahbakhsh2012detecting}
\bysame, \emph{Detecting, representing and querying collusion in online rating
  systems}, arXiv preprint arXiv:1211.0963 (2012).

\bibitem{allahbakhsh2014representation}
Mohammad Allahbakhsh, Aleksandar Ignjatovic, Boualem Benatallah,
  Seyed{-}Mehdi{-}Reza Beheshti, Norman Foo, and Elisa Bertino,
  \emph{Representation and querying of unfair evaluations in social rating
  systems}, Comput. Secur. \textbf{41} (2014), 68--88.

\bibitem{allahbakhsh2012reputation}
Mohammad Allahbakhsh, Aleksandar Ignjatovic, Boualem Benatallah, Elisa Bertino,
  Norman Foo, et~al., \emph{Reputation management in crowdsourcing systems},
  8th International Conference on Collaborative Computing: Networking,
  Applications and Worksharing (CollaborateCom), IEEE, 2012, pp.~664--671.

\bibitem{allan2004hard}
James Allan, \emph{Hard track overview in trec 2004 (notebook), high accuracy
  retrieval from documents}, The Thirteenth Text Retrieval Conference (TREC
  2004) Notebook, 2004, pp.~226--235.

\bibitem{iSheets}
Farhad Amouzgar, Amin Beheshti, Samira Ghodratnama, Boualem Benatallah, Jian
  Yang, and Quan~Z. Sheng, \emph{isheets: {A} spreadsheet-based machine
  learning development platform for data-driven process analytics},
  Service-Oriented Computing - {ICSOC} 2018 Workshops - ADMS, ASOCA, ISYyCC,
  CloTS, DDBS, and NLS4IoT, Hangzhou, China, November 12-15, 2018, Revised
  Selected Papers, Lecture Notes in Computer Science, vol. 11434, Springer,
  2018, pp.~453--457.

\bibitem{brainwash}
Michael Anderson and et~al., \emph{Brainwash: A data system for feature
  engineering.}, CIDR, 2013.

\bibitem{anderson2014integrated}
Michael~R Anderson, Michael Cafarella, Yixing Jiang, Guan Wang, and Bochun
  Zhang, \emph{An integrated development environment for faster feature
  engineering}, Proceedings of the VLDB Endowment \textbf{7} (2014), no.~13,
  1657--1660.

\bibitem{anick2008longitudinal}
Peter Anick and Raj~Gopal Kantamneni, \emph{A longitudinal study of real-time
  search assistance adoption}, Proceedings of the 31st annual international ACM
  SIGIR conference on Research and development in information retrieval, 2008,
  pp.~701--702.

\bibitem{araque2006application}
Francisco Araque, Alberto Salguero, and Maria~M Abad, \emph{Application of data
  warehouse and decision support system in soaring site recommendation},
  Information and Communication Technologies in Tourism 2006, Springer, 2006,
  pp.~308--319.

\bibitem{soundex2007}
National Archives and Records Administration, \emph{The soundex indexing
  system}, he Soundex Indexing System (2007-05-30).

\bibitem{asopen}
CESSDA AS, \emph{The open archival information system (oais) reference model},
  (2004).

\bibitem{auer2007dbpedia}
S{\"o}ren Auer, Christian Bizer, Georgi Kobilarov, Jens Lehmann, Richard
  Cyganiak, and Zachary Ives, \emph{Dbpedia: A nucleus for a web of open data},
  The semantic web, Springer, 2007, pp.~722--735.

\bibitem{baeza2004query}
Ricardo Baeza-Yates, Carlos Hurtado, and Marcelo Mendoza, \emph{Query
  recommendation using query logs in search engines}, International conference
  on extending database technology, Springer, 2004, pp.~588--596.

\bibitem{bak2014rule}
Peter Bak, Dotan Dolev, and Tali Yatzkar-Haham, \emph{Rule adjustment by
  visualization of physical location data}, September~11 2014, US Patent App.
  14/483,158.

\bibitem{yeastmind2012}
Rama Balakrishnan, Julie Park, Kalpana Karra, Benjamin~C. Hitz, Gail Binkley,
  Eurie~L. Hong, Julie Sullivan, Gos Micklem, and J.~Michael~Cherry,
  \emph{{YeastMine—an integrated data warehouse for Saccharomyces cerevisiae
  data as a multipurpose tool-kit}}, Database \textbf{2012} (2012), bar062.

\bibitem{graphperformance}
Ahmed Barnawi, Omar Batarfi, Seyed{-}Mehdi{-}Reza Beheshti, Radwa~El Shawi,
  Ayman~G. Fayoumi, Reza Nouri, and Sherif Sakr, \emph{On characterizing the
  performance of distributed graph computation platforms}, Performance
  Characterization and Benchmarking. Traditional to Big Data - 6th {TPC}
  Technology Conference, {TPCTC} 2014, Hangzhou, China, September 1-5, 2014.
  Revised Selected Papers, Lecture Notes in Computer Science, vol. 8904,
  Springer, 2014, pp.~29--43.

\bibitem{graphSurvey}
Omar Batarfi, Radwa~El Shawi, Ayman~G. Fayoumi, Reza Nouri,
  Seyed{-}Mehdi{-}Reza Beheshti, Ahmed Barnawi, and Sherif Sakr, \emph{Large
  scale graph processing systems: survey and an experimental evaluation},
  Cluster Computing \textbf{18} (2015), no.~3, 1189--1213.

\bibitem{bates1989design}
Marcia~J Bates et~al., \emph{The design of browsing and berrypicking techniques
  for the online search interface}, Online review \textbf{13} (1989), no.~5,
  407--424.

\bibitem{beheshti2018processatlas}
Amin Beheshti, Boualem Benatallah, and Hamid~Reza Motahari-Nezhad,
  \emph{Processatlas a scalable and extensible platform for business process
  analytics}, Software Practice and Experience (2018), 842--866.

\bibitem{ProcessAtlas}
Amin Beheshti, Boualem Benatallah, and Hamid~Reza Motahari{-}Nezhad,
  \emph{Processatlas: {A} scalable and extensible platform for business process
  analytics}, Softw., Pract. Exper. \textbf{48} (2018), no.~4, 842--866.

\bibitem{beheshti2017coredb}
Amin Beheshti, Boualem Benatallah, Reza Nouri, Van~Munin Chhieng, HuangTao
  Xiong, and Xu~Zhao, \emph{Coredb: a data lake service}, Proceedings of the
  2017 ACM on Conference on Information and Knowledge Management, ACM, 2017,
  pp.~2451--2454.

\bibitem{beheshti2018corekg}
Amin Beheshti, Boualem Benatallah, Reza Nouri, and Alireza Tabebordbar,
  \emph{Corekg: a knowledge lake service}, Proceedings of the VLDB Endowment
  \textbf{11} (2018), no.~12, 1942--1945.

\bibitem{IntelligentKG}
Amin Beheshti, Boualem Benatallah, Quan~Z. Sheng, and Francesco Schiliro,
  \emph{Intelligent knowledge lakes: The age of artificial intelligence and big
  data}, Web Information Systems Engineering - {WISE} 2019 Workshop, Demo, and
  Tutorial, Hong Kong and Macau, China, January 19-22, 2020, Revised Selected
  Papers, Communications in Computer and Information Science, vol. 1155,
  Springer, 2019, pp.~24--34.

\bibitem{beheshti2018datasynapse}
Amin Beheshti, Boualem Benatallah, Alireza Tabebordbar, Hamid~Reza
  Motahari-Nezhad, Moshe~Chai Barukh, and Reza Nouri, \emph{Datasynapse: A
  social data curation foundry}, Distributed and Parallel Databases (2018),
  1--34.

\bibitem{BehavioralAnalytics}
Amin Beheshti, Vahid~Moraveji Hashemi, and Shahpar Yakhchi, \emph{Towards
  context-aware social behavioral analytics}, Proceedings of the 17th
  International Conference on Advances in Mobile Computing \& Multimedia, 2019,
  pp.~28--35.

\bibitem{personality2vec}
Amin Beheshti, Vahid~Moraveji Hashemi, Shahpar Yakhchi, Hamid~Reza
  Motahari{-}Nezhad, Seyed~Mohssen Ghafari, and Jian Yang,
  \emph{personality2vec: Enabling the analysis of behavioral disorders in
  social networks}, {WSDM} '20: The Thirteenth {ACM} International Conference
  on Web Search and Data Mining, Houston, TX, USA, February 3-7, 2020, {ACM},
  2020, pp.~825--828.

\bibitem{iProcess}
Amin Beheshti, Francesco Schiliro, Samira Ghodratnama, Farhad Amouzgar, Boualem
  Benatallah, Jian Yang, Quan~Z. Sheng, Fabio Casati, and Hamid~Reza
  Motahari{-}Nezhad, \emph{iprocess: Enabling iot platforms in data-driven
  knowledge-intensive processes}, Business Process Management Forum - {BPM}
  Forum 2018, Sydney, NSW, Australia, September 9-14, 2018, Proceedings,
  Lecture Notes in Business Information Processing, vol. 329, Springer, 2018,
  pp.~108--126.

\bibitem{iStory}
Amin Beheshti, Alireza Tabebordbar, and Boualem Benatallah, \emph{istory:
  Intelligent storytelling with social data}, The World Wide Web Conference,
  {WWW} 2020, {ACM}, 2020.

\bibitem{beheshti2018crowdcorrect}
Amin Beheshti, Kushal Vaghani, Boualem Benatallah, and Alireza Tabebordbar,
  \emph{Crowdcorrect: {A} curation pipeline for social data cleansing and
  curation}, Information Systems in the Big Data Era - CAiSE Forum 2018,
  Tallinn, Estonia, June 11-15, 2018, Proceedings, Lecture Notes in Business
  Information Processing, vol. 317, Springer, 2018, pp.~24--38.

\bibitem{caise13}
Seyed{-}Mehdi{-}Reza Beheshti, Boualem Benatallah, and Hamid~R. {Motahari
  Nezhad}, \emph{Enabling the analysis of cross-cutting aspects in ad-hoc
  processes}, Advanced Information Systems Engineering - 25th International
  Conference, CAiSE 2013, Valencia, Spain, June 17-21, 2013. Proceedings,
  Lecture Notes in Computer Science, vol. 7908, Springer, 2013, pp.~51--67.

\bibitem{WISE12}
Seyed{-}Mehdi{-}Reza Beheshti, Boualem Benatallah, Hamid~R. {Motahari Nezhad},
  and Mohammad Allahbakhsh, \emph{A framework and a language for on-line
  analytical processing on graphs}, Web Information Systems Engineering -
  {WISE} 2012 - 13th International Conference, Paphos, Cyprus, November 28-30,
  2012. Proceedings, Lecture Notes in Computer Science, vol. 7651, Springer,
  2012, pp.~213--227.

\bibitem{BPM11}
Seyed{-}Mehdi{-}Reza Beheshti, Boualem Benatallah, Hamid~R. {Motahari Nezhad},
  and Sherif Sakr, \emph{A query language for analyzing business processes
  execution}, Business Process Management - 9th International Conference, {BPM}
  2011, Clermont-Ferrand, France, August 30 - September 2, 2011. Proceedings,
  Lecture Notes in Computer Science, vol. 6896, Springer, 2011, pp.~281--297.

\bibitem{beheshti2016galaxy}
Seyed{-}Mehdi{-}Reza Beheshti, Boualem Benatallah, and Hamid~Reza
  Motahari{-}Nezhad, \emph{Galaxy: {A} platform for explorative analysis of
  open data sources}, Proceedings of the 19th International Conference on
  Extending Database Technology, {EDBT} 2016, Bordeaux, France, March 15-16,
  2016, Bordeaux, France, March 15-16, 2016, OpenProceedings.org, 2016,
  pp.~640--643.

\bibitem{POLAP}
\bysame, \emph{Scalable graph-based {OLAP} analytics over process execution
  data}, Distributed and Parallel Databases \textbf{34} (2016), no.~3,
  379--423.

\bibitem{ProcessAnalytics}
Seyed{-}Mehdi{-}Reza Beheshti, Boualem Benatallah, Sherif Sakr, Daniela
  Grigori, Hamid~Reza Motahari{-}Nezhad, Moshe~Chai Barukh, Ahmed Gater, and
  Seung~Hwan Ryu, \emph{Process analytics - concepts and techniques for
  querying and analyzing process data}, Springer, 2016.

\bibitem{beheshti2017systematic}
Seyed{-}Mehdi{-}Reza Beheshti, Boualem Benatallah, Srikumar Venugopal,
  Seung~Hwan Ryu, Hamid~Reza Motahari{-}Nezhad, and Wei Wang, \emph{A
  systematic review and comparative analysis of cross-document coreference
  resolution methods and tools}, Computing \textbf{99} (2017), no.~4, 313--349.

\bibitem{Beheshti2016}
\bysame, \emph{A systematic review and comparative analysis of cross-document
  coreference resolution methods and tools}, Computing \textbf{99} (2017),
  no.~4, 313--349.

\bibitem{Provenance}
Seyed{-}Mehdi{-}Reza Beheshti, Hamid~R. {Motahari Nezhad}, and Boualem
  Benatallah, \emph{Temporal provenance model {(TPM):} model and query
  language}, CoRR \textbf{abs/1211.5009} (2012).

\bibitem{Extendingsparql}
Seyed{-}Mehdi{-}Reza Beheshti, Sherif Sakr, Boualem Benatallah, and Hamid~R.
  {Motahari Nezhad}, \emph{Extending {SPARQL} to support entity grouping and
  path queries}, CoRR \textbf{abs/1211.5817} (2012).

\bibitem{beheshti2017automating}
Seyed{-}Mehdi{-}Reza Beheshti, Alireza Tabebordbar, Boualem Benatallah, and
  Reza Nouri, \emph{On automating basic data curation tasks}, Proceedings of
  the 26th International Conference on World Wide Web Companion, Perth,
  Australia, April 3-7, 2017, {ACM}, 2017, pp.~165--169.

\bibitem{BigCDCR}
Seyed{-}Mehdi{-}Reza Beheshti, Srikumar Venugopal, Seung~Hwan Ryu, Boualem
  Benatallah, and Wei Wang, \emph{Big data and cross-document coreference
  resolution: Current state and future opportunities}, CoRR
  \textbf{abs/1311.3987} (2013).

\bibitem{CurationAPIs}
Seyed-Mehdi-Rezaand et~al. Beheshti, \emph{Data curation {APIs}}, Tech. Report
  UNSW-CSE-TR-201617, The University of New South Wales, Sydney, Australia,
  2016.

\bibitem{bergsma2007wikimedia}
Mark Bergsma, \emph{Wikimedia architecture}, Wikimedia Foundation Inc (2007).

\bibitem{biem2010ibm}
Alain Biem, Eric Bouillet, Hanhua Feng, Anand Ranganathan, Anton Riabov,
  Olivier Verscheure, Haris Koutsopoulos, and Carlos Moran, \emph{Ibm
  infosphere streams for scalable, real-time, intelligent transportation
  services}, Proceedings of the 2010 ACM SIGMOD International Conference on
  Management of data, ACM, 2010, pp.~1093--1104.

\bibitem{bird2009natural}
Steven Bird, Ewan Klein, and Edward Loper, \emph{Natural language processing
  with python: analyzing text with the natural language toolkit}, " O'Reilly
  Media, Inc.", 2009.

\bibitem{birjali2017analyzing}
Marouane Birjali, Abderrahim Beni-Hssane, and Mohammed Erritali,
  \emph{Analyzing social media through big data using infosphere biginsights
  and apache flume}, Procedia computer science \textbf{113} (2017), 280--285.

\bibitem{blanco20118th}
Roi Blanco, B~Barla Cambazoglu, and Claudio Lucchese, \emph{The 8th workshop on
  large-scale distributed systems for information retrieval (lsds-ir'10)}, ACM
  SIGIR Forum, vol.~44, ACM New York, NY, USA, 2011, pp.~54--58.

\bibitem{blei2003latent}
David~M Blei, Andrew~Y Ng, and Michael~I Jordan, \emph{Latent dirichlet
  allocation}, Journal of machine Learning research \textbf{3} (2003), no.~Jan,
  993--1022.

\bibitem{boudoukh2013news}
Jacob Boudoukh, Ronen Feldman, Shimon Kogan, and Matthew Richardson,
  \emph{Which news moves stock prices? a textual analysis}, Tech. report,
  National Bureau of Economic Research, 2013.

\bibitem{brodie2010power}
Michael~L Brodie and Jason~T Liu, \emph{The power and limits of relational
  technology in the age of information ecosystems}, Keynote at On The Move
  Federated Conferences, 2010.

\bibitem{brooke1996sus}
John Brooke et~al., \emph{Sus-a quick and dirty usability scale}, Usability
  evaluation in industry \textbf{189} (1996), no.~194, 4--7.

\bibitem{brooks2015featureinsight}
Michael Brooks, Saleema Amershi, Bongshin Lee, Steven~M Drucker, Ashish Kapoor,
  and Patrice Simard, \emph{Featureinsight: Visual support for error-driven
  feature ideation in text classification}, Visual Analytics Science and
  Technology (VAST), 2015 IEEE Conference on, IEEE, 2015, pp.~105--112.

\bibitem{brown2014finding}
Eli~T Brown, Alvitta Ottley, Helen Zhao, Quan Lin, Richard Souvenir, Alex
  Endert, and Remco Chang, \emph{Finding waldo: Learning about users from their
  interactions}, IEEE Transactions on visualization and computer graphics
  \textbf{20} (2014), no.~12, 1663--1672.

\bibitem{brusilovsky2018social}
Peter Brusilovsky, Barry Smyth, and Bracha Shapira, \emph{Social search},
  Social Information Access, Springer, 2018, pp.~213--276.

\bibitem{burigat2013effectiveness}
Stefano Burigat and Luca Chittaro, \emph{On the effectiveness of overview+
  detail visualization on mobile devices}, Personal and ubiquitous computing
  \textbf{17} (2013), no.~2, 371--385.

\bibitem{burke2013introduction}
Mary~C Burke, \emph{Introduction to human trafficking: definitions and
  prevalence}, Human trafficking: interdisciplinary perspectives. New York:
  Routledge (2013), 3--23.

\bibitem{burtini2015improving}
Giuseppe Burtini, Jason Loeppky, and Ramon Lawrence, \emph{Improving online
  marketing experiments with drifting multi-armed bandits.}, ICEIS (1), 2015,
  pp.~630--636.

\bibitem{buys2015data}
Cunera~M Buys and Pamela~L Shaw, \emph{Data management practices across an
  institution: Survey and report.}, Journal of Librarianship \& Scholarly
  Communication \textbf{3} (2015), no.~2.

\bibitem{card1999readings}
Mackinlay Card, \emph{Readings in information visualization: using vision to
  think}, Morgan Kaufmann, 1999.

\bibitem{carterette2008here}
Ben Carterette, Paul~N Bennett, David~Maxwell Chickering, and Susan~T Dumais,
  \emph{Here or there preference judgement for relevance}, European Conference
  on Information Retrieval, Springer, 2008, pp.~16--27.

\bibitem{cavanillas2016new}
Jos{\'e}~M Cavanillas, Edward Curry, and Wolfgang Wahlster, \emph{New horizons
  for a data-driven economy: a roadmap for usage and exploitation of big data
  in europe}, Springer, 2016.

\bibitem{cayrol1982fuzzy}
M~Cayrol, H~Farreny, and H~Prade, \emph{Fuzzy pattern matching}, Kybernetes
  \textbf{11} (1982), no.~2, 103--116.

\bibitem{chen2012business}
Hsinchun Chen, Roger~HL Chiang, and Veda~C Storey, \emph{Business intelligence
  and analytics: From big data to big impact.}, MIS quarterly \textbf{36}
  (2012), no.~4.

\bibitem{chen2016deep}
Yushi Chen, Hanlu Jiang, Chunyang Li, Xiuping Jia, and Pedram Ghamisi,
  \emph{Deep feature extraction and classification of hyperspectral images
  based on convolutional neural networks}, IEEE Transactions on Geoscience and
  Remote Sensing \textbf{54} (2016), no.~10, 6232--6251.

\bibitem{cheng2015flock}
Justin Cheng and Michael~S Bernstein, \emph{Flock: Hybrid crowd-machine
  learning classifiers}, Proceedings of the 18th ACM conference on computer
  supported cooperative work \& social computing, ACM, 2015, pp.~600--611.

\bibitem{chiticariu2010systemt}
Laura Chiticariu, Rajasekar Krishnamurthy, Yunyao Li, Sriram Raghavan,
  Frederick~R Reiss, and Shivakumar Vaithyanathan, \emph{Systemt: an algebraic
  approach to declarative information extraction}, Proceedings of the 48th
  Annual Meeting of the Association for Computational Linguistics, Association
  for Computational Linguistics, 2010, pp.~128--137.

\bibitem{chiticariu2013rule}
Laura Chiticariu, Yunyao Li, and Frederick~R Reiss, \emph{Rule-based
  information extraction is dead! long live rule-based information extraction
  systems!}, EMNLP, no. October, 2013, pp.~827--832.

\bibitem{chute2010enterprise}
Christopher~G Chute, Scott~A Beck, Thomas~B Fisk, and David~N Mohr, \emph{The
  enterprise data trust at mayo clinic: a semantically integrated warehouse of
  biomedical data}, Journal of the American Medical Informatics Association
  \textbf{17} (2010), no.~2, 131--135.

\bibitem{clement2014online}
Benjamin Clement, Pierre-Yves Oudeyer, Didier Roy, and Manuel Lopes,
  \emph{Online optimization of teaching sequences with multi-armed bandits},
  Educational Data Mining 2014, 2014.

\bibitem{cont2011statistical}
Rama Cont, \emph{Statistical modeling of high-frequency financial data}, IEEE
  Signal Processing Magazine \textbf{28} (2011), no.~5, 16--25.

\bibitem{cox2015research}
Andrew Cox, Rosie Higman, and Stephen Pinfield, \emph{Research data management
  and openness}, Program: electronic library and information systems (2015).

\bibitem{coyle2004searchguide}
Maurice Coyle and Barry Smyth, \emph{Searchguide: Beyond the results page},
  International Conference on Adaptive Hypermedia and Adaptive Web-Based
  Systems, Springer, 2004, pp.~296--299.

\bibitem{criado2013government}
J~Ignacio Criado, Rodrigo Sandoval-Almazan, and J~Ramon Gil-Garcia,
  \emph{Government innovation through social media}, 2013.

\bibitem{croset2016flexible}
Samuel Croset, Joachim Rupp, and Martin Romacker, \emph{Flexible data
  integration and curation using a graph-based approach}, Bioinformatics
  \textbf{32} (2016), no.~6, 918--925.

\bibitem{cucerzan2004spelling}
Silviu Cucerzan and Eric Brill, \emph{Spelling correction as an iterative
  process that exploits the collective knowledge of web users}, Proceedings of
  the 2004 Conference on Empirical Methods in Natural Language Processing,
  2004, pp.~293--300.

\bibitem{culliss2004personalized}
Gary~A Culliss, \emph{Personalized search methods including combining index
  entries for catagories of personal data}, November~9 2004, US Patent
  6,816,850.

\bibitem{curry2010role}
Edward Curry, Andre Freitas, and Sean O’Ri{\'a}in, \emph{The role of
  community-driven data curation for enterprises}, Linking enterprise data,
  Springer, 2010, pp.~25--47.

\bibitem{cutting2017scatter}
Douglass~R Cutting, David~R Karger, Jan~O Pedersen, and John~W Tukey,
  \emph{Scatter/gather: A cluster-based approach to browsing large document
  collections}, ACM SIGIR Forum, vol.~51, ACM New York, NY, USA, 2017,
  pp.~148--159.

\bibitem{danielsson1980euclidean}
Per-Erik Danielsson, \emph{Euclidean distance mapping}, Computer Graphics and
  image processing \textbf{14} (1980), no.~3, 227--248.

\bibitem{de2016deepdive}
Christopher De~Sa, Alex Ratner, Christopher R{\'e}, Jaeho Shin, Feiran Wang,
  Sen Wu, and Ce~Zhang, \emph{Deepdive: Declarative knowledge base
  construction}, ACM SIGMOD Record \textbf{45} (2016), no.~1, 60--67.

\bibitem{MapReduce}
Jeffrey Dean and Sanjay Ghemawat, \emph{Mapreduce: simplified data processing
  on large clusters}, Commun. ACM \textbf{51} (2008), no.~1, 107--113.

\bibitem{dennis2002web}
Simon Dennis, Peter Bruza, and Robert McArthur, \emph{Web searching: A
  process-oriented experimental study of three interactive search paradigms},
  Journal of the American Society for Information Science and Technology
  \textbf{53} (2002), no.~2, 120--133.

\bibitem{di2018study}
Cecilia di~Sciascio, Peter Brusilovsky, and Eduardo Veas, \emph{A study on
  user-controllable social exploratory search}, 23rd International Conference
  on Intelligent User Interfaces, ACM, 2018, pp.~353--364.

\bibitem{di2016rank}
Cecilia di~Sciascio, Vedran Sabol, and Eduardo~E Veas, \emph{Rank as you go:
  User-driven exploration of search results}, Proceedings of the 21st
  International Conference on Intelligent User Interfaces, ACM, 2016,
  pp.~118--129.

\bibitem{dinet2004searching}
Jerome Dinet, Monik Favart, and Jean-Michel Passerault, \emph{Searching for
  information in an online public access catalogue (opac): the impacts of
  information search expertise on the use of boolean operators}, Journal of
  Computer Assisted Learning \textbf{20} (2004), no.~5, 338--346.

\bibitem{domingos2012few}
Pedro~M Domingos, \emph{A few useful things to know about machine learning.},
  Commun. acm \textbf{55} (2012), no.~10, 78--87.

\bibitem{dou2012kurator}
L~Dou, G~Cao, Paul~J Morris, Robert~A Morris, Bertram Lud{\"a}scher, James~A
  Macklin, and James Hanken, \emph{Kurator: A kepler package for data curation
  workflows}, Procedia Computer Science \textbf{9} (2012), 1614--1619.

\bibitem{dumais2001optimizing}
Susan Dumais, Edward Cutrell, and Hao Chen, \emph{Optimizing search by showing
  results in context}, Proceedings of the SIGCHI conference on Human factors in
  computing systems, 2001, pp.~277--284.

\bibitem{eades2017shape}
Peter Eades, Seok-Hee Hong, An~Nguyen, and Karsten Klein, \emph{Shape-based
  quality metrics for large graph visualization.}, J. Graph Algorithms Appl.
  \textbf{21} (2017), no.~1, 29--53.

\bibitem{ellis2007taxonomy}
Geoffrey Ellis and Alan Dix, \emph{A taxonomy of clutter reduction for
  information visualisation}, IEEE transactions on visualization and computer
  graphics \textbf{13} (2007), no.~6, 1216--1223.

\bibitem{esuli2006sentiwordnet}
Andrea Esuli and Fabrizio Sebastiani, \emph{Sentiwordnet: A publicly available
  lexical resource for opinion mining}, Proceedings of LREC, vol.~6, Citeseer,
  2006, pp.~417--422.

\bibitem{ab2017angling}
Aberdeen et~al, \emph{Angling for insight in today’s data lake},  (2017).

\bibitem{tabebordbar2019adaptive}
XX~et~al, \emph{Anonymised}, Anonymised.

\bibitem{etzioni2004web}
Oren Etzioni, Michael Cafarella, Doug Downey, Stanley Kok, Ana-Maria Popescu,
  Tal Shaked, Stephen Soderland, Daniel~S Weld, and Alexander Yates,
  \emph{Web-scale information extraction in knowitall: (preliminary results)},
  Proceedings of the 13th international conference on World Wide Web, 2004,
  pp.~100--110.

\bibitem{fader2014open}
Anthony Fader, Luke Zettlemoyer, and Oren Etzioni, \emph{Open question
  answering over curated and extracted knowledge bases}, Proceedings of the
  20th ACM SIGKDD international conference on Knowledge discovery and data
  mining, ACM, 2014, pp.~1156--1165.

\bibitem{fast2016empath}
Ethan Fast, Binbin Chen, and Michael~S Bernstein, \emph{Empath: Understanding
  topic signals in large-scale text}, Proceedings of the 2016 CHI Conference on
  Human Factors in Computing Systems, ACM, 2016, pp.~4647--4657.

\bibitem{fatemi2009using}
Nastaran Fatemi, Florian Poulin, Laura~E Raileany, and Alan~F Smeaton,
  \emph{Using association rule mining to enrich semantic concepts for video
  retrieval},  (2009).

\bibitem{ferrucci2012introduction}
David~A Ferrucci, \emph{Introduction to 'this is watson'}, IBM Journal of
  Research and Development \textbf{56} (2012), no.~3.4.

\bibitem{fonseca2003discovering}
Bruno~M Fonseca, Paulo~B Golgher, Edleno~S De~Moura, Bruno P{\^o}ssas, and
  Nivio Ziviani, \emph{Discovering search engine related queries using
  association rules}, Journal of Web Engineering \textbf{2} (2003), no.~4,
  215--227.

\bibitem{forehand2010bloom}
Mary Forehand, \emph{Bloom’s taxonomy}, Emerging perspectives on learning,
  teaching, and technology \textbf{41} (2010), no.~4, 47--56.

\bibitem{freyne2004experiment}
Jill Freyne and Barry Smyth, \emph{An experiment in social search},
  International Conference on Adaptive Hypermedia and Adaptive Web-Based
  Systems, Springer, 2004, pp.~95--103.

\bibitem{friedman1974projection}
Jerome~H Friedman and John~W Tukey, \emph{A projection pursuit algorithm for
  exploratory data analysis}, IEEE Transactions on computers \textbf{100}
  (1974), no.~9, 881--890.

\bibitem{garfieldnegative}
Eugene Garfield, \emph{When is a negative search result positive? essays of an
  information scientist vol. 1, 12 august 1970}.

\bibitem{gattani2013entity}
Abhishek Gattani, Digvijay~S Lamba, Nikesh Garera, Mitul Tiwari, Xiaoyong Chai,
  Sanjib Das, Sri Subramaniam, Anand Rajaraman, Venky Harinarayan, and AnHai
  Doan, \emph{Entity extraction, linking, classification, and tagging for
  social media: a wikipedia-based approach}, Proceedings of the VLDB Endowment
  \textbf{6} (2013), no.~11, 1126--1137.

\bibitem{gc2015big}
Paul~Suganthan GC, Chong Sun, Haojun Zhang, Frank Yang, Narasimhan Rampalli,
  Shishir Prasad, Esteban Arcaute, Ganesh Krishnan, Rohit Deep, Vijay
  Raghavendra, et~al., \emph{Why big data industrial systems need rules and
  what we can do about it}, Proceedings of the 2015 ACM SIGMOD International
  Conference on Management of Data, ACM, 2015, pp.~265--276.

\bibitem{girden1992anova}
Ellen~R Girden, \emph{Anova: Repeated measures}, no.~84, Sage, 1992.

\bibitem{gomez2014similarity}
Erick Gomez-Nieto, Frizzi San~Roman, Paulo Pagliosa, Wallace Casaca, Elias~S
  Helou, Maria Cristina~F de~Oliveira, and Luis~Gustavo Nonato,
  \emph{Similarity preserving snippet-based visualization of web search
  results}, IEEE transactions on visualization and computer graphics
  \textbf{20} (2014), no.~3, 457--470.

\bibitem{gratzl2013lineup}
Samuel Gratzl, Alexander Lex, Nils Gehlenborg, Hanspeter Pfister, and Marc
  Streit, \emph{Lineup: Visual analysis of multi-attribute rankings}, IEEE
  transactions on visualization and computer graphics \textbf{19} (2013),
  no.~12, 2277--2286.

\bibitem{gretarsson2012topicnets}
Brynjar Gretarsson, John O’donovan, Svetlin Bostandjiev, Tobias H{\"o}llerer,
  Arthur Asuncion, David Newman, and Padhraic Smyth, \emph{Topicnets: Visual
  analysis of large text corpora with topic modeling}, ACM Transactions on
  Intelligent Systems and Technology (TIST) \textbf{3} (2012), no.~2, 23.

\bibitem{DREAM}
Mohammad Hammoud, Dania~Abed Rabbou, Reza Nouri, Seyed{-}Mehdi{-}Reza Beheshti,
  and Sherif Sakr, \emph{{DREAM:} distributed {RDF} engine with adaptive query
  planner and minimal communication}, {PVLDB} \textbf{8} (2015), no.~6,
  654--665.

\bibitem{hargittai2004classifying}
Eszter Hargittai, \emph{Classifying and coding online actions}, Social Science
  Computer Review \textbf{22} (2004), no.~2, 210--227.

\bibitem{harrison2014ranking}
Lane Harrison, Fumeng Yang, Steven Franconeri, and Remco Chang, \emph{Ranking
  visualizations of correlation using weber's law}, IEEE transactions on
  visualization and computer graphics \textbf{20} (2014), no.~12, 1943--1952.

\bibitem{he2016interactive}
Jian He, Enzo Veltri, Donatello Santoro, Guoliang Li, Giansalvatore Mecca,
  Paolo Papotti, and Nan Tang, \emph{Interactive and deterministic data
  cleaning}, Proceedings of the 2016 International Conference on Management of
  Data, 2016, pp.~893--907.

\bibitem{hearst2009search}
Marti Hearst, \emph{Search user interfaces}, Cambridge university press, 2009.

\bibitem{hearst1995tilebars}
Marti~A Hearst, \emph{Tilebars: visualization of term distribution information
  in full text information access}, Chi, vol.~95, 1995, pp.~59--66.

\bibitem{heer2015predictive}
Jeffrey Heer, Joseph~M Hellerstein, and Sean Kandel, \emph{Predictive
  interaction for data transformation.}, CIDR, 2015.

\bibitem{hertzum1996browsing}
Morten Hertzum and Erik Fr{\o}kj{\ae}r, \emph{Browsing and querying in online
  documentation: a study of user interfaces and the interaction process}, ACM
  Transactions on Computer-Human Interaction (TOCHI) \textbf{3} (1996), no.~2,
  136--161.

\bibitem{higgins2008dcc}
Sarah Higgins, \emph{The dcc curation lifecycle model}, International journal
  of digital curation \textbf{3} (2008), no.~1.

\bibitem{hildreth1989general}
Charles~R Hildreth, \emph{General introduction; opac research: laying the
  groundwork for future opac design}, The online catalogue: developments and
  directions. London: Library Association (1989), 1--24.

\bibitem{howe2008big}
Doug Howe, Maria Costanzo, Petra Fey, Takashi Gojobori, Linda Hannick, Winston
  Hide, David~P Hill, Renate Kania, Mary Schaeffer, Susan St~Pierre, et~al.,
  \emph{Big data: The future of biocuration}, Nature \textbf{455} (2008),
  no.~7209, 47.

\bibitem{hu2014interactive}
Yuening Hu, Jordan Boyd-Graber, Brianna Satinoff, and Alison Smith,
  \emph{Interactive topic modeling}, Machine learning \textbf{95} (2014),
  no.~3, 423--469.

\bibitem{huang2003relevant}
Chien-Kang Huang, Lee-Feng Chien, and Yen-Jen Oyang, \emph{Relevant term
  suggestion in interactive web search based on contextual information in query
  session logs}, Journal of the American Society for Information Science and
  Technology \textbf{54} (2003), no.~7, 638--649.

\bibitem{hunt2001stratified}
Neville Hunt and Sidney Tyrrell, \emph{Stratified sampling}, Retrieved November
  \textbf{10} (2001), 2012.

\bibitem{igarashi2000speed}
Takeo Igarashi and Ken Hinckley, \emph{Speed-dependent automatic zooming for
  browsing large documents}, UIST, Citeseer, 2000, pp.~139--148.

\bibitem{jajuga2012classification}
Krzystof Jajuga, Andrzej Sokolowski, and Hans-Hermann Bock,
  \emph{Classification, clustering, and data analysis: recent advances and
  applications}, Springer Science \& Business Media, 2012.

\bibitem{jansen2007web}
Bernard~J Jansen, Amanda Spink, and Sherry Koshman, \emph{Web searcher
  interaction with the dogpile. com metasearch engine}, Journal of the American
  Society for Information Science and Technology \textbf{58} (2007), no.~5,
  744--755.

\bibitem{jansen2005temporal}
Bernard~J Jansen, Amanda Spink, and Jan Pedersen, \emph{A temporal comparison
  of altavista web searching}, Journal of the American Society for Information
  Science and Technology \textbf{56} (2005), no.~6, 559--570.

\bibitem{javed2010stack}
Waqas Javed and Niklas Elmqvist, \emph{Stack zooming for multi-focus
  interaction in time-series data visualization}, 2010 IEEE Pacific
  Visualization Symposium (PacificVis), IEEE, 2010, pp.~33--40.

\bibitem{javed2012polyzoom}
Waqas Javed, Sohaib Ghani, and Niklas Elmqvist, \emph{Polyzoom: multiscale and
  multifocus exploration in 2d visual spaces}, Proceedings of the SIGCHI
  Conference on Human Factors in Computing Systems, ACM, 2012, pp.~287--296.

\bibitem{jurafsky2000speech}
Dan Jurafsky, \emph{Speech \& language processing}, Pearson Education India,
  2000.

\bibitem{kamel2016instagram}
Maged Kamel~Boulos, Dean Giustini, and Steve Wheeler, \emph{Instagram and
  whatsapp in health and healthcare: An overview}, Future Internet \textbf{8}
  (2016), no.~3, 37.

\bibitem{karlgren2014semantic}
Jussi Karlgren, Martin Bohman, Ariel Ekgren, Gabriel Isheden, Emelie Kullmann,
  and David Nilsson, \emph{Semantic topology}, Proceedings of the 23rd ACM
  International Conference on Conference on Information and Knowledge
  Management, 2014, pp.~1939--1942.

\bibitem{kaufmann2006querix}
Esther Kaufmann, Abraham Bernstein, and Renato Zumstein, \emph{Querix: A
  natural language interface to query ontologies based on clarification
  dialogs}, 5th International Semantic Web Conference (ISWC 2006), Citeseer,
  2006, pp.~980--981.

\bibitem{kelly2005loquacious}
Diane Kelly, Vijay~Deepak Dollu, and Xin Fu, \emph{The loquacious user: a
  document-independent source of terms for query expansion}, Proceedings of the
  28th annual international ACM SIGIR conference on Research and development in
  information retrieval, 2005, pp.~457--464.

\bibitem{kidwell2008visualizing}
Paul Kidwell, Guy Lebanon, and William Cleveland, \emph{Visualizing incomplete
  and partially ranked data}, IEEE Transactions on visualization and computer
  graphics \textbf{14} (2008), no.~6, 1356--1363.

\bibitem{kim2017topiclens}
Minjeong Kim, Kyeongpil Kang, Deokgun Park, Jaegul Choo, and Niklas Elmqvist,
  \emph{Topiclens: Efficient multi-level visual topic exploration of
  large-scale document collections}, IEEE transactions on visualization and
  computer graphics \textbf{23} (2017), no.~1, 151--160.

\bibitem{BudgetMap}
Nam~Wook Kim and et~al., \emph{Budgetmap: Engaging taxpayers in the
  issue-driven classification of a government budget}, CSCW, 2016,
  pp.~1026--1037.

\bibitem{kimball2000data}
Ralph Kimball and Richard Merz, \emph{The data webhouse toolkit: Building the
  web-enabled data warehouse}, Industrial Management \& Data Systems (2000).

\bibitem{klouche2015designing}
Khalil Klouche, Tuukka Ruotsalo, Diogo Cabral, Salvatore Andolina, Andrea
  Bellucci, and Giulio Jacucci, \emph{Designing for exploratory search on touch
  devices}, Proceedings of the 33rd annual ACM conference on human factors in
  computing systems, ACM, 2015, pp.~4189--4198.

\bibitem{klouche2017visual}
Khalil Klouche, Tuukka Ruotsalo, Luana Micallef, Salvatore Andolina, and Giulio
  Jacucci, \emph{Visual re-ranking for multi-aspect information retrieval},
  Proceedings of the 2017 Conference on Conference Human Information
  Interaction and Retrieval, ACM, 2017, pp.~57--66.

\bibitem{koenemann1996case}
J{\"u}rgen Koenemann and Nicholas~J Belkin, \emph{A case for interaction: A
  study of interactive information retrieval behavior and effectiveness},
  Proceedings of the SIGCHI conference on human factors in computing systems,
  1996, pp.~205--212.

\bibitem{kohavi2009controlled}
Ron Kohavi, Roger Longbotham, Dan Sommerfield, and Randal~M Henne,
  \emph{Controlled experiments on the web: survey and practical guide}, Data
  mining and knowledge discovery \textbf{18} (2009), no.~1, 140--181.

\bibitem{krishnan2016towards}
Sanjay Krishnan and et~al., \emph{Towards reliable interactive data cleaning: a
  user survey and recommendations.}, HILDA@ SIGMOD, 2016, p.~9.

\bibitem{kukich1992techniques}
Karen Kukich, \emph{Techniques for automatically correcting words in text}, Acm
  Computing Surveys (CSUR) \textbf{24} (1992), no.~4, 377--439.

\bibitem{kules2008users}
Bill Kules and Ben Shneiderman, \emph{Users can change their web search
  tactics: Design guidelines for categorized overviews}, Information Processing
  \& Management \textbf{44} (2008), no.~2, 463--484.

\bibitem{kwak2010twitter}
Haewoon Kwak, Changhyun Lee, Hosung Park, and Sue Moon, \emph{What is twitter,
  a social network or a news media?}, Proceedings of the 19th international
  conference on World wide web, AcM, 2010, pp.~591--600.

\bibitem{lai2018robust}
Zhihui Lai, Dongmei Mo, Wai~Keung Wong, Yong Xu, Duoqian Miao, and David Zhang,
  \emph{Robust discriminant regression for feature extraction}, IEEE
  transactions on cybernetics \textbf{48} (2018), no.~8, 2472--2484.

\bibitem{w3c-dataextraction}
W3C Data~Extraction Language, \emph{https://www.w3.org/tr/data-extraction}.

\bibitem{lee2013real}
Kathy Lee and et~al., \emph{Real-time disease surveillance using twitter data:
  demonstration on flu and cancer}, SIGKDD, 2013.

\bibitem{li2006exploring}
Mu~Li, Yang Zhang, Muhua Zhu, and Ming Zhou, \emph{Exploring distributional
  similarity based models for query spelling correction}, Proceedings of the
  21st International Conference on Computational Linguistics and the 44th
  annual meeting of the Association for Computational Linguistics, Association
  for Computational Linguistics, 2006, pp.~1025--1032.

\bibitem{liu2010refining}
Bin Liu, Laura Chiticariu, Vivian Chu, HV~Jagadish, and Frederick Reiss,
  \emph{Refining information extraction rules using data provenance.}, IEEE
  Data Eng. Bull. \textbf{33} (2010), no.~3, 17--24.

\bibitem{liu2009interactive}
Shixia Liu, Michelle~X Zhou, Shimei Pan, Weihong Qian, Weijia Cai, and Xiaoxiao
  Lian, \emph{Interactive, topic-based visual text summarization and analysis},
  Proceedings of the 18th ACM conference on Information and knowledge
  management, ACM, 2009, pp.~543--552.

\bibitem{liu2014trading}
Yun-En Liu, Travis Mandel, Emma Brunskill, and Zoran Popovic, \emph{Trading off
  scientific knowledge and user learning with multi-armed bandits.}, EDM, 2014,
  pp.~161--168.

\bibitem{locher2016starting}
Anita~E Locher, \emph{Starting points for lowering the barrier to spatial data
  preservation}, Journal of Map \& Geography Libraries \textbf{12} (2016),
  no.~1, 28--51.

\bibitem{lohr2012age}
Steve Lohr, \emph{The age of big data}, New York Times \textbf{11} (2012).

\bibitem{lopez2012poweraqua}
Vanessa Lopez, Miriam Fern{\'a}ndez, Enrico Motta, and Nico Stieler,
  \emph{Poweraqua: Supporting users in querying and exploring the semantic
  web}, Semantic Web \textbf{3} (2012), no.~3, 249--265.

\bibitem{ma2008learning}
Hao Ma, Haixuan Yang, Irwin King, and Michael~R Lyu, \emph{Learning latent
  semantic relations from clickthrough data for query suggestion}, Proceedings
  of the 17th ACM conference on Information and knowledge management, 2008,
  pp.~709--718.

\bibitem{tagging}
Zakaria Maamar, Sherif Sakr, Ahmed Barnawi, and Seyed{-}Mehdi{-}Reza Beheshti,
  \emph{A framework of enriching business processes life-cycle with tagging
  information}, Databases Theory and Applications - 26th Australasian Database
  Conference, {ADC} 2015, Melbourne, VIC, Australia, June 4-7, 2015.
  Proceedings, Lecture Notes in Computer Science, vol. 9093, Springer, 2015,
  pp.~309--313.

\bibitem{macmillan2014data}
Don MacMillan, \emph{Data sharing and discovery: What librarians need to know},
  The Journal of Academic Librarianship \textbf{40} (2014), no.~5, 541--549.

\bibitem{manning-EtAl:2014:P14-5}
Christopher~D. Manning, Mihai Surdeanu, John Bauer, Jenny Finkel, Steven~J.
  Bethard, and David McClosky, \emph{The {Stanford} {CoreNLP} natural language
  processing toolkit}, Association for Computational Linguistics (ACL) System
  Demonstrations, 2014, pp.~55--60.

\bibitem{marchionini2006exploratory}
Gary Marchionini, \emph{Exploratory search: from finding to understanding},
  Communications of the ACM \textbf{49} (2006), no.~4, 41--46.

\bibitem{marchionini1988finding}
Gary Marchionini and Ben Shneiderman, \emph{Finding facts vs. browsing
  knowledge in hypertext systems}, Computer \textbf{21} (1988), no.~1, 70--80.

\bibitem{marietto2013artificial}
Maria das Gra{\c{c}}as~Bruno Marietto, Rafael~Varago de~Aguiar, Gislene
  de~Oliveira Barbosa, Wagner~Tanaka Botelho, Edson Pimentel, Robson dos~Santos
  Fran{\c{c}}a, and Vera~L{\'u}cia da~Silva, \emph{Artificial intelligence
  markup language: A brief tutorial}, arXiv preprint arXiv:1307.3091 (2013).

\bibitem{marquez2013coreference}
Llu{\'\i}s M{\`a}rquez, Marta Recasens, and Emili Sapena, \emph{Coreference
  resolution: an empirical study based on semeval-2010 shared task 1}, Language
  resources and evaluation \textbf{47} (2013), no.~3, 661--694.

\bibitem{massis2011serendipitous}
Bruce~E Massis, \emph{“serendipitous” browsing versus library space}, New
  Library World \textbf{112} (2011), no.~3/4, 178--182.

\bibitem{mayfield2009cross}
James Mayfield, David Alexander, Bonnie~J Dorr, Jason Eisner, Tamer Elsayed,
  Tim Finin, Clayton Fink, Marjorie Freedman, Nikesh Garera, Paul McNamee,
  et~al., \emph{Cross-document coreference resolution: A key technology for
  learning by reading.}, AAAI Spring Symposium: Learning by Reading and
  Learning to Read, vol.~9, 2009, pp.~65--70.

\bibitem{mikolov2013distributed}
Tomas Mikolov, Ilya Sutskever, Kai Chen, Greg~S Corrado, and Jeff Dean,
  \emph{Distributed representations of words and phrases and their
  compositionality}, Advances in neural information processing systems, 2013,
  pp.~3111--3119.

\bibitem{miller2014big}
Ren{\'e}e~J Miller, \emph{Big data curation.}, COMAD, 2014, p.~4.

\bibitem{milo2016rudolf}
Tova Milo, Slava Novgorodov, and Wang-Chiew Tan, \emph{Rudolf: interactive rule
  refinement system for fraud detection}, Proceedings of the VLDB Endowment
  \textbf{9} (2016), no.~13, 1465--1468.

\bibitem{milo2018interactive}
\bysame, \emph{Interactive rule refinement for fraud detection}, EDBT, 2018.

\bibitem{mirza2016data}
Ali Mirza and Imran Siddiqi, \emph{Data level conflicts resolution for
  multi-sources heterogeneous databases}, 2016 Sixth International Conference
  on Innovative Computing Technology (INTECH), IEEE, 2016, pp.~36--40.

\bibitem{mohr2015data}
Alicia~Hofelich Mohr, Josh Bishoff, Carolyn Bishoff, Steven Braun, Christine
  Storino, and Lisa~R Johnston, \emph{When data is a dirty word: A survey to
  understand data management needs across diverse research disciplines},
  Bulletin of the Association for Information Science and Technology
  \textbf{42} (2015), no.~1, 51--53.

\bibitem{ShareInsights}
Deshpande Mukund, \emph{Shareinsights an unified approach to full-stack data
  processing}, SIGMOD, 2015, p.~9.

\bibitem{navigli2012babelnet}
Roberto Navigli and Simone~Paolo Ponzetto, \emph{Babelnet: The automatic
  construction, evaluation and application of a wide-coverage multilingual
  semantic network}, Artificial Intelligence \textbf{193} (2012), 217--250.

\bibitem{nguyen2006novel}
Tien Nguyen and Jin Zhang, \emph{A novel visualization model for web search
  results}, IEEE transactions on visualization and computer graphics
  \textbf{12} (2006), no.~5, 981--988.

\bibitem{niwattanakul2013using}
Suphakit Niwattanakul, Jatsada Singthongchai, Ekkachai Naenudorn, and
  Supachanun Wanapu, \emph{Using of jaccard coefficient for keywords
  similarity}, Proceedings of the international multiconference of engineers
  and computer scientists, vol.~1, 2013, pp.~380--384.

\bibitem{norouzi2012hamming}
Mohammad Norouzi, David~J Fleet, and Russ~R Salakhutdinov, \emph{Hamming
  distance metric learning}, Advances in neural information processing systems,
  2012, pp.~1061--1069.

\bibitem{olendorf2012beyond}
Robert Olendorf and Steve Koch, \emph{Beyond the low hanging fruit: Archving
  complex data and data services at university of new mexico}, Journal of
  Digital Information (2012).

\bibitem{pasupuleti2015data}
Pradeep Pasupuleti and Beulah~Salome Purra, \emph{Data lake development with
  big data}, Packt Publishing Ltd, 2015.

\bibitem{patel2011using}
Kayur Patel, Steven~M Drucker, James Fogarty, Ashish Kapoor, and Desney~S Tan,
  \emph{Using multiple models to understand data}, IJCAI
  Proceedings-International Joint Conference on Artificial Intelligence,
  vol.~22, 2011, p.~1723.

\bibitem{patterson2001predicting}
Emily~S Patterson, Emilie~M Roth, and David~D Woods, \emph{Predicting
  vulnerabilities in computer-supported inferential analysis under data
  overload}, Cognition, Technology \& Work \textbf{3} (2001), no.~4, 224--237.

\bibitem{peltonen2017topic}
Jaakko Peltonen, Kseniia Belorustceva, and Tuukka Ruotsalo,
  \emph{Topic-relevance map: Visualization for improving search result
  comprehension}, Proceedings of the 22nd International Conference on
  Intelligent User Interfaces, ACM, 2017, pp.~611--622.

\bibitem{pennington2014glove}
Jeffrey Pennington, Richard Socher, and Christopher~D Manning, \emph{Glove:
  Global vectors for word representation}, Proceedings of the 2014 conference
  on empirical methods in natural language processing (EMNLP), 2014,
  pp.~1532--1543.

\bibitem{pham2009time}
Quang-Khai Pham, Guillaume Raschia, Noureddine Mouaddib, Regis Saint-Paul, and
  Boualem Benatallah, \emph{Time sequence summarization to scale up
  chronology-dependent applications}, Proceedings of the 18th ACM conference on
  Information and knowledge management, ACM, 2009, pp.~1137--1146.

\bibitem{pietriga2008sigma}
Emmanuel Pietriga and Caroline Appert, \emph{Sigma lenses: focus-context
  transitions combining space, time and translucence}, Proceedings of the
  SIGCHI Conference on Human Factors in Computing Systems, ACM, 2008,
  pp.~1343--1352.

\bibitem{pirolli2005sensemaking}
Peter Pirolli and Stuart Card, \emph{The sensemaking process and leverage
  points for analyst technology as identified through cognitive task analysis},
  Proceedings of international conference on intelligence analysis, vol.~5,
  McLean, VA, USA, 2005, pp.~2--4.

\bibitem{pu2015topic}
Xiaojia Pu, Rong Jin, Gangshan Wu, Dingyi Han, and Gui-Rong Xue, \emph{Topic
  modeling in semantic space with keywords}, Proceedings of the 24th ACM
  International on Conference on Information and Knowledge Management, 2015,
  pp.~1141--1150.

\bibitem{rajaonarivo2017inline}
Landy Rajaonarivo, Matthieu Courgeon, Eric Maisel, and Pierre De~Loor,
  \emph{Inline co-evolution between users and information presentation for data
  exploration}, Proceedings of the 22nd International Conference on Intelligent
  User Interfaces, ACM, 2017, pp.~215--219.

\bibitem{ratner2017snorkel}
Alexander Ratner, Stephen~H Bach, Henry Ehrenberg, Jason Fries, Sen Wu, and
  Christopher R{\'e}, \emph{Snorkel: Rapid training data creation with weak
  supervision}, arXiv preprint arXiv:1711.10160 (2017).

\bibitem{ratner2017snorkel2}
Alexander~J Ratner, Stephen~H Bach, Henry~R Ehrenberg, and Chris R{\'e},
  \emph{Snorkel: Fast training set generation for information extraction},
  Proceedings of the 2017 ACM International Conference on Management of Data,
  ACM, 2017, pp.~1683--1686.

\bibitem{10.1093/bioinformatics/btw579}
Carlo Ravagli, Francois Pognan, and Philippe Marc, \emph{{OntoBrowser: a
  collaborative tool for curation of ontologies by subject matter experts}},
  Bioinformatics \textbf{33} (2016), no.~1, 148--149.

\bibitem{rebele2016yago}
Thomas Rebele, Fabian Suchanek, Johannes Hoffart, Joanna Biega, Erdal Kuzey,
  and Gerhard Weikum, \emph{Yago: A multilingual knowledge base from wikipedia,
  wordnet, and geonames}, International Semantic Web Conference, Springer,
  2016, pp.~177--185.

\bibitem{ritter2011named}
Alan Ritter, Sam Clark, Oren Etzioni, et~al., \emph{Named entity recognition in
  tweets: an experimental study}, Proceedings of the conference on empirical
  methods in natural language processing, Association for Computational
  Linguistics, 2011, pp.~1524--1534.

\bibitem{rocchio1971relevance}
Joseph~John Rocchio, \emph{Relevance feedback in information retrieval}, The
  SMART retrieval system: experiments in automatic document processing,
  313--323.

\bibitem{ruder2011suicide}
Thomas~D Ruder, Gary~M Hatch, Garyfalia Ampanozi, Michael~J Thali, and Nadja
  Fischer, \emph{Suicide announcement on facebook}, Crisis (2011).

\bibitem{ruotsalo2018interactive}
Tuukka Ruotsalo, Jaakko Peltonen, Manuel~JA Eugster, Dorota G{\l}owacka, Patrik
  Flor{\'e}en, Petri Myllym{\"a}ki, Giulio Jacucci, and Samuel Kaski,
  \emph{Interactive intent modeling for exploratory search}, ACM Transactions
  on Information Systems (TOIS) \textbf{36} (2018), no.~4, 44.

\bibitem{russell2006being}
Daniel~M Russell, Malcolm Slaney, Yan Qu, and Mave Houston, \emph{Being
  literate with large document collections: Observational studies and cost
  structure tradeoffs}, Proceedings of the 39th Annual Hawaii International
  Conference on System Sciences (HICSS'06), vol.~3, IEEE, 2006, pp.~55--55.

\bibitem{russell1993cost}
Daniel~M Russell, Mark~J Stefik, Peter Pirolli, and Stuart~K Card, \emph{The
  cost structure of sensemaking}, Proceedings of the INTERACT'93 and CHI'93
  conference on Human factors in computing systems, 1993, pp.~269--276.

\bibitem{russo2017tutorial}
D~Russo, B~Van~Roy, A~Kazerouni, and I~Osband, \emph{A tutorial on thompson
  sampling}, arXiv preprint arXiv:1707.02038 (2017).

\bibitem{russom2011big}
Philip Russom et~al., \emph{Big data analytics}, TDWI Best Practices Report,
  Fourth Quarter (2011), 1--35.

\bibitem{ruthven2003survey}
Ian Ruthven and Mounia Lalmas, \emph{A survey on the use of relevance feedback
  for information access systems}, The Knowledge Engineering Review \textbf{18}
  (2003), no.~2, 95--145.

\bibitem{saint2005general}
R{\'e}gis Saint-Paul, Guillaume Raschia, and Noureddine Mouaddib, \emph{General
  purpose database summarization}, Proceedings of the 31st international
  conference on Very large data bases, VLDB Endowment, 2005, pp.~733--744.

\bibitem{salton1985advanced}
Gerard Salton, Edward~A Fox, and Ellen Voorhees, \emph{Advanced feedback
  methods in information retrieval}, Journal of the American Society for
  Information Science \textbf{36} (1985), no.~3, 200--210.

\bibitem{sarkar2011community}
Somwrita Sarkar and Andy Dong, \emph{Community detection in graphs using
  singular value decomposition}, Physical Review E \textbf{83} (2011), no.~4,
  046114.

\bibitem{sarkar2009design}
Somwrita Sarkar, Andy Dong, and John~S Gero, \emph{Design optimization problem
  reformulation using singular value decomposition}, Journal of Mechanical
  Design \textbf{131} (2009), no.~8.

\bibitem{sauro2012standardized}
Jeff Sauro and James~R Lewis, \emph{Standardized usability questionnaires},
  Quantifying the user experience (2012), 185--240.

\bibitem{iCOP}
Francesco Schiliro, Amin Beheshti, Samira Ghodratnama, Farhad Amouzgar, Boualem
  Benatallah, Jian Yang, Quan~Z. Sheng, Fabio Casati, and Hamid~Reza
  Motahari{-}Nezhad, \emph{icop: Iot-enabled policing processes},
  Service-Oriented Computing - {ICSOC} 2018 Workshops - ADMS, ASOCA, ISYyCC,
  CloTS, DDBS, and NLS4IoT, Hangzhou, China, November 12-15, 2018, Revised
  Selected Papers, Lecture Notes in Computer Science, vol. 11434, Springer,
  2018, pp.~447--452.

\bibitem{sebastiani2002machine}
Fabrizio Sebastiani, \emph{Machine learning in automated text categorization},
  ACM computing surveys (CSUR) \textbf{34} (2002), no.~1, 1--47.

\bibitem{sellam2015semi}
Thibault Sellam, Emmanuel M{\"u}ller, and Martin Kersten, \emph{Semi-automated
  exploration of data warehouses}, Proceedings of the 24th ACM International on
  Conference on Information and Knowledge Management, 2015, pp.~1321--1330.

\bibitem{iRecruit}
Usman Shahbaz, Amin Beheshti, Sadegh Nobari, Qiang Qu, Hye{-}Young Paik, and
  Mehregan Mahdavi, \emph{irecruit: Towards automating the recruitment
  process}, Service Research and Innovation - 7th Australian Symposium, {ASSRI}
  2018, Sydney, NSW, Australia, September 6, 2018, and Wollongong, NSW,
  Australia, December 14, 2018, Revised Selected Papers, Lecture Notes in
  Business Information Processing, vol. 367, Springer, 2018, pp.~139--152.

\bibitem{shani2013displaying}
Guy Shani and Noam Tractinsky, \emph{Displaying relevance scores for search
  results}, Proceedings of the 36th international ACM SIGIR conference on
  Research and development in information retrieval, 2013, pp.~901--904.

\bibitem{shi2012rankexplorer}
Conglei Shi, Weiwei Cui, Shixia Liu, Panpan Xu, Wei Chen, and Huamin Qu,
  \emph{Rankexplorer: Visualization of ranking changes in large time series
  data}, IEEE Transactions on Visualization and Computer Graphics \textbf{18}
  (2012), no.~12, 2669--2678.

\bibitem{singh2018learn}
Vikram Singh and Ajay Singh, \emph{Learn-as-you-go: Feedback-driven result
  ranking and query refinement for interactive data exploration}, Procedia
  Computer Science \textbf{125} (2018), 550--559.

\bibitem{singhal2012introducing}
Amit Singhal, \emph{Introducing the knowledge graph: things, not strings},
  Official google blog \textbf{16} (2012).

\bibitem{song2014towards}
Tianhong Song, Sven kohler, Bertram Ludscher, James Hanken, Maureen Kelly,
  David Lowery, James~A Macklin, Paul~J Morris, and Robert~A Morris,
  \emph{Towards automated design, analysis and optimization of declarative
  curation workflows},  (2014).

\bibitem{song2012query}
Yang Song, Dengyong Zhou, and Li-wei He, \emph{Query suggestion by constructing
  term-transition graphs}, Proceedings of the fifth ACM international
  conference on Web search and data mining, 2012, pp.~353--362.

\bibitem{spasic2010medication}
Irena Spasi{\'c}, Farzaneh Sarafraz, John~A Keane, and Goran Nenadi{\'c},
  \emph{Medication information extraction with linguistic pattern matching and
  semantic rules}, Journal of the American Medical Informatics Association
  \textbf{17} (2010), no.~5, 532--535.

\bibitem{speer2017conceptnet}
Robyn Speer, Joshua Chin, and Catherine Havasi, \emph{Conceptnet 5.5: An open
  multilingual graph of general knowledge}, Thirty-First AAAI Conference on
  Artificial Intelligence, 2017.

\bibitem{stoffel2015feature}
Florian Stoffel, Lucie Flekova, Daniela Oelke, Iryna Gurevych, and Daniel~A
  Keim, \emph{Feature-based visual exploration of text classification},
  Symposium on Visualization in Data Science at IEEE VIS, 2015.

\bibitem{stonebraker2013data}
Michael Stonebraker and et~al., \emph{Data curation at scale: The data tamer
  system.}, CIDR, 2013.

\bibitem{sultanum2019doccurate}
Nicole Sultanum, Devin Singh, Michael Brudno, and Fanny Chevalier,
  \emph{Doccurate: A curation-based approach for clinical text visualization},
  IEEE transactions on visualization and computer graphics \textbf{25} (2019),
  no.~1, 142--151.

\bibitem{sun2014chimera}
Chong Sun, Narasimhan Rampalli, Frank Yang, and AnHai Doan, \emph{Chimera:
  Large-scale classification using machine learning, rules, and crowdsourcing},
  VLDB Endowment \textbf{7} (2014), no.~13, 1529--1540.

\bibitem{casewall}
Yu{-}Jen~John Sun, Moshe~Chai Barukh, Boualem Benatallah, and
  Seyed{-}Mehdi{-}Reza Beheshti, \emph{Scalable saas-based process
  customization with casewalls}, Service-Oriented Computing - 13th
  International Conference, {ICSOC} 2015, Goa, India, November 16-19, 2015,
  Proceedings, Lecture Notes in Computer Science, vol. 9435, Springer, 2015,
  pp.~218--233.

\bibitem{tabebordbar2018adaptive}
Alireza Tabebordbar and Amin Beheshti, \emph{Adaptive rule monitoring system},
  Proceedings of the 1st International Workshop on Software Engineering for
  Cognitive Services, SE4COG@ICSE 2018, Gothenburg, Sweden, May 28-2, 2018,
  {ACM}, 2018, pp.~45--51.

\bibitem{tabebordbar2019conceptmap}
Alireza Tabebordbar, Amin Beheshti, and Boualem Benatallah, \emph{Conceptmap: A
  conceptual approach for formulating user preferences in large information
  spaces}, International Conference on Web Information Systems Engineering,
  Springer, 2019, pp.~779--794.

\bibitem{tabebordbar2020feature}
Alireza Tabebordbar, Amin Beheshti, Boualem Benatallah, and Moshe~Chai Barukh,
  \emph{Feature-based and adaptive rule adaptation in dynamic environments},
  Data Science and Engineering (2020), 1--17.

\bibitem{tatu2010automated}
Andrada Tatu, Georgia Albuquerque, Martin Eisemann, Peter Bak, Holger Theisel,
  Marcus Magnor, and Daniel Keim, \emph{Automated analytical methods to support
  visual exploration of high-dimensional data}, IEEE Transactions on
  Visualization and Computer Graphics \textbf{17} (2010), no.~5, 584--597.

\bibitem{tene2012big}
Omer Tene and Jules Polonetsky, \emph{Big data for all: Privacy and user
  control in the age of analytics}, Nw. J. Tech. \& Intell. Prop. \textbf{11}
  (2012), xxvii.

\bibitem{terrizzano2015data}
Ignacio Terrizzano, Peter~M Schwarz, Mary Roth, and John~E Colino, \emph{Data
  wrangling: The challenging yourney from the wild to the lake.}, CIDR, 2015.

\bibitem{towns2014xsede}
John Towns, Timothy Cockerill, Maytal Dahan, Ian Foster, Kelly Gaither, Andrew
  Grimshaw, Victor Hazlewood, Scott Lathrop, Dave Lifka, Gregory~D Peterson,
  et~al., \emph{Xsede: accelerating scientific discovery}, Computing in Science
  \& Engineering \textbf{16} (2014), no.~5, 62--74.

\bibitem{troncy2016linking}
Rapha{\"e}l Troncy, \emph{Linking entities for enriching and structuring social
  media content}, WWW, 2016, pp.~597--597.

\bibitem{uzuner2010extracting}
{\"O}zlem Uzuner, Imre Solti, and Eithon Cadag, \emph{Extracting medication
  information from clinical text}, Journal of the American Medical Informatics
  Association \textbf{17} (2010), no.~5, 514--518.

\bibitem{vahabi2013orthogonal}
Hossein Vahabi, Margareta Ackerman, David Loker, Ricardo Baeza-Yates, and
  Alejandro Lopez-Ortiz, \emph{Orthogonal query recommendation}, Proceedings of
  the 7th ACM conference on Recommender systems, 2013, pp.~33--40.

\bibitem{valenzuela2016odin}
Marco~A Valenzuela-Escarcega, Gus Hahn-Powell, and Mihai Surdeanu,
  \emph{Odin’s runes: A rule language for information extraction},
  Proceedings of the Tenth International Conference on Language Resources and
  Evaluation (LREC'16), 2016, pp.~322--329.

\bibitem{van2014datafication}
Jos{\'e} Van~Dijck, \emph{Datafication, dataism and dataveillance: Big data
  between scientific paradigm and ideology}, Surveillance \& Society
  \textbf{12} (2014), no.~2, 197--208.

\bibitem{veeramachaneni2014towards}
Kalyan Veeramachaneni, Una-May O'Reilly, and Colin Taylor, \emph{Towards
  feature engineering at scale for data from massive open online courses},
  arXiv preprint arXiv:1407.5238 (2014).

\bibitem{volkovs2014continuous}
Maksims Volkovs, Fei Chiang, Jaroslaw Szlichta, and Ren{\'e}e~J Miller,
  \emph{Continuous data cleaning}, Data Engineering (ICDE), 2014 IEEE 30th
  International Conference on, IEEE, 2014, pp.~244--255.

\bibitem{vrandevcic2014wikidata}
Denny Vrande{\v{c}}i{\'c} and Markus Kr{\"o}tzsch, \emph{Wikidata: a free
  collaborative knowledgebase}, Communications of the ACM \textbf{57} (2014),
  no.~10, 78--85.

\bibitem{wall2018podium}
Emily Wall, Subhajit Das, Ravish Chawla, Bharath Kalidindi, Eli~T Brown, and
  Alex Endert, \emph{Podium: Ranking data using mixed-initiative visual
  analytics}, IEEE transactions on visualization and computer graphics
  \textbf{24} (2018), no.~1, 288--297.

\bibitem{wang2011fast}
Jiannan Wang, Guoliang Li, and Jianhua Fe, \emph{Fast-join: An efficient method
  for fuzzy token matching based string similarity join}, 2011 IEEE 27th
  International Conference on Data Engineering, IEEE, 2011, pp.~458--469.

\bibitem{wang2016linked}
Suhang Wang, Jiliang Tang, Charu Aggarwal, and Huan Liu, \emph{Linked document
  embedding for classification}, Proceedings of the 25th ACM international on
  conference on information and knowledge management, 2016, pp.~115--124.

\bibitem{white2007investigating}
Ryen~W White and Dan Morris, \emph{Investigating the querying and browsing
  behavior of advanced search engine users}, Proceedings of the 30th annual
  international ACM SIGIR conference on Research and development in information
  retrieval, 2007, pp.~255--262.

\bibitem{wiatowski2018mathematical}
Thomas Wiatowski and Helmut B{\"o}lcskei, \emph{A mathematical theory of deep
  convolutional neural networks for feature extraction}, IEEE Transactions on
  Information Theory \textbf{64} (2018), no.~3, 1845--1866.

\bibitem{wick2009entity}
M.~Wick, A.~Culotta, K.~Rohanimanesh, and A.~McCallum, \emph{An entity based
  model for coreference resolution}, SDM, 2009.

\bibitem{williams2016axis}
Joseph~Jay Williams, Juho Kim, Anna Rafferty, Samuel Maldonado, Krzysztof~Z
  Gajos, Walter~S Lasecki, and Neil Heffernan, \emph{Axis: Generating
  explanations at scale with learnersourcing and machine learning}, ACM
  Conference on Learning@ Scale, ACM, 2016, pp.~379--388.

\bibitem{xie2017automatic}
Jun Xie, Chong Sun, Fan Yang, and Narasimhan Rampalli, \emph{Automatic rule
  coaching}, September~5 2017, US Patent 9,754,208.

\bibitem{zamanirad2017programming}
Shayan Zamanirad, Boualem Benatallah, Moshe Chai~Barukh, Fabio Casati, and
  Carlos Rodriguez, \emph{Programming bots by synthesizing natural language
  expressions into api invocations}, Proceedings of the 32nd IEEE/ACM
  International Conference on Automated Software Engineering, IEEE Press, 2017,
  pp.~832--837.

\end{thebibliography}

\end{document}